\documentclass[traditabstract,longauth]{aa}
\usepackage{graphicx}
\usepackage{amsmath}     
\usepackage{amssymb}
\usepackage{mathrsfs}
\usepackage{xspace}
\usepackage{placeins}

\usepackage{subcaption}  
\usepackage{changepage}  
\usepackage{booktabs}    
\usepackage[varg]{txfonts} 
\usepackage{textcomp}    
\usepackage{gensymb}
\usepackage[bookmarks, colorlinks, breaklinks]{hyperref}
\hypersetup{
  linkcolor=blue,
  citecolor=blue,
  filecolor=blue,
  urlcolor=blue
}
\usepackage{natbib}

\newcommand{\simgt}{\lower.5ex\hbox{$\; \buildrel > \over \sim \;$}}
\newcommand{\simlt}{\lower.5ex\hbox{$\; \buildrel < \over \sim \;$}}

\newcommand{\percent}{\ensuremath{\%}}
\newcommand{\Om}{\Omega_\mathrm{m}}
\newcommand{\OL}{\Omega_\Lambda}
\newcommand{\Ob}{\Omega_\mathrm{b}}
\newcommand{\zs}{z_s}
\newcommand{\zl}{z_l}
\newcommand{\zlmed}{z_{l,\mathrm{med}}}
\newcommand{\zref}{z_\mathrm{ref}}
\newcommand{\zph}{z_\mathrm{ph}}
\newcommand{\Mwl}{M_\mathrm{WL}}
\newcommand{\Mtrue}{M_\mathrm{true}}
\newcommand{\Msz}{M_\mathrm{SZ}}
\newcommand{\cwl}{c_{200,\mathrm{WL}}}

\newcommand{\bsz}{b_\mathrm{SZ}}
\newcommand{\ngal}{n_\mathrm{g}}
\newcommand{\Nbin}{N}
\newcommand{\Ncl}{N_\mathrm{cl}}
\newcommand{\SigmaCrit}{\Sigma_\mathrm{cr}}
\newcommand{\Mpch}{\,h^{-1}\,\mathrm{cMpc}}

\newcommand{\Msunh}{\,h^{-1}\,M_\odot}
\newcommand{\Msun}{\,M_\odot}
\newcommand{\CBI}{$C_\mathrm{BI}$\xspace}
\newcommand{\SBI}{$S_\mathrm{BI}$\xspace}
\newcommand{\smem}{$\mathrm{S}_\mathrm{mem}$\xspace}
\newcommand{\LCDM}{$\Lambda$CDM\xspace}
\newcommand{\amalgam}{AMALGAM\xspace}
\newcommand{\lira}{\texttt{LIRA}\xspace}
\newcommand{\WL}{WL\xspace}
\newcommand{\WMAP}{WMAP\xspace}
\newcommand{\Planck}{\textit{Planck}\xspace}
\newcommand{\BRz}{$B_\mathrm{J} R_\mathrm{C} z^+$\xspace}

\def\btheta{\mbox{\boldmath $\theta$}} 

\def\bp{\mbox{\boldmath $p$}}

\def\le{\leqslant}

\graphicspath{{./}}

\begin{document}

\title{CHEX-MATE: AMALGAM weak-lensing analysis of 41 Planck Sunyaev--Zel'dovich-selected galaxy clusters}
\titlerunning{CHEX-MATE: AMALGAM weak-lensing analysis of 41 Planck clusters}

\institute{Academia Sinica Institute of Astronomy and Astrophysics (ASIAA),
           No.~1, Section~4, Roosevelt Road, Taipei 106319, Taiwan\label{asiaa}
           \and Laboratoire d'Astrophysique de Marseille, Aix-Marseille Univ., CNRS, CNES, Marseille, France\label{lam}
           \and Institut d'Astrophysique de Paris, UMR 7095, CNRS \& Sorbonne Universit\'e, 98 bis Boulevard Arago, 75014 Paris, France\label{iap}
           \and INAF -- Osservatorio di Astrofisica e Scienza dello Spazio di Bologna, via Piero Gobetti 93/3, I-40129 Bologna, Italy\label{oas_bologna}
           \and INFN, Sezione di Bologna, viale Berti Pichat 6/2, I-40127 Bologna, Italy\label{infn_bologna}
           \and Department of Physical Science, Hiroshima University, 1-3-1 Kagamiyama, Higashi-Hiroshima, Hiroshima 739-8526, Japan\label{hiroshima_phys}
           \and Hiroshima Astrophysical Science Center, Hiroshima University, 1-3-1 Kagamiyama, Higashi-Hiroshima, Hiroshima 739-8526, Japan\label{hiroshima_asc}
           \and Core Research for Energetic Universe, Hiroshima University, 1-3-1 Kagamiyama, Higashi-Hiroshima, Hiroshima 739-8526, Japan\label{hiroshima_core}
           \and Universit\'e Paris-Saclay, Universit\'e Paris Cit\'e, CEA, CNRS, AIM, 91191, Gif-sur-Yvette, France\label{aim_cea}
           \and INAF -- IASF Milano, via A. Corti 12, I-20133 Milano, Italy\label{iasf_milano}
           \and INAF -- Osservatorio Astronomico di Trieste, via G. Tiepolo 11, I-34143 Trieste, Italy\label{oats}
           \and Department of Physics, University of Michigan, Ann Arbor, MI 48109, USA\label{umich}
           \and Dipartimento di Fisica G. Occhialini, Universit\`a di Milano-Bicocca, Piazza della Scienza 3, I-20126 Milano, Italy\label{bicocca}
           \and California Institute of Technology, 1200 East California Boulevard, Pasadena, CA 91125, USA\label{caltech}
           \and Jodrell Bank Centre for Astrophysics, Department of Physics and Astronomy, The University of Manchester, Manchester M13 9PL, UK\label{manchester}
           \and Department of Physics, Korea Advanced Institute of Science and Technology (KAIST), 291 Daehak-ro, Yuseong-gu, Daejeon 34141, Republic of Korea\label{kaist}
           \and School of Physics \& Astronomy, University of Nottingham, Nottingham, UK\label{nottingham}
           \and IRAP, CNRS, Universit\'e de Toulouse, CNES, Toulouse, France\label{irap}
           \and HH Wills Physics Laboratory, University of Bristol, Bristol, UK\label{bristol}
           \and INAF -- Osservatorio Astronomico di Padova, via dell'Osservatorio 5, I-35122 Padova, Italy\label{inaf_padova}
          }

\author{Keiichi Umetsu \inst{\ref{asiaa}}\thanks{Corresponding author: keiichi@asiaa.sinica.edu.tw}
        \and Raphael Gavazzi \inst{\ref{lam},\ref{iap}}
        \and Mauro Sereno \inst{\ref{oas_bologna},\ref{infn_bologna}}
        \and Nobuhiro Okabe \inst{\ref{hiroshima_phys},\ref{hiroshima_asc},\ref{hiroshima_core}}
        \and Emmanuel Bertin \inst{\ref{aim_cea},\ref{iap}} 
        \and Gianluca Castignani \inst{\ref{oas_bologna}} 
        \and Stefano Ettori \inst{\ref{oas_bologna},\ref{infn_bologna}} 
        \and Fabio Gastaldello \inst{\ref{iasf_milano}} 
        \and Carlo Giocoli \inst{\ref{oas_bologna},\ref{infn_bologna}} 
        \and Scott T. Kay \inst{\ref{manchester}} 
        \and Junhan Kim \inst{\ref{kaist}} 
        \and Maggie Lieu \inst{\ref{nottingham}}
        \and Lorenzo Lovisari \inst{\ref{iasf_milano}} 
        \and Ben J. Maughan \inst{\ref{bristol}} 
        \and Mario Nonino \inst{\ref{oats}}\thanks{Deceased} 
        \and Lorenzo Pizzuti \inst{\ref{bicocca},\ref{oats}} 
        \and Etienne Pointecouteau \inst{\ref{irap}} 
	\and Gabriel W. Pratt \inst{\ref{aim_cea}}
        \and Mario Radovich \inst{\ref{inaf_padova}} 
        \and Elena Rasia \inst{\ref{oats},\ref{umich}} 
        \and Mariachiara Rossetti \inst{\ref{iasf_milano}} 
        \and Harshda Saxena \inst{\ref{caltech}} 
        \and Jack Sayers \inst{\ref{caltech}} 
        }

\authorrunning{K. Umetsu et al.}

\date{}

\abstract
{We  present a weak-lensing shear analysis of 41 \Planck Sunyaev--Zel'dovich (SZ)-selected galaxy clusters at $0.11 \le z \le 0.55$ from the CHEX-MATE sample, using wide-field Subaru/Suprime-Cam and CFHT/MegaPrime imaging from the \amalgam project. The azimuthally averaged weak-lensing signal, measured around the X-ray peak of each cluster, has a median signal-to-noise ratio of $6.5$ per cluster. The $45^\circ$-rotated component has a median signal-to-noise ratio of $-0.1$ and ranges from $-1.8$ to $+1.8$, consistent with zero. We modelled the excess surface mass density profile of each cluster with a Navarro--Frenk--White profile to infer the weak-lensing mass and concentration constraints. The estimated total systematic uncertainty in the weak-lensing mass calibration is $8\percent$. Using a hierarchical Bayesian framework, we then derived weak-lensing-calibrated scaling relations for the halo concentration, $c_{200}$, as a function of $M_{200}$ and redshift, and for the \Planck SZ mass proxy, $\Msz$, as a function of $M_{500}$ and redshift. In both regressions, we modelled the latent true-mass distribution of the selected sample and accounted for weak-lensing modelling biases and residual calibration uncertainty. For the $\Msz$--$M_{500}$--$z$ relation, we also included an effective treatment of the CHEX-MATE tier selection. At the pivot mass $M_{200}=10^{15}\Msun$ and redshift $z=0.25$, we find $c_{200}=3.53\pm0.71$ with an intrinsic scatter of $0.22\pm0.04$~dex. The inferred normalisation and scatter are consistent with recent \LCDM predictions for massive haloes, with no significant mass or redshift dependence over the probed range. For the \Planck mass proxy, our baseline regression yields $\Msz/M_{500}=0.83\pm0.09$ at $M_{500}=7\times10^{14}\Msun$ and $z=0.25$, with an intrinsic scatter of $0.10\pm0.02$~dex. A restricted model with the mass slope fixed to unity and no redshift evolution yields $1-\bsz=0.72\pm0.11$. We also provide weak-lensing-calibrated posterior estimates of $M_{500}$ for the sample based on the baseline $\Msz$--$M_{500}$--$z$ relation. These results provide an initial weak-lensing mass calibration for CHEX-MATE multi-probe cluster studies.}


\maketitle
\nolinenumbers

\section{Introduction}
\label{sec:intro}

Galaxy clusters are powerful probes of cosmology and structure formation. As the most massive self-gravitating systems in the Universe, they occupy a distinctive regime in mass and physical scale, where their abundance and internal structure encode information on the background cosmology, the growth of cosmic structure, and halo assembly. Their mass budget is dominated by dark matter, while most of their baryons reside in the hot intracluster medium. Galaxy clusters thus serve as astrophysical laboratories for studying the interplay between dark matter and baryons and as sensitive probes of the underlying cosmological model \citep{Voit2005,Allen2011,Kravtsov+Borgani2012}.

Their cosmological sensitivity primarily arises because cluster haloes populate the exponential tail of the halo mass function \citep{Haiman2001,Watson2014}. The abundance of rare massive clusters is therefore highly sensitive to cosmological parameters such as the matter density, $\Om$, and the amplitude of matter fluctuations, $\sigma_8$, within the standard $\Lambda$ cold dark matter (\LCDM) framework \citep{Mantz2015}. Large cluster samples spanning a broad range of masses and redshifts thus provide an important avenue for cosmological tests \citep{Vikhlinin2009,Mantz2010,Bocquet2019,Chiu2023}, complementary to early-Universe probes such as cosmic microwave background (CMB) anisotropies, to geometric probes such as type Ia supernovae and baryon acoustic oscillations, and to other low-redshift large-scale-structure probes, including cosmic shear and galaxy clustering. Realising this potential, however, requires accurate cluster mass measurements \citep{Pratt2019}.

Weak gravitational lensing (\WL) provides the most direct observational route to cluster mass calibration. By inducing coherent distortions in the shapes of background galaxies, it probes the projected mass distribution of clusters without relying on assumptions about the dynamical state or equilibrium structure of the systems \citep{Bartelmann+Schneider2001,Umetsu2020rev}. In practice, however, \WL mass measurements are affected by observational systematics and astrophysical projection effects. On the observational side, the dominant sources of uncertainty arise from the source selection, shape measurement, and photometric-redshift (photo-$z$) calibration. In addition, the interpretation of the measured lensing signal in terms of halo mass is affected by halo triaxiality, substructures, and projected large-scale structure, which can introduce intrinsic scatter and systematic bias \citep{Becker+Kravtsov2011,Gruen2015,Umetsu2020xxl,Grandis2024,Saxena2025}. These effects can be quantified and statistically calibrated using synthetic shear catalogues drawn from cosmological simulations, enabling \WL masses to serve as an essential calibration anchor for precision cluster cosmology \citep{Dietrich2019,Chiu2022,Aymerich2025}. This calibration is particularly important in the context of the long-standing tension between low-redshift structure-growth measurements and cosmological parameters inferred from the CMB \citep{vonderLinden2014calib,Hoekstra2015,Planck2016XXIV,Pratt2019}.

The internal structure of dark matter haloes provides an additional and complementary probe of non-linear structure formation. In the standard halo model, the concentration parameter describes the shape of the halo density profile $\rho(r)$ through the ratio $c_\Delta \equiv r_\Delta/r_\mathrm{s}$ of an outer halo radius, defined at a specified overdensity, to an inner scale radius. In hierarchical structure formation, the halo concentration is linked to the assembly history of the inner halo, so the halo population is expected to follow a concentration--mass--redshift ($c$--$M$--$z$) relation with a dependence on halo mass, redshift, and assembly history \citep{Bullock2001,Wechsler2002,Diemer2015}. Numerical simulations in the \LCDM framework generally predict that at fixed redshift, more massive haloes are less concentrated on average and that the relation exhibits non-negligible intrinsic scatter driven by variations in assembly history and mass accretion \citep[][]{Bhattacharya2013,Correa2015cm,Diemer2019}. For massive galaxy clusters, the normalisation, slope, redshift dependence, and intrinsic scatter of the $c$--$M$--$z$ relation therefore provide observational diagnostics of the halo structure in the high-mass regime and enable direct tests of theoretical predictions \citep{Merten2015clash,Okabe+Smith2016}. Observational constraints on the $c$--$M$--$z$ relation are thus important for testing theoretical models of the halo structure and for quantifying the effect of the sample selection, projection effects, and residual measurement systematics. 

The Cluster HEritage project with XMM-Newton: Mass Assembly and Thermodynamics at the Endpoint of Structure Formation collaboration \citep[CHEX-MATE;][]{Chexmate2021} is designed to exploit this potential by studying a minimally biased sample of 118 galaxy clusters detected by \Planck through the thermal Sunyaev--Zel'dovich (SZ) effect.\footnote{\protect\url{http://xmm-heritage.oas.inaf.it/}} 
The sample is defined through two selection tiers. Here, the subscript MMF3 denotes quantities based on or derived from the \Planck multi-frequency matched-filter MMF3 catalogues \citep{Planck2014XXIX,Planck2016XXVII}. In this notation, $(\mathrm{S/N})_\mathrm{MMF3}$ denotes the signal-to-noise ratio (S/N), or detection significance, and $\Msz \equiv M_{500,\mathrm{MMF3}}$ is the \Planck SZ-based mass proxy for $M_{500}$. Tier-1 comprises northern-hemisphere \Planck clusters with $\mathrm{Dec}>0$, $(\mathrm{S/N})_\mathrm{MMF3}>6.5$, and $0.05<z<0.2$. Tier-2 comprises \Planck clusters with $(\mathrm{S/N})_\mathrm{MMF3}>6.5$, $z<0.6$, and $\Msz>7.25\times10^{14}\Msun$. Tier-2 contains the most massive systems by design, whereas, owing to the limited local volume, Tier-1 mostly consists of lower-mass clusters. The two tiers are not mutually exclusive, and four clusters are common to both.

The rich multi-probe dataset assembled for CHEX-MATE enables a detailed characterisation of the physical state and structure of baryons and total matter in individual clusters \citep{Bartalucci2023chex,Rossetti2024chex,Sereno2025chex,Pizzuti2025chex,Chappuis2025chex,Gavidia2026chex}. Accurate and precise mass measurements are therefore essential not only for establishing CHEX-MATE as an astrophysical laboratory, but also for calibrating its cluster sample for cosmological applications in the era of large surveys.

\begin{figure}[tbp]
 \begin{center}
  \includegraphics[width=0.9\columnwidth,angle=0,clip]{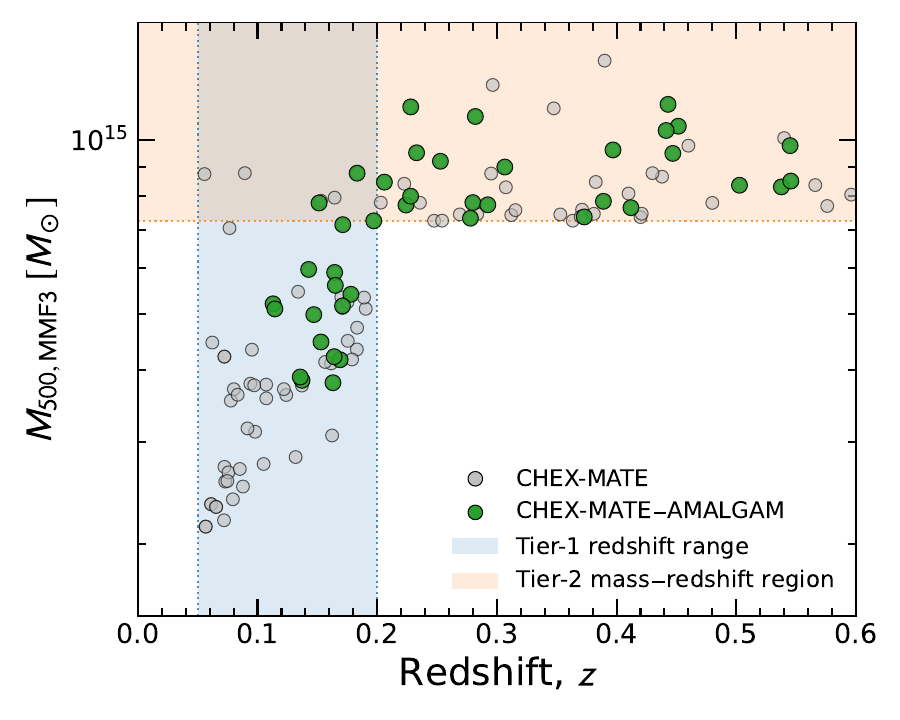} 
 \end{center}
 \caption{Distribution of the CHEX-MATE parent sample in the $M_{500,\mathrm{MMF3}}$--redshift plane, where $\Msz \equiv M_{500,\mathrm{MMF3}}$. The light grey circles show all CHEX-MATE clusters, and the green circles mark the CHEX-MATE--\amalgam subsample we analysed. The subsample spans the low-redshift Tier-1 regime and the high-mass Tier-2 regime. The shaded blue region indicates the Tier-1 redshift range, $0.05<z<0.2$, and the shaded orange region indicates the Tier-2 mass--redshift selection, $z<0.6$ and $M_{500,\mathrm{MMF3}}>7.25\times10^{14}\Msun$. Only the mass--redshift part of the CHEX-MATE selection is shown; the \Planck MMF3 detection significance and sky-selection criteria are not represented in this plane.
}
 \label{fig:mz_selection}
\end{figure}

We present a \WL analysis of 41 \Planck SZ-selected clusters at $0.11 \le z \le 0.55$, drawn from the CHEX-MATE sample. This subsample is covered by the \amalgam project \citep{Gavazzi2026}, which provides wide-field imaging from Subaru/Suprime-Cam and MegaPrime on the Canada--France--Hawaii Telescope (CFHT) for galaxy shape measurements and multiband photometry. Figure~\ref{fig:mz_selection} shows the CHEX-MATE--\amalgam subsample relative to the CHEX-MATE parent sample in the $\Msz$--redshift plane. The subsample includes systems in the low-redshift Tier-1 regime and the high-mass Tier-2 regime, spanning the two main regimes covered by the parent sample. The basic properties of the CHEX-MATE--\amalgam sample, together with the relevant filter information, are summarised in Table~\ref{tab:sample}. 

By combining the \amalgam data with CHEX-MATE observations, we measured the \WL signal around individual clusters and used it to constrain their mass distributions. Our primary goal was to derive \WL-calibrated scaling relations for the halo concentration and for the \Planck SZ mass proxy through Bayesian population modelling of this cluster sample. This analysis provides an initial \WL calibration for multi-probe cluster studies within the CHEX-MATE programme.

\begin{table}
\caption{Broad-band filter description.}
\label{tab:filters}
\centering
\begin{tabular}{lll}
\hline\hline
Telescope/instrument & Filter        & Filter description \\
\hline
Subaru/Suprime-Cam  & $B_\mathrm{J}$ & Johnson $B$-band \\
                    & $V_\mathrm{J}$ & Johnson $V$-band \\
                    & $R_\mathrm{C}$ & Cousins $R$-band \\
                    & $I_\mathrm{C}$ & Cousins $I$-band \\
                    & $g^+$          & Suprime-Cam $g$-band \\
                    & $i^+$          & Suprime-Cam $i$-band \\
                    & $z^+$          & Suprime-Cam $z$-band \\
\hline
CFHT/MegaPrime      & $u^\star$      & MegaPrime $u$-band \\
                    & $g^\star$      & MegaPrime $g$-band \\
                    & $r^\star$      & MegaPrime $r$-band \\
                    & $i^\star$      & MegaPrime $i$-band \\
                    & $z^\star$      & MegaPrime $z$-band \\
\hline
\end{tabular}
\tablefoot{The filter notation follows the CHEX-MATE overview paper \citep{Chexmate2021}. The simplified notation listed here is used for presentation. In the \amalgam photometric matching to the COSMOS reference catalogue, the relevant instrumental filter definitions, including different filter generations and bandpasses where applicable, are accounted for.}
\end{table}

The paper is organised as follows. In Section~\ref{sec:wl} we summarise the cluster \WL formalism and profile estimators we adopted. In Section~\ref{sec:data} we describe the cluster sample, imaging data, shape measurements, and source selection. In Section~\ref{sec:mass} we present the individual-cluster and stacked-lensing analyses, including the mass modelling and assessment of systematic uncertainties. In Section~\ref{sec:scaling} we introduce the hierarchical Bayesian framework and present the population-level constraints on the concentration--mass--redshift relation and the \Planck SZ mass calibration. In Section~\ref{sec:discussion} we discuss the implications of our results. Finally, in Section~\ref{sec:summary}, we summarise our main conclusions.

Throughout this paper, we assume a spatially flat \LCDM cosmology with $\Om=0.3$, $\OL=0.7$, and a Hubble constant of $H_0=100\,h$~km~s$^{-1}$~Mpc$^{-1}$ with $h=0.7$. The critical density of the Universe at redshift $z$ is defined as $\rho_\mathrm{c}(z)=3H^2(z)/(8\pi G)$, where $H(z)$ is the Hubble function. For each cluster, we adopt as the reference centre the X-ray peak position defined by the CHEX-MATE collaboration \citep{Bartalucci2023chex}. We denote spherical and projected radii from the cluster centre by $r$ and $R$, respectively. We define $M_\Delta$ as the mass enclosed within the overdensity radius $r_\Delta$, inside which the mean density is $\Delta$ times the critical density $\rho_\mathrm{c}(z)$. We use $\log$ for base-10 logarithms and $\ln$ for natural logarithms. Intrinsic scatters are quoted primarily in dex, consistent with the $\log_{10}$-space parametrisation of the regression model, and the corresponding fractional scatter is given where useful for intuition. Comoving and physical (proper) quantities are distinguished by the prefixes c and p, respectively (e.g. cMpc and pMpc).\footnote{Explicit $h^{-1}$ units are used only where tied to standard choices in previous cluster \WL analyses: the fixed fitting range $R=[0.3,3]\Mpch$, offset scales compared directly with this range, and the broad log-uniform prior $10^{13} \leq M_{200}/(\Msunh) \leq 10^{16}$. Other quoted masses and radii are expressed in the fiducial cosmology we adopted.}
Unless otherwise stated, all quoted uncertainties are $1\sigma$. All magnitudes are given in the AB system.

\section{Cluster weak-lensing formalism}
\label{sec:wl}

\subsection{Weak-lensing basics}
\label{subsec:basics}

\WL arises from the deflection of light by intervening matter overdensities, such as galaxy clusters, inducing small but coherent distortions in the observed images of background galaxies. Cluster--galaxy \WL is characterised by two fundamental quantities: the convergence, $\kappa$, and the complex shear, $\gamma=|\gamma|e^{2i\phi_\gamma}$, which has spin-2 rotational symmetry in the projected lens plane \citep[for a review, see][]{Umetsu2020rev}. The convergence produces an isotropic magnification, while the shear generates anisotropic shape distortions.

The convergence is defined as the surface mass density $\Sigma$ of the lensing mass distribution normalised by the critical surface mass density, $\SigmaCrit$,
\begin{equation}
  \kappa(\btheta) = \frac{\Sigma(\btheta)}{\SigmaCrit},
\end{equation}
where $\SigmaCrit$ depends on the geometric configuration of the observer, lens, and source,
\begin{equation}
 \SigmaCrit(\zl, \zs) = \frac{c^2}{4\pi G} \frac{D_s}{D_l D_{ls}} \frac{1}{(1+\zl)^2}
 = \frac{c^2}{4\pi G}\frac{1}{(1+\zl)^2 D_l\beta}.
\end{equation}
Here, $c$ is the speed of light, $G$ is the gravitational constant, and $D_l(\zl)$, $D_s(\zs)$, and $D_{ls}(\zl,\zs)$ are the angular diameter distances from observer to lens, observer to source, and lens to source, respectively. These distances depend on the background cosmology. The factor of $(1+\zl)^2$ appears because we use comoving, rather than physical, surface mass densities. The geometric lensing efficiency, $\beta(\zl,\zs)=D_{ls}/D_s$, quantifies the lensing strength as a function of lens redshift $\zl$ and source redshift $\zs$, with $\beta(\zl,\zs)=0$ for $\zs \le \zl$.

The shear field can be written as $\gamma(\btheta)=\gamma_1(\btheta)+i\gamma_2(\btheta)$, with Cartesian components $(\gamma_1,\gamma_2)$. It is often convenient to decompose $\gamma(\btheta)$ into a tangential component, $\gamma_+$, and a $45^\circ$-rotated cross component, $\gamma_\times$, defined with respect to a polar coordinate system $(\theta,\phi)$ centred on a chosen reference point,
\begin{equation}
 \begin{aligned}
  \gamma_+(\theta,\phi) &= -\left(\gamma_1\cos 2\phi + \gamma_2\sin 2\phi\right),\\
  \gamma_\times(\theta,\phi) &= -\left(-\gamma_1\sin 2\phi + \gamma_2\cos 2\phi\right).
 \end{aligned}
\end{equation}
For a given choice of centre, the azimuthally averaged shear components at angular radius $\theta$ satisfy
\begin{equation}
\label{eq:gammatx}
  \begin{aligned}
  \gamma_+(\theta) &= \overline{\kappa}(<\theta)-\kappa(\theta)
   \equiv \frac{\Delta\Sigma(R)}{\SigmaCrit},\\
  \gamma_\times(\theta) &= 0,
  \end{aligned}
\end{equation}
where $R=(1+\zl)D_l\theta$ is the comoving transverse radius. Here, $\kappa(\theta)=\Sigma(R)/\SigmaCrit$ denotes the azimuthally averaged convergence at radius $\theta$, $\overline{\kappa}(<\theta)=\overline{\Sigma}(<R)/\SigmaCrit$ is the mean interior convergence, and $\Delta\Sigma(R)=\overline{\Sigma}(<R)-\Sigma(R)$ is the excess surface mass density. The azimuthally averaged cross component, $\gamma_\times(\theta)$, is expected to vanish for the \WL signal, so deviations from zero provide a useful null test for residual systematic errors.

In \WL shape measurements, the primary observable in subcritical regions is the reduced shear field, $g(\btheta)=g_1+ig_2$, defined as
\begin{equation}
 \label{eq:g}
 g(\btheta)=\frac{\gamma(\btheta)}{1-\kappa(\btheta)},
\end{equation}
which is estimated statistically from the observed ellipticities of background galaxies. For a given reference point, the reduced shear can likewise be decomposed into tangential and cross components, $g_+=\gamma_+/(1-\kappa)$ and $g_\times=\gamma_\times/(1-\kappa)$.

The azimuthally averaged reduced tangential shear, $g_+(\theta)$, measured as a function of cluster-centric radius $\theta$, is related to $\Sigma(R)$ and $\Delta\Sigma(R)$ by \citep{Umetsu2020rev}
\begin{equation}
 \label{eq:deltaSigma}
 \Delta\Sigma_+(R) \equiv \SigmaCrit g_+(\theta) = \frac{\Delta\Sigma(R)}{1-\SigmaCrit^{-1}\Sigma(R)},
\end{equation}
where $\Delta\Sigma_+(R)$ denotes the excess surface mass density observable inferred from the reduced tangential shear.\footnote{The second equality in Equation~\eqref{eq:deltaSigma} corresponds to the standard mean-field description of azimuthally averaged \WL observables. It is exact for axisymmetric mass distributions and provides a useful approximation for mildly non-axisymmetric lenses \citep{Umetsu2020rev,Umetsu2026}.}

\subsection{Weak-lensing observables and estimators}
\label{subsec:DSigma}

The X-ray-emitting intracluster gas is often a useful tracer of the central gravitational potential of a galaxy cluster \citep{Donahue2014clash,Umetsu2018clump3d,OkabeT2018}, although this correspondence can break down in strongly disturbed systems, particularly dissociative mergers \citep{Clowe2006,Okabe+Umetsu2008}. We measured the \WL signal around the X-ray peak of each cluster \citep{Bartalucci2023chex} and adopted it as the reference centre.

We computed $\Delta\Sigma_+$ in $\Nbin=11$ radial bins, equally spaced in logarithmic radius, with $\Delta\ln R=\ln(R_\mathrm{max}/R_\mathrm{min})/N\approx 0.21$, spanning $R_\mathrm{min}=0.3\Mpch$ to $R_\mathrm{max}=3\Mpch$ \citep[e.g.][]{Medezinski2018planck,Miyatake2019,Umetsu2020xxl}. The inner limit $R_\mathrm{min}$ is sufficiently large that shape and photo-$z$ measurements are not expected to be significantly affected by masking or blending from bright cluster galaxies \citep[][]{Medezinski2018src}. Moreover, $R_\mathrm{min}$ is much larger than the typical offset between the brightest cluster galaxy (BCG) and the X-ray peak measured for \Planck SZ-selected clusters, whose median value is $D_{\mathrm{X\text{--}BCG}}\approx 0.017\,r_{500}$ ($\approx 21.5$~kpc; \citealt{Rossetti2016}). Because this offset scale is much smaller than the innermost radius used in the profile analysis, it is unlikely to cause substantial smoothing of the azimuthally averaged \WL signal for most clusters when adopting X-ray centring, although larger X-ray--mass offsets may occur in dynamically disturbed systems. Possible residual X-ray--mass offsets are assessed qualitatively using the two-dimensional mass reconstructions in Section~\ref{subsec:kmap}.

We estimated $\Delta\Sigma_+$ in each radial bin for an individual cluster $l$ using the estimator
\begin{equation}
\label{eq:DSigma}
\Delta\Sigma_+(R_i) = \frac{\sum_{s\in i} w_{ls}\, \Sigma_{\mathrm{cr},ls}\, g_{+,ls}} {\sum_{s\in i} w_{ls}},
\end{equation}
where the sum runs over all source galaxies $s$ lying in the $i$th radial bin around cluster $l$. Here, $g_{+,ls}$ is the tangential reduced-shear estimator measured with respect to the X-ray peak of cluster $l$, computed from the source-galaxy shape components in Cartesian sky coordinates provided by the \amalgam pipeline (Section~\ref{sec:data}).

The critical surface mass density for each lens--source pair is given by
\begin{equation}
\label{eq:Sigma_cr}
\Sigma_{\mathrm{cr},ls}^{-1} = \frac{4\pi G}{c^2}(1+\zl)^2 D_l \, \beta_{ls},
\end{equation}
where $\beta_{ls}$ denotes the estimated lensing efficiency for source galaxy $s$ relative to lens $l$. The \amalgam pipeline provides an estimate of $\beta_{ls}$ for each lens--source pair by matching the source multiband photometry to the Cosmic Evolution Survey (COSMOS) reference catalogue \citep{Laigle2016cosmos} (see Section~\ref{subsec:amalgam}).

The statistical weight factor is taken to be
\begin{equation}
\label{eq:wls}
w_{ls} = \Sigma_{\mathrm{cr},ls}^{-2}\, \frac{1}{\sigma_{\epsilon,s}^2 + \sigma_0^2},
\end{equation}
which corresponds to an inverse-variance weighting based on the lensing efficiency and the per-component shape noise. Here, $\sigma_{\epsilon,s}$ is the shape measurement uncertainty per ellipticity component provided by the \amalgam pipeline, and $\sigma_0$ is the adopted root-mean-square (rms) intrinsic ellipticity per component entering the statistical weight. We fixed $\sigma_0=0.4/\sqrt{2} \approx 0.28$,\footnote{The adopted value of $\sigma_0$ enters the inverse-variance weights used to compute the azimuthally averaged $\Delta\Sigma_+$ profiles. Since the statistical uncertainties are estimated empirically from the measured cross-component signal, as described below, rather than propagated directly from the weight factor itself, the profile estimates and their uncertainties depend only weakly on the exact choice of $\sigma_0$.}
which is consistent with the level of galaxy shape dispersions typically measured in Subaru-based \WL analyses \citep[e.g.][]{Tam2026}. The adopted value is also within the range $0.1 \simlt\sigma_0\simlt 1$ tested in the \amalgam shear-calibration analysis; over this range, multiplicative shape-measurement biases were found to remain at the sub-percent level \citep{GREAT3results,Gavazzi2026}. The ellipticity is defined as $\epsilon=(a-b)/(a+b)$, where $a$ and $b$ are the major and minor axes, respectively.

Similarly, we define the $\times$-component surface mass density, $\Delta\Sigma_\times$, by replacing $g_{+,ls}$ in Equation~\eqref{eq:DSigma} with the 45$^\circ$-rotated component $g_{\times,ls}$. The azimuthally averaged $\times$ component, or $B$-mode signal, is expected to be statistically consistent with zero for a pure \WL signal.

When interpreting the observed lensing profile $\{\Delta\Sigma_+(R_i)\}_{i=1}^{\Nbin}$, it is important to define the corresponding bin radii $\{R_i\}_{i=1}^{\Nbin}$ accurately so as to minimise systematic bias in cluster mass measurements. We define the effective bin radius $R_i$ using the weighted harmonic mean of the lens--source transverse separations $R_{ls}$ as
\begin{equation}
\label{eq:binradius}
 R_i \equiv \frac{\sum_{s\in i}w_{ls}}{\sum_{s\in i} w_{ls} R_{ls}^{-1}},
\end{equation}
which allows an unbiased estimate of the underlying cluster lensing profile \citep{Okabe+Smith2016,Sereno2017psz2lens}.

To quantify the significance of the tangential-component profile measurements, $\{\Delta\Sigma_+(R_i)\}_{i=1}^{\Nbin}$, we adopted the linear S/N estimator of \citet{Sereno2017psz2lens},\footnote{This $\mathrm{S/N}$ estimator differs from the conventional quadratic definition, $\mathrm{S/N}_{+,\mathrm{q}} = \left[\sum_{i=1}^{\Nbin} (\Delta\Sigma_{+,i})^2 / \sigma_{\mathrm{shape},i}^2\right]^{1/2} > 0$. The quadratic form tends to overestimate the detection significance in the noise-dominated regime, where the per-bin $\mathrm{S/N}$ is lower than unity \citep{Umetsu2020rev}.}
defined as $\mathrm{S/N}_{+} = S_+ / \sigma_S$, with
\begin{equation}
 \label{eq:snr}
  \begin{aligned}
   S_+ &= \frac{\sum_{i=1}^{\Nbin}
   \Delta\Sigma_{+}(R_i)/\sigma^2_\mathrm{shape}(R_i)}
   {\sum_{i=1}^{\Nbin} 1/\sigma^2_\mathrm{shape}(R_i)},\\
   \sigma_S &= \left[\sum_{i=1}^{\Nbin} 1/\sigma^2_\mathrm{shape}(R_i)\right]^{-1/2},
  \end{aligned}
\end{equation}
where $\sigma_\mathrm{shape}(R_i)$ denotes the statistical uncertainty of the $\Delta\Sigma_+$ estimator (Equation~\ref{eq:DSigma}), estimated as
\begin{equation}
 \label{eq:sigma_stat}
 \sigma_\mathrm{shape}^2(R_i) =
 \frac{\sum_{s\in i} w_{ls}^2\, \Sigma_{\mathrm{cr},ls}^{2}\, g_{\times, ls}^2 }
  {\left(\sum_{s\in i} w_{ls}\right)^2}.
\end{equation}
By evaluating this empirical noise estimator directly from the measured cross-component signal, we obtained a conservative estimate of the diagonal noise that captures shape noise together with additional scatter present in the data, including variance induced by azimuthal asymmetry of the lensing field and residual systematics.

\subsection{Error covariance matrix}
\label{subsec:cmat}

For robust statistical inference of cluster mass and concentration from the \WL signal, it is essential to account for relevant sources of uncertainty in the likelihood analysis \citep{Hoekstra2003,Gruen2015,Umetsu2020rev}. For an individual cluster, we express the error covariance matrix of the measured excess surface mass density profile, $\{\Delta\Sigma_+(R_i)\}_{i=1}^{\Nbin}$, as
\begin{equation}
\label{eq:cmat}
C = C^\mathrm{shape} + C^\mathrm{lss},
\end{equation}
where $C^\mathrm{shape}$ denotes the statistical covariance due to shape noise, and $C^\mathrm{lss}$ is the covariance arising from uncorrelated large-scale structure projected along the line of sight.

For an individual-cluster $\Delta\Sigma_+$ profile, we estimated the statistical uncertainty in each radial bin by the shape-noise term $\sigma_\mathrm{shape}(R_i)$ (Equation~\ref{eq:sigma_stat}). The corresponding shape-noise covariance was assumed to be diagonal,
\begin{equation}
\label{eq:Cshape}
(C^\mathrm{shape})_{ij}=\sigma^2_{\mathrm{shape}}(R_i)\,\delta_{ij},
\end{equation}
with $\delta_{ij}$ the Kronecker delta.

The elements of the $C^\mathrm{lss}$ matrix are given by \citep{Hoekstra2003}
\begin{equation}
 \label{eq:Clss}
  \begin{aligned}
 (C^\mathrm{lss})_{ij} &=
   \langle\SigmaCrit^{-1}\rangle_i^{-1}
   \langle\SigmaCrit^{-1}\rangle_j^{-1}\\
   &\times \int\!\frac{\ell d\ell}{2\pi}P_\kappa(\ell)
   J_2(\ell\theta_i) J_2(\ell\theta_{j}),
  \end{aligned}
\end{equation}
where $P_\kappa(\ell)$ is the convergence power spectrum as a function of angular multipole $\ell$, evaluated for the adopted source population, and $J_2$ is the second-order Bessel function of the first kind. Here $\theta_i=(1+\zl)^{-1}R_i/D_l$ denotes the effective angular radius of the $i$th annulus. The quantity $\langle\Sigma_{\mathrm{cr},i}^{-1}\rangle$ is the sensitivity-weighted inverse critical surface mass density evaluated in the $i$th radial bin, defined as
\begin{equation}
 \label{eq:inv_sigma_crit}
  \langle\Sigma_{\mathrm{cr},i}^{-1}\rangle = \frac{\sum_{s\in i}
   w_{ls}\Sigma_{\mathrm{cr},ls}^{-1}}{\sum_{s\in i}
   w_{ls}}.
\end{equation}

We computed $(C^\mathrm{lss})_{ij}$ for each cluster by closely following the procedure outlined in \citet{Miyaoka2018} \citep[see also][]{Umetsu2020xxl}. Specifically, we placed the effective source plane at the median photo-$z$ of the selected background galaxies for each cluster. We employed the non-linear matter power spectrum of \citet{Smith+2003halofit}, assuming a \textit{Wilkinson Microwave Anisotropy Probe} (\WMAP) nine-year flat $\Lambda$CDM cosmology characterised by a cosmological constant density $\OL=0.718$, baryon density $\Ob = 0.0461$, Hubble constant $H_0=69.7$~km~s$^{-1}$~Mpc$^{-1}$, power-spectrum normalisation $\sigma_8=0.817$, and scalar spectral index $n_s=0.9646$.

As found by \citet{Miyatake2019}, the total uncertainty per cluster is dominated by shape noise ($C^\mathrm{shape}$) at projected radii $R\simlt 3\Mpch$ (see their Figure~4), beyond which the contribution from cosmic noise ($C^\mathrm{lss}$) becomes increasingly important. We modelled the per-cluster covariance as $C=C^\mathrm{shape}+C^\mathrm{lss}$. Additional profile-to-profile fluctuations associated with intrinsic halo structure, such as correlated substructure and cluster asphericity \citep{Umetsu2020rev}, were not included explicitly as a separate covariance term at this stage. Instead, their impact on the inferred \WL mass was absorbed effectively into the simulation-based calibration of the \WL mass bias relation described in Section~\ref{sec:scaling}. Given that the total per-cluster uncertainty is dominated by shape noise over the radial range used in our analysis, the precise cosmological choice adopted for the $C^\mathrm{lss}$ calculation has a negligible impact on our mass inference.

\section{Sample and data}
\label{sec:data}

We used the \WL data products derived from the \amalgam dataset \citep{Gavazzi2026}, a broader archival \WL programme comprising 122 massive cluster fields assembled from Subaru/Suprime-Cam and CFHT/MegaPrime imaging. Our analysis used the 41 CHEX-MATE clusters covered by the current \amalgam data products, including 15 Tier-1, 24 Tier-2, and 2 Tier-1+2 systems (Table~\ref{tab:sample}). Suprime-Cam and MegaPrime/MegaCam provide fields of view of approximately $34\arcmin\times27\arcmin$ and $1\deg^2$, respectively, making them well suited for wide-field cluster \WL studies. Each cluster in the sample has \amalgam imaging in at least two optical bands. For the \amalgam stacks, the typical $80\percent$-completeness limiting magnitudes are $m_\mathrm{lim}\approx 24$--$25$~mag in the broad optical bands used for photometry and shape measurements, with a cluster-to-cluster dispersion of about 0.5 mag, while the $z^+/z^\star$-band stacks are shallower, with $m_\mathrm{lim}\approx 22.3$~mag. The broad-band filter notation adopted for the \amalgam imaging data is summarised in Table~\ref{tab:filters}.

\subsection{\amalgam data products}
\label{subsec:amalgam}

The \amalgam project assembles and homogenises archival wide-field imaging from Subaru/Suprime-Cam and CFHT/MegaPrime for \WL studies of massive galaxy clusters \citep{Gavazzi2026}. For our CHEX-MATE analysis, we used the \amalgam processing only through the final source catalogues and imaging products needed for cluster shear measurements: coadded images in the available bands, spatially varying point-spread-function (PSF) models, multiband model photometry, galaxy shape measurements corrected for the effects of atmospheric seeing, telescope optics, and instrumental response through the PSF models in the available bands, and photometry-based source-redshift products derived from an empirical reference catalogue. A full description of the end-to-end reduction pipeline is given by \citet{Gavazzi2026}. Here we summarise only the elements required to define the source samples and calibration quantities we used.

Source detection and model fitting are performed on the coadded images with the AstrOmatic software suite \citep{Bertin1996,Bertin2002,Bertin2006}. Combined multiband catalogues are constructed by running \textsc{SExtractor}, a source-detection and photometry package, in dual-image mode, using a $\chi^2$ coadd as the common detection image so that sources share the same identification across filters. In each band, spatially varying PSF models based on \textsc{PSFEx} \citep{Bertin2011} are used to fit a single S\'ersic \citep{Sersic1968} model to the light distribution of each detected source. The resulting catalogues provide model magnitudes and corrected shape measurements in the available filters. For our \WL analysis, the lensing band was selected from the red optical bands available for a given system, most commonly among $r/R$ or $i/I$, so as to provide the best combination of image quality and depth for shape measurement. These lensing-band shape measurements, combined with the multiband photometry, form the basis of the source selection and shear-profile construction used below.

The shape-measurement component of the \amalgam pipeline, based on the \textsc{SExtractor}+\textsc{PSFEx} modelling described above, was assessed in the third GRavitational lEnsing Accuracy Testing challenge (GREAT3; \citealt{GREAT3results}). In the high-S/N regime, the \amalgam validation work reports multiplicative shear biases at the level of $m \sim 3\times10^{-3}$ for galaxies with $\texttt{snr\_win}>20$, and PSF-aligned additive residuals of order $c \sim 3\times10^{-4}$ \citep{Gavazzi2026}, where \texttt{snr\_win} denotes the \textsc{SExtractor}-based S/N estimator used in the \amalgam source catalogues. The multiplicative and additive shear biases are defined by $g_\alpha^\mathrm{est} = (1+m_\alpha)g_\alpha^\mathrm{true} + c_\alpha$ for each reduced-shear component $\alpha$. These results demonstrate the sub-percent-level shear-calibration performance of the shape-measurement pipeline under conservative high-S/N conditions.

For our analysis, we used source samples selected with $\texttt{snr\_win}>15$ in the lensing band, in order to increase the usable background-source density for cluster \WL measurements. This relaxed threshold reflects the different trade-off between source density and calibration conservatism in cluster \WL. Additional \amalgam validation tests show that residual PSF-correlated additive systematics remain very small over the source S/N range relevant here, with PSF-leakage residuals in the mean galaxy ellipticity limited to a few $10^{-4}$ \citep{Gavazzi2026}. These tests do not show evidence for a significant increase in PSF-related shape systematics when extending the source selection to $\texttt{snr\_win}>15$ for our cluster-\WL analysis. Possible residual multiplicative shear-calibration uncertainty associated with this source selection is included in the systematic budget described in Section~\ref{subsubsec:residual}.

The source redshift information is inferred empirically from the multiband model photometry by matching each source to galaxies in the COSMOS2015 reference catalogue \citep{Laigle2016cosmos} in the space of observed magnitudes. In the \amalgam framework, this is implemented with a nearest-neighbour search in the magnitude space defined by the filters available for each cluster. To account for the heterogeneous depth and photometric uncertainties of the \amalgam data, the COSMOS reference photometry is noise-degraded to match the photometric uncertainties and depth of each cluster field and filter set before performing the magnitude-space matching. In this procedure, the relevant instrumental filter transmissions are used, including differences among filter generations where applicable, rather than relying only on the simplified filter notation listed in Table~\ref{tab:filters}. 

The resulting ensemble of matched COSMOS galaxies provides an approximate redshift probability distribution $p(z)$ for each source. From this distribution, the \amalgam catalogues store summary quantities we used, including a point estimate of the photo-$z$ and the first two moments of the lensing-efficiency factor, $\beta$ and $\beta^2$. In our analysis, these quantities were used as source-level redshift summaries, with $\beta$ entering the conversion of the measured shear signal into $\Delta\Sigma_+$.

The same empirical redshift framework also provides the cluster membership score, \smem \citep{Gavazzi2026}. This is an empirical COSMOS-based score that quantifies the relative enhancement, in multiband magnitude space, of galaxies associated with the cluster redshift. In practice, \smem is evaluated using COSMOS neighbours within a narrow redshift interval around the cluster redshift, $|z-\zl|<0.05$. It should therefore be interpreted as a cluster-membership indicator rather than as a calibrated membership probability. Operationally, high values of \smem identify regions of multiband magnitude space preferentially associated with galaxies at the cluster redshift, whereas low values, combined with positive lensing efficiency, preferentially select background sources.

For clusters with suitable filter coverage, we adopted colour--colour (CC) cuts to define background samples, because they provide a transparent and empirically well-tested separation between unlensed galaxies and lensed sources. For the remaining systems, where the available photometry does not support the adopted CC selection, we instead used threshold cuts based on the \amalgam redshift products, namely a requirement of positive lensing efficiency $\beta$ together with a low cluster membership score \smem. As shown below, these two approaches suppress similar regions of foreground and cluster-member contamination, although they do not select identical source populations.

The \amalgam catalogues we used thus provide for each source the multiband model photometry, PSF-model-corrected shape measurements in the available bands, and empirical redshift summary quantities, including a photo-$z$ point estimate, $\beta$, $\beta^2$, and the cluster-membership score \smem. For our \WL analysis, we used the shape measurement in the adopted lensing band, typically chosen to have the best combination of depth and image quality and listed in Table~\ref{tab:sample}. Our analysis used these delivered quantities directly, without reprocessing the imaging data or re-deriving the source-redshift products from the raw photometry.

\subsection{Background galaxy selection}
 \label{subsec:back}

Contamination of background-galaxy samples by unlensed objects, when not accounted for, leads to a systematic suppression of the true lensing signal. Inclusion of foreground galaxies produces a dilution that is approximately independent of cluster radius, whereas contamination by cluster members causes a stronger dilution toward smaller cluster radii. A secure selection of background galaxies is therefore essential for obtaining accurate cluster mass measurements from \WL \citep{Medezinski2010,Medezinski2018src,Gruen2014}.

For the \WL analysis, we adopted a homogeneous bright-end magnitude selection in the shape-measurement band, requiring $m_\mathrm{shape} > 21$ for all clusters (Table~\ref{tab:sample}). This cut excludes very bright objects from the source sample. At the faint end, the effective selection was set primarily by the \texttt{snr\_win} threshold, together with the requirement that the source photometry can be matched to the noise-degraded COSMOS2015 reference catalogue used to derive the \amalgam source-redshift products (Section~\ref{subsec:amalgam}). 
In practice, the resulting source samples are therefore S/N limited rather than limited by a common faint-end magnitude threshold. We nevertheless imposed an additional conservative cut of $m_\mathrm{shape} < 27$ to remove any remaining very faint detections with high \texttt{snr\_win}, including possible artefacts, and to keep the selection uniform across the sample. In addition to these magnitude and S/N requirements, we applied object-quality cuts in the shape-measurement band. Stars were rejected using the morphological criterion $\texttt{SPREAD\_MODEL}>0.005$ \citep{Desai2012}, and objects falling in masked regions around bright stars, diffraction spikes, saturation trails, and related imaging artefacts were excluded by requiring $\texttt{FLAG\_MASKS}=0$.

The CHEX-MATE--\amalgam analysis employed two source-selection routes to identify background galaxies and exclude unlensed foreground and cluster populations. For clusters with three or four available bands that define a suitable CC space, we adopted the CC selection method \citep{Medezinski2010,Medezinski2018src}, which provides a stringent empirical separation of background galaxies from the main locus of foreground and cluster-member galaxies. For the remaining systems, the available filters do not provide a suitable set of distinct colours for robust CC selection, for example when only two effective bands are available. For these systems, we adopted a \smem-threshold selection, requiring $\mathrm{S}_\mathrm{mem} < 0.25$ and $\beta > 0$, where the latter condition excludes sources with non-positive inferred lensing efficiency. Thus, the CC selection uses observed colour space directly, whereas the \smem-threshold selection uses the \amalgam COSMOS-matched redshift products to exclude sources lying in regions of magnitude space associated with the cluster redshift.

\begin{figure*}[tbp] 
 \begin{center}
  \includegraphics[width=0.9\textwidth,angle=0,clip]{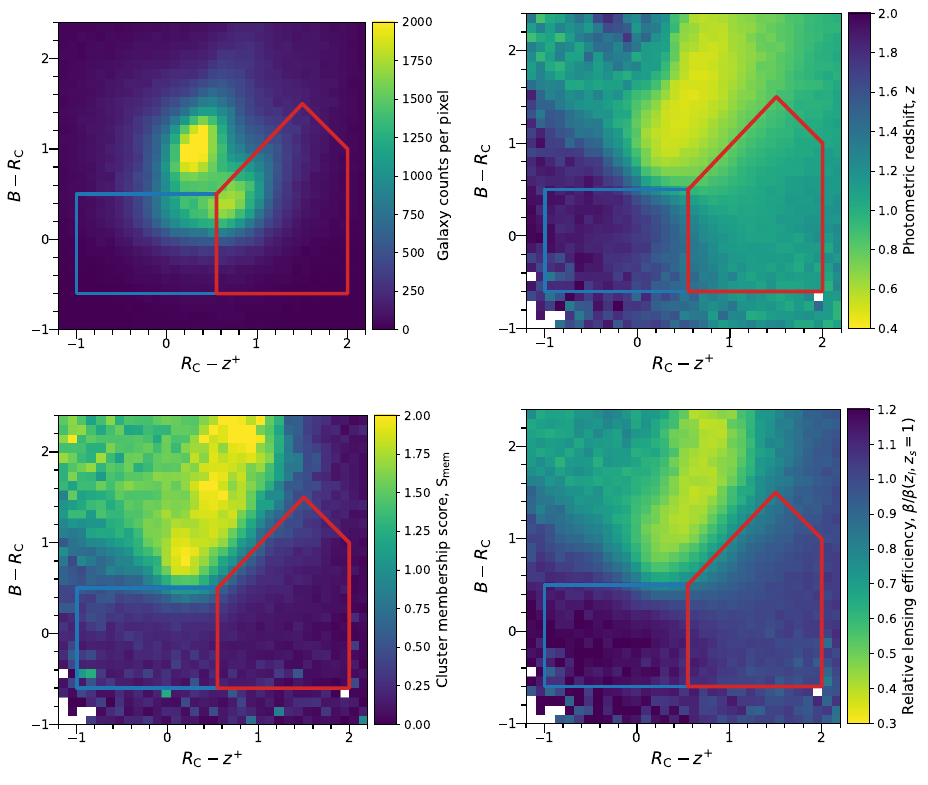} 
 \end{center}
 \caption{Stacked distribution of 339,014 magnitude- and S/N-selected galaxies in the $B_\mathrm{J}-R_\mathrm{C}$ vs. $R_\mathrm{C}-z^+$ diagram for a subsample of 13 clusters, where Subaru \BRz photometry is available for CC background selection. The upper left panel displays the distribution of galaxy counts in CC space. The blue and red polygons define the CC-selection boundaries of the blue and red background samples, respectively. The upper right, lower left, and lower right panels show the distributions of bin-averaged values of photo-$z$, cluster membership score \smem, and relative lensing efficiency $\beta/\beta(\zl,\zs=1)$, respectively.}
 \label{fig:stack_CC}
\end{figure*}

The threshold $\mathrm{S}_\mathrm{mem}<0.25$ was chosen as a conservative cut to reject the high-\smem cluster-redshift locus, which typically appears at $\mathrm{S}_\mathrm{mem} \simgt 1$ (see Figure~\ref{fig:stack_CC}), while the requirement $\beta>0$ excludes sources with non-positive inferred lensing efficiency. The operational background-galaxy selection used in this analysis was thus based on either the CC selection or the \smem-threshold selection, depending on the available photometric information.

For low-redshift cluster lenses at $\zl<0.4$, \citet{Medezinski2018src} found that CC-selected source samples are largely free from foreground contamination, with a foreground contamination level of $(2.8 \pm 0.4)\percent$ based on reweighted spectroscopic-redshift samples in the Hyper Suprime-Cam (HSC) Subaru Strategic Program (HSC-SSP) footprint \citep{Tanaka2018}. This value characterises the residual contamination of the selected source sample and should not be interpreted as implying the same fractional bias in the inferred cluster mass, which is constrained mainly by the \WL signal at larger radii where cluster-member contamination is substantially reduced. Since our sample has a median lens redshift of $\zlmed=0.23$, this result provides empirical support for the low contamination level of the adopted CC-selection methodology for the bulk of our clusters.

Accordingly, the CC selection was adopted for 30 clusters, while the \smem-threshold selection was used for the remaining 11 clusters. For the 30 CC-selected clusters, the \smem-threshold selection can also be applied, allowing an empirical comparison of the \WL masses obtained with the two source-selection methods (Section~\ref{subsec:single}). Across the full sample of 41 clusters, the resulting source population has a mean surface number density of $\langle\ngal\rangle \approx 7.7$~arcmin$^{-2}$ and a mean photo-$z$ of $\langle\zs\rangle \approx 0.99$. The cluster-by-cluster background-selection method, together with the corresponding \WL band and the available photometric filters, is summarised in Table~\ref{tab:sample}. For CC-selected clusters, the filters used to define the selection are underlined.

For clusters analysed with the CC-selection method, the background-galaxy colour cuts were defined consistently for each common broad-band filter combination. Although the observed colours of cluster red-sequence galaxies evolve with lens redshift, the aim of the CC selection is not to trace the red sequence of each individual cluster, but to define conservative background regions that exclude the foreground and cluster-galaxy locus in the relevant observed CC space. For our cluster sample, which is limited to $\zl \leq 0.55$, this locus remains sufficiently well separated from the adopted blue and red background regions for each filter-set subsample. We therefore defined common CC-selection criteria for clusters sharing the same filter set, such as \BRz, following the methodology established by \citet{Medezinski2010,Medezinski2018src}. The relevant filter combinations are summarised in Table~\ref{tab:sample}, which also identifies the cluster subsamples analysed with a common CC-selection scheme. When multiple filters were available, the CC filter combination was chosen to provide effective wavelength coverage for separating the foreground and cluster-galaxy locus from the background population, while accounting for the relative depth and quality of the available bands.

Figure~\ref{fig:stack_CC} illustrates this strategy for a subsample of 13 clusters with Subaru \BRz photometry. The stacked galaxy counts in CC space (upper left panel) show that the foreground and cluster-galaxy population forms an extended overdensity, approximately aligned along a diagonal locus \citep[see also][]{Medezinski2011}. The main peak of this overdensity corresponds to foreground galaxies, and its extension toward redder colours is due to red-sequence cluster galaxies. The regions with elevated cluster membership score \smem (lower left panel) closely follow the same diagonal locus, while the adopted blue and red CC cuts isolate conservative background regions with lower \smem and nonzero lensing efficiency. This behaviour supports the use of common CC-selection boundaries for clusters sharing the same filter combination over the redshift range probed by our sample.

\begin{figure}[tbp]
 \begin{center}
  \includegraphics[width=0.9\columnwidth,angle=0,clip]{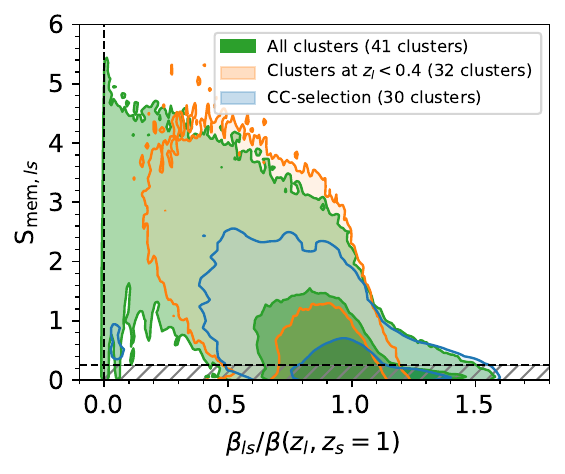} 
 \end{center}
 \caption{Stacked distribution of magnitude- and S/N-selected galaxies in the cluster membership score (\smem) vs. geometric lensing efficiency ($\beta$) plane for different cluster subsamples. For each subsample, the inner and outer contours enclose $68\percent$ and $95\percent$ of the stacked lens--source ($ls$) pairs, where $\beta_{ls}$ is normalised by $\beta(\zl,\zs=1)$ for each cluster. The green, orange, and blue contours represent the full sample of 41 clusters, the subsample of 32 clusters at $\zl<0.4$, and the CC-selected source population for the subsample of 30 clusters, respectively. The hatched region denotes the \smem-threshold source-selection criteria, $\beta_{ls}>0$ and $\mathrm{S}_{\mathrm{mem},ls}<0.25$.}
 \label{fig:smem}
\end{figure}

Figure~\ref{fig:smem} illustrates the stacked distribution of magnitude- and S/N-selected galaxies in the \smem--$\beta$ plane for different cluster subsamples. The CC-selected source population avoids much of the low-$\beta$, high-\smem region populated by unlensed foreground and cluster galaxies. At the same time, it remains distributed over a broad range in \smem and is not confined to the narrow \smem-threshold selection band defined by $\mathrm{S}_\mathrm{mem} < 0.25$ and $\beta > 0$. Thus, the CC and \smem-threshold selection methods suppress similar contaminants at a broad level, while selecting substantially different source populations. This interpretation is consistent with the mean overlap fraction of only $\sim 60\percent$ between the two selections for the 30 clusters where both methods can be applied.

\subsection{Tangential shear profiles around CHEX-MATE clusters}

Using the \amalgam shape measurements for multiband-selected background galaxies, we derived azimuthally averaged excess surface mass density profiles over the comoving radial range $R\in[0.3,3]\Mpch$, centred on the adopted X-ray peak position of each cluster in our sample (Table~\ref{tab:sample}). The azimuthally averaged tangential- and cross-component profiles, $\Delta\Sigma_+$ and $\Delta\Sigma_\times$, for individual clusters are shown as thumbnails in Appendix~\ref{appendix:individual} (Figure~\ref{fig:gt_all}).

\begin{figure}[tbp]
 \begin{center}
  \includegraphics[width=0.9\columnwidth,angle=0,clip]{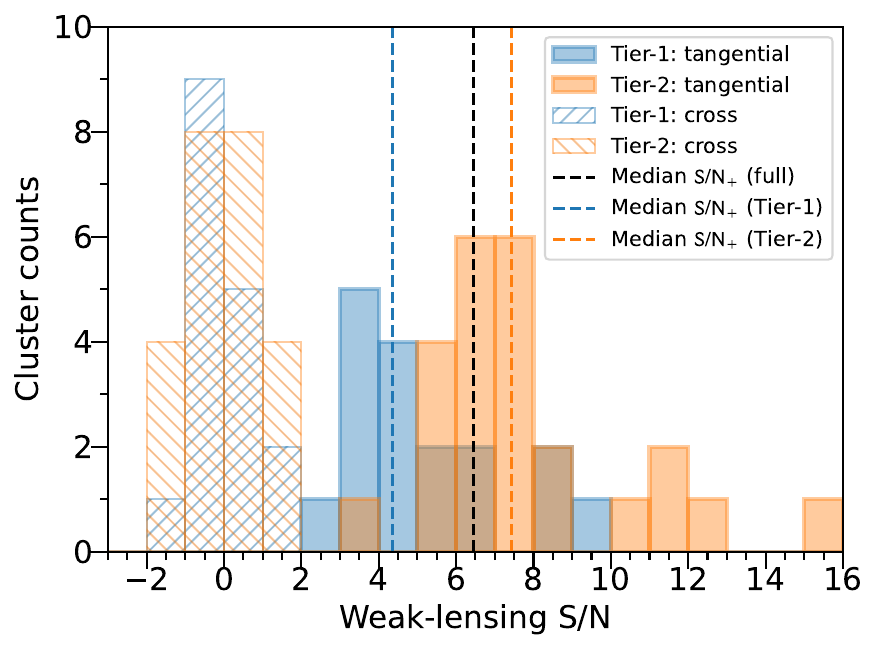} 
 \end{center}
 \caption{Distributions of the \WL S/N, $\mathrm{S/N}_+$, shown separately for the Tier-1 and Tier-2 subsamples; the tier-based subsamples are defined to be mutually exclusive, as described in the text. The shaded blue and orange histograms show the Tier-1 and Tier-2 distributions, respectively. The vertical dashed lines in black, blue, and orange indicate the median $\mathrm{S/N}_+$ values for the full sample, Tier-1 subsample, and Tier-2 subsample, respectively. The hatched blue and orange histograms show the corresponding distributions of the cross-component S/N, $\mathrm{S/N}_\times$.}
 \label{fig:WLSN} 
\end{figure}

For a meaningful ensemble analysis based on individual-cluster \WL measurements, the integrated S/N defined by Equation~\eqref{eq:snr} should be at least of order unity. Figure~\ref{fig:WLSN} shows the distributions of the tangential- and cross-component S/N values, $\mathrm{S/N}_+$ and $\mathrm{S/N}_\times$, for the full sample. All clusters have positive values of $\mathrm{S/N}_+$. For the full sample, the tangential component has a median $\mathrm{S/N}_+=6.5$ and spans the range $2.3$--$15.4$, comfortably satisfying this requirement. The corresponding median values are $4.4$ for the Tier-1 plus Tier-1+2 subsample and $7.4$ for the Tier-2-only subsample, consistent with the higher-$\Msz$ selection of the latter. By contrast, the cross component has a median $\mathrm{S/N}_\times=-0.1$, with values ranging from $-1.8$ to $+1.8$, fully consistent with zero, as expected for a null signal.

In the following, we use the \amalgam \WL measurements of $\Delta\Sigma_+(R)$ together with their covariance matrices (Section~\ref{subsec:cmat}) to perform mass modelling of the individual clusters (Section~\ref{subsec:single}). These cluster-by-cluster constraints then form the basis for the stacked-lensing and population-level analyses presented in Sections~\ref{subsec:stack} and~\ref{sec:scaling}, respectively.

\subsection{Weak-lensing mass maps around CHEX-MATE clusters}
\label{subsec:kmap}

The \amalgam shear data are also presented in the form of two-dimensional projected mass maps for individual clusters, shown as thumbnails in Appendix~\ref{appendix:individual} (Figure~\ref{fig:kmap_all}). These reconstructions are intended primarily for visualisation and qualitative assessment of the projected mass distribution, rather than for the quantitative mass measurements we used.

For a shear field generated by a lensing potential, the $E$ mode is the physical lensing signal and is directly related to the convergence $\kappa(\btheta)$, or equivalently proportional to the projected surface mass density of the lens, $\Sigma(\btheta)$. By contrast, the corresponding $B$-mode signal vanishes identically in the ideal \WL limit and therefore serves as a null test for residual systematics and noise.

For all clusters, we constructed smoothed reduced-shear fields, $(g_1(\btheta),g_2(\btheta))$, on a regular grid covering a fixed $26\arcmin\times26\arcmin$ field centred on the adopted X-ray peak position (Table~\ref{tab:sample}). This common field size was chosen to match the Suprime-Cam field-of-view scale and to avoid using shape measurements close to observed-image boundaries. The reduced-shear field was smoothed with a circular Gaussian kernel with a full width at half maximum (FWHM) of $4.0\arcmin$.

The projected mass maps were obtained from these smoothed reduced-shear grids using the linear Kaiser--Squires inversion \citep{KS1993}, implemented in Fourier space.\footnote{We did not apply an iterative non-linear $g\rightarrow\gamma$ correction for these visual maps. Such corrections introduce higher-order terms in the observed reduced-shear field \citep[see Equation~52 of][]{Umetsu2020rev} and hence modify the noise propagation relative to the linear reconstruction. Since the maps were used only for visualisation and qualitative quality assessment, we kept the reconstruction linear.}
In the map-making procedure, the pixelised reduced-shear field was weighted by the inverse of its error-variance map \citep[for details, see][]{Umetsu2009}. To mitigate spurious boundary artefacts associated with the periodic boundary condition imposed by the Fourier inversion, the input grids were zero-padded to twice the original length in each spatial dimension before reconstruction.

In Figure~\ref{fig:kmap_all}, the adopted smoothing highlights the dominant projected mass concentrations on cluster scales while suppressing small-scale noise fluctuations in the reconstructed field. The lowest contour level and the contour spacing are both set to $2\sigma_B$, where $\sigma_B$ is the rms of the corresponding $B$-mode map. The red circle centred on each image indicates the cluster radius $r_{500,\mathrm{WL}}$ estimated from mass modelling of the azimuthally averaged $\Delta\Sigma_+(R)$ profile (see Section~\ref{subsec:single}). The clusters are displayed in descending order of $\mathrm{S/N}_+$ for the $\Delta\Sigma_+(R)$ measurements shown in Figure~\ref{fig:gt_all}.

\begin{figure*}[t]
 \begin{center}
  \includegraphics[width=0.9\textwidth,angle=0,clip]{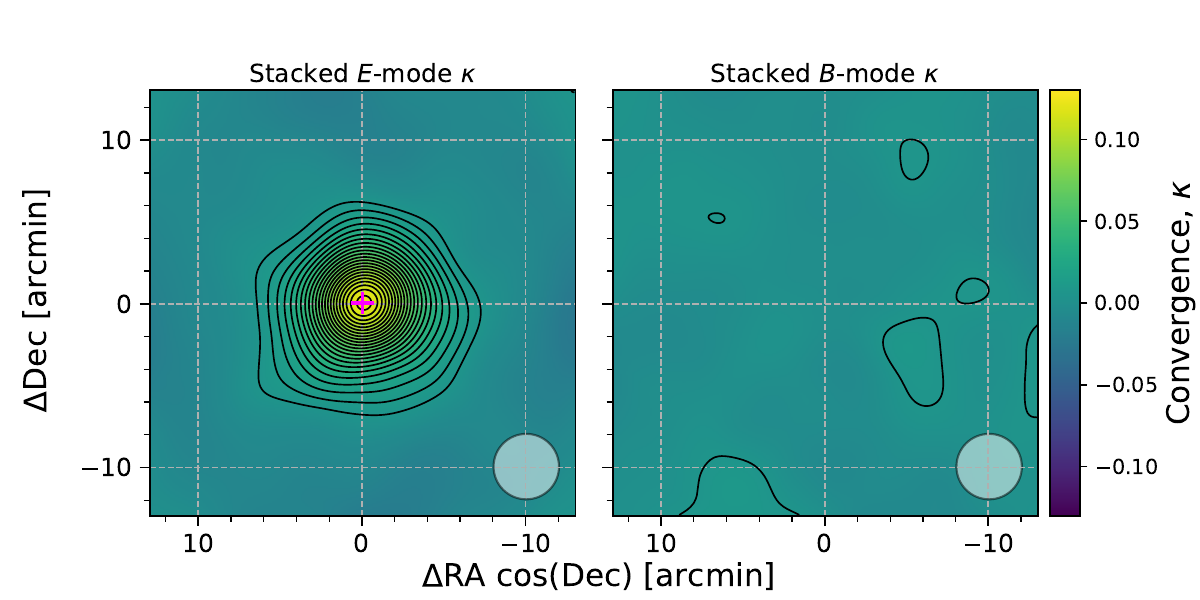} 
 \end{center}
 \caption{Averaged projected mass distribution of the CHEX-MATE--\amalgam sample of 41 galaxy clusters (left panel), derived from a weighted stack of $E$-mode $\kappa$ maps, each centred on the corresponding X-ray peak position (see Figure~\ref{fig:kmap_all}). The contours begin at an S/N of 2 and increase in increments of 2. The average cluster mass distribution reaches a peak significance of $47\sigma$. As a null test, the right panel shows the weighted stack of $B$-mode $\kappa$ maps for the same sample, where no significant signal is expected if the observed signal arises from \WL. The magenta plus sign marks the location of the maximum reconstructed $\kappa$ peak in the stacked $E$-mode map. No peak marker is shown for the stacked $B$-mode map. The shaded circle indicates the FWHM of the Gaussian window ($4\arcmin$).}
 \label{fig:kmap_stack}
\end{figure*}

Although these mass maps were not used directly in the subsequent mass modelling, they provide useful diagnostics of data quality, observational systematics, and possible offsets between the reconstructed mass distribution and the adopted X-ray centre \citep{Clowe2006,Okabe+Umetsu2008,Jee2009}. In particular, the maximum of the reconstructed E-mode map provides a qualitative estimate of the apparent X-ray--\WL peak separation, $D_{\mathrm{X\text{--}WL}}$. We caution, however, that this quantity was measured from maps smoothed with FWHM $4\arcmin$, and that no uncertainty on the \WL peak position was assigned here. The apparent peak offsets were therefore used only as qualitative diagnostics of the projected mass distribution, and not to define the centring model or enter the quantitative mass analysis.

\begin{figure}[tbp]
 \begin{center}
  \includegraphics[width=0.9\columnwidth,angle=0,clip]{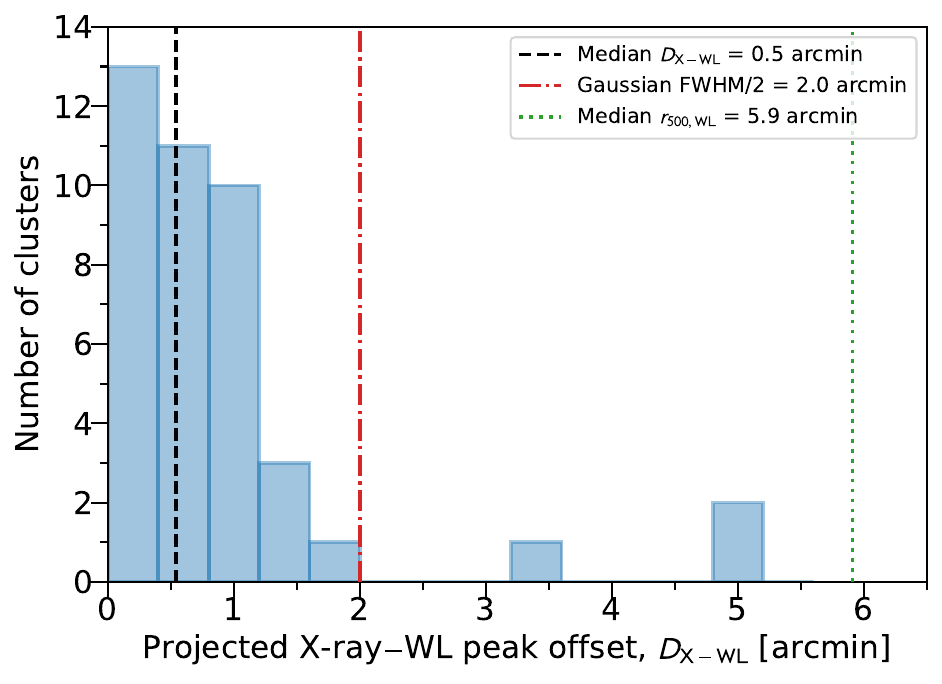}
 \end{center}
 \caption{Histogram of the projected separation $D_{\mathrm{X\text{--}WL}}$ between the X-ray peak and the maximum reconstructed $E$-mode mass peak for the 41 clusters in our sample. The vertical dashed black line marks the sample median, $D_{\mathrm{X\text{--}WL}}=0.54\arcmin$. The vertical dash-dotted red line marks $\mathrm{FWHM}/2=2.0\arcmin$, corresponding to the one-sided half-width of the Gaussian smoothing kernel used in the mass reconstruction. The vertical dotted green line marks the median $r_{500,\mathrm{WL}}$ of the sample. Only three systems show offsets larger than $2\arcmin$.}
 \label{fig:offset}
\end{figure}

As shown in Figure~\ref{fig:offset}, the reconstructed $E$-mode mass peaks are generally well aligned with the X-ray peaks. The distribution of projected peak offsets, $D_{\mathrm{X\text{--}WL}}$, is strongly concentrated toward small values, with a median of $0.54\arcmin$, corresponding to $0.10\,r_{500,\mathrm{WL}}$, where $r_{500,\mathrm{WL}}$ is the value of $r_{500}$ inferred for each cluster from our \WL mass modelling in Section~\ref{sec:mass}. In comoving projected units, the median offset is $\approx 0.14\Mpch$, below the inner radial limit of the azimuthally averaged shear-profile analysis, $R_\mathrm{min}=0.3\Mpch$. Five systems have offsets exceeding $R_\mathrm{min}$, but only two are substantially above this scale, with $D_{\mathrm{X\text{--}WL}}\approx 0.76\Mpch$ (PSZ2~G041.45$+$29.10) and $0.83\Mpch$ (PSZ2~G124.20$-$36.48). Approximately $80\percent$ of the clusters have offsets within $0.2\,r_{500,\mathrm{WL}}$, and $90\percent$ lie within $0.3\,r_{500,\mathrm{WL}}$.

The high-offset tail visible in Figure~\ref{fig:offset} is associated with low-$\mathrm{S/N}_+$ systems. The three largest angular-offset systems have $\mathrm{S/N}_+=4.3$ (PSZ2~G124.20$-$36.48), $3.1$ (PSZ2~G041.45$+$29.10), and $2.3$ (PSZ2~G046.88$+$56.48). These systems are among the nine lowest-ranked clusters in $\mathrm{S/N}_+$ (see Figure~\ref{fig:kmap_all} and Table~\ref{tab:sample}). These large apparent offsets should therefore be interpreted as qualitative map-based diagnostics, not as inputs to the centring model or quantitative mass analysis.

For the full sample, inverse-variance-weighted stacks of the $E$- and $B$-mode maps are shown in Figure~\ref{fig:kmap_stack}, providing a complementary summary of the mean projected mass signal and of the null $B$-mode behaviour. The stacked $E$-mode map exhibits a symmetric, single-peaked mass distribution with a peak significance of $47\sigma$, where $\sigma=(\sum_{l=1}^{\Ncl} \sigma_{B,l}^{-2})^{-1/2}$ is the inverse-variance-weighted noise scale of the stack, constructed from the rms values of the individual $B$-mode maps over the $\Ncl$ clusters. The offset between the reconstructed mass peak and the adopted X-ray peak in the stacked $E$-mode map is smaller than the map pixel scale of $0.1\arcmin$. By contrast, the stacked $B$-mode map shows no significant signal, as expected for a null test.

\section{Weighing CHEX-MATE clusters}
\label{sec:mass}

In this section, we use the \amalgam \WL data to infer the mass and concentration parameters of our cluster sample. In Section~\ref{subsec:modelling}, we describe our \WL modelling procedure and summarise the calibration tests of the shear-to-mass modelling pipeline based on simulations. Section~\ref{subsec:single} presents the \WL mass estimates for individual clusters. Section~\ref{subsec:stack} presents the results of the stacked lensing analysis. In Section~\ref{subsec:syst}, we discuss and summarise the systematic uncertainties in the ensemble \WL modelling of the sample.

\subsection{Cluster mass modelling}
\label{subsec:modelling}

We modelled the radial mass distribution of galaxy clusters with a Navarro--Frenk--White (\citealt{NFW1996,NFW1997}; NFW) density profile, motivated by cosmological $N$-body simulations \citep[][]{Oguri+Hamana2011,Child2018cm} as well as by direct lensing measurements \citep[][]{Umetsu2016clash,Niikura2015,Okabe+Smith2016}. The radial dependence of the NFW density profile is given by \citep{NFW1996}
\begin{equation}
 \label{eq:NFW}
 \rho_\mathrm{NFW}(r)=\frac{\rho_\mathrm{s}}{(r/r_\mathrm{s})(1+r/r_\mathrm{s})^2},
\end{equation}
where $\rho_\mathrm{s}$ is the characteristic density, and $r_\mathrm{s}$ is the scale radius at which the logarithmic density slope equals $-2$.

The overdensity mass $M_\Delta$ is defined by integrating Equation~\eqref{eq:NFW} out to the corresponding overdensity radius $r_\Delta$, within which the mean interior density is $\Delta\,\rho_\mathrm{c}(\zl)$ (Section~\ref{sec:intro}), so 
\begin{equation}
 M_\Delta=\frac{4\pi\Delta}{3}\rho_{\mathrm{c}}(\zl)r_\Delta^3.
\end{equation}
We specified the NFW model by the mass, $M_{200}$, and the concentration parameter, $c_{200}=r_{200}/r_\mathrm{s}$. The corresponding characteristic density is then
\begin{equation}
 \rho_\mathrm{s}=
 \frac{\Delta}{3}
 \frac{c_{\Delta}^3}{\ln(1+c_{\Delta})-c_{\Delta}/(1+c_{\Delta})}\rho_\mathrm{c}(\zl).
\end{equation}

We used a Markov chain Monte Carlo (MCMC) method to obtain well-characterised posterior constraints on the mass and concentration parameters from the \amalgam \WL data. We adopted log-uniform priors on $M_{200}$ and $c_{200}$, equivalently uniform priors on $\log(M_{200}/\Msunh)$ and $\log c_{200}$, over the ranges $10^{13} \leq M_{200}/(\Msunh) \leq 10^{16}$ and $1 \leq c_{200} \leq 20$.

A log-uniform prior, rather than a uniform prior, is appropriate for a positive-definite quantity that spans a wide dynamic range \citep{Sereno+Covone2013,Umetsu2014clash,Umetsu2020xxl}. This choice is also consistent with our scaling-relation analysis, where we work with the logarithmic quantities $\log M_{\Delta}$ and $\log c_{200}$ (Section~\ref{sec:scaling}). Since the corresponding prior densities for $M_{200}$ and $c_{200}$ scale as $1/M_{200}$ and $1/c_{200}$, respectively, the choice of the lower prior bounds is important. The adopted priors span a sufficiently broad range of mass and concentration values relevant to the \Planck SZ-selected CHEX-MATE cluster sample.

The log-likelihood function for the observed profile $\{\Delta\Sigma_+(R_i)\}_{i=1}^{\Nbin}$ is given, up to an additive constant, by
\begin{equation}
\begin{aligned}
-\ln{\mathcal{L}}(\bp) &= \frac{1}{2}\sum_{i,j=1}^{N}
\left[\Delta\Sigma_+(R_i) - f_\mathrm{model}(R_i|\bp)\right] \\
&\times (C^{-1})_{ij}
\left[\Delta\Sigma_+(R_j) - f_\mathrm{model}(R_j|\bp)\right],
\end{aligned}
\end{equation}
where $C^{-1}$ is the inverse covariance matrix, and $f_\mathrm{model}(R_i|\bp)$ is the theoretical prediction for the parameter vector $\bp = (M_{200}, c_{200})$.

The radial dependence of the projected NFW profiles, $\Sigma_\mathrm{NFW}(R|\bp)$ and $\Delta\Sigma_\mathrm{NFW}(R|\bp)$, was computed using the analytic expressions of \citet{2000ApJ...534...34W}. The contribution of the 2-halo term to $\Delta\Sigma$ is expected to become important mainly at scales of several virial radii \citep{Oguri+Hamana2011}, which are beyond our outer radial limit of $R_\mathrm{max}=3\Mpch$. We therefore fitted the data vector, $\{\Delta\Sigma_+(R_i)\}_{i=1}^{\Nbin}$, over the full radial range $R \in [0.3,3]\Mpch$ in comoving units.

Using the leading non-linear correction given by Equation~\eqref{eq:deltaSigma}, we modelled the excess surface mass density profile as
\begin{equation}
 f_\mathrm{model}(R_i|\bp) =
 \frac{\Delta\Sigma_\mathrm{NFW}(R_i|\bp)}
 {1-\langle\Sigma_{\mathrm{cr},i}^{-1}\rangle\,\Sigma_\mathrm{NFW}(R_i|\bp)},
\end{equation}
where $\langle\Sigma_{\mathrm{cr},i}^{-1}\rangle$ is the sensitivity-weighted inverse critical surface mass density in the $i$th radial bin, as given by Equation~\eqref{eq:inv_sigma_crit}.

As summary statistics for the posterior distributions, we used the biweight estimator of \citet{1990AJ....100...32B} to characterise the central location (\CBI) and scale (\SBI) of the marginalised one-dimensional posterior distributions \citep[e.g.][]{Stanford1998,Sereno+Umetsu2011,Biviano2013,Umetsu2020xxl}. Biweight statistics are robust to noisy outliers because they assign larger weights to samples closer to the centre of the distribution. 
For a symmetric distribution, \CBI approaches the centre of the distribution; for a moderately skewed distribution such as a lognormal distribution, it typically remains close to the median.
From the posterior samples, we derived marginalised constraints on the total mass $M_\Delta$ and concentration $c_\Delta$ at several characteristic overdensities $\Delta$.

For clarity, we denote the cluster mass and concentration inferred from the \WL analysis described above by $M_{\Delta,\mathrm{WL}}$ and $c_{\Delta,\mathrm{WL}}$. These quantities represent the observationally inferred NFW parameters derived from the $\Delta\Sigma_+$ profile measurements. When comparing with latent or simulation-input halo properties, we denote the corresponding true quantities by $M_{\Delta,\mathrm{true}}$ and $c_{\Delta,\mathrm{true}}$.

\subsection{Individual and average cluster-lensing measurements}
\label{subsec:single}

\begin{figure}[tbp]
 \begin{center}
  \includegraphics[width=0.9\columnwidth,angle=0,clip]{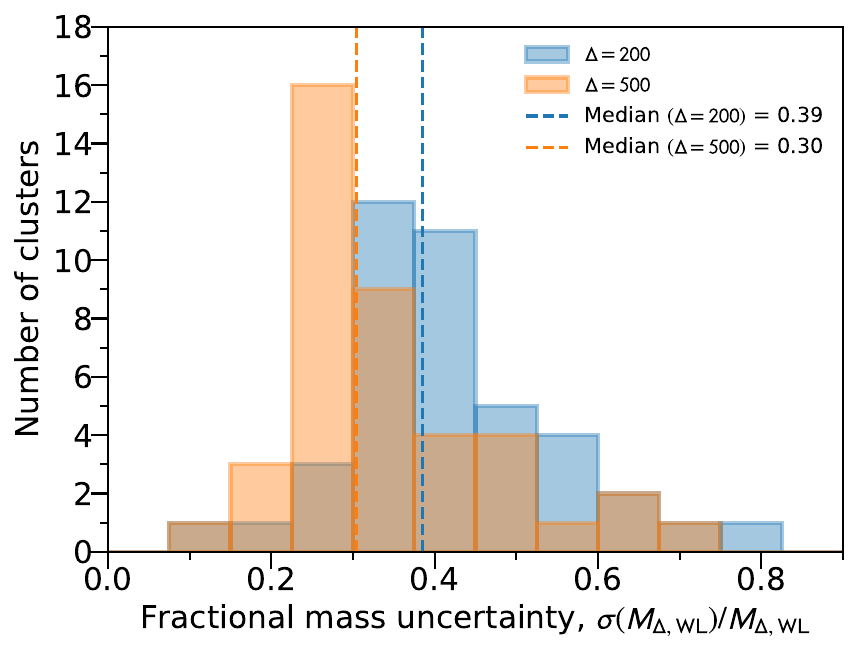} 
 \end{center}
 \caption{Distributions of the fractional \WL mass uncertainty, $\sigma(M_{\Delta,\mathrm{WL}})/M_{\Delta,\mathrm{WL}}$, for the full sample of 41 clusters, shown separately for $\Delta=200$ (shaded blue) and $\Delta=500$ (shaded orange). The vertical dashed blue and orange lines mark the median values of $\sigma(M_{200,\mathrm{WL}})/M_{200,\mathrm{WL}}$ and $\sigma(M_{500,\mathrm{WL}})/M_{500,\mathrm{WL}}$, respectively.}
 \label{fig:mprec} 
\end{figure}

In Table~\ref{tab:mass} we list symmetrised posterior summaries ($C_\mathrm{BI}\pm S_\mathrm{BI}$) for the mass and concentration parameters $(c_{200,\mathrm{WL}}, M_{200,\mathrm{WL}}, M_{500,\mathrm{WL}})$ of all clusters in the full sample, derived from NFW modelling of the \WL data, in logarithmic and linear variables. The logarithmic summaries were adopted in the hierarchical regression analysis of Section~\ref{sec:scaling}, while the linear summaries are provided for direct physical interpretation. Figure~\ref{fig:mprec} shows the distributions of the fractional \WL mass uncertainty, $\sigma(M_{\Delta,\mathrm{WL}})/M_{\Delta,\mathrm{WL}}$, for the full sample, separately for $\Delta=200$ and $\Delta=500$. The median fractional uncertainties are $0.39$ for $M_{200,\mathrm{WL}}$ and $0.30$ for $M_{500,\mathrm{WL}}$. These fractional uncertainties were evaluated from the symmetrised posterior summaries in the linear mass variables.

\begin{figure*}[tbp]
 \begin{center}
  \includegraphics[width=0.9\textwidth,angle=0,clip]{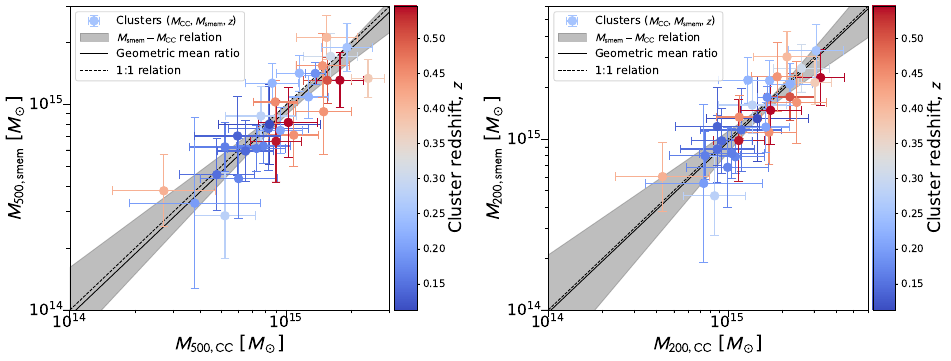}
 \end{center}
 \caption{Consistency of the \WL mass estimates derived from two different source-selection methods for a subset of 30 CHEX-MATE--\amalgam clusters for which the \smem-threshold and CC selection methods are available (see Figure~\ref{fig:smem}). The left and right panels show the cluster mass $M_\Delta$ at overdensities of $\Delta=500$ and $\Delta=200$, respectively. The circles with error bars represent the measured masses $(M_{\Delta,\mathrm{CC}}, M_{\Delta,\mathrm{smem}})$ along with their $1\sigma$ uncertainties for individual clusters. The cluster redshifts are colour-coded according to the colour bar. In each panel, the solid black line corresponds to the weighted geometric mean ratio, $\langle M_{\mathrm{smem}}/M_{\mathrm{CC}} \rangle_\mathrm{g}$, while the shaded grey region represents the marginalised $1\sigma$ credible interval for the mean $M_\mathrm{smem}$--$M_\mathrm{CC}$ relation. The covariance between the measured masses, arising from the partial overlap of background samples selected by the two methods, is not shown but was accounted for in the regression. The average overlap fraction of the source samples from the \smem-threshold and CC selection methods is $\approx 60\percent$.}
 \label{fig:msys}
\end{figure*}

To characterise the average \WL-inferred mass or concentration of a given cluster sample, we used geometric means rather than arithmetic means \citep[][]{Umetsu2014clash}. This choice is particularly natural for our scaling-relation analysis, which is formulated in logarithmic space (Section~\ref{sec:scaling}). Specifically, for a cluster parameter such as $\Mwl$, we define the error-weighted geometric mean of a sample of $\Ncl$ clusters as
\begin{equation}
\label{eq:geom}
 \langle \Mwl\rangle_\mathrm{g} \equiv e^{\langle \ln \Mwl\rangle}
 = \exp\left(
	\frac{\sum_{l=1}^{\Ncl} u_l \ln M_{\mathrm{WL},l}}{\sum_{l=1}^{\Ncl} u_l}
       \right),
\end{equation}
with the symmetrised uncertainty
\begin{equation}
\sigma_{\langle \Mwl \rangle_\mathrm{g}} =
\frac{\langle \Mwl \rangle_\mathrm{g}}{2}
\left[
\exp\left(\frac{1}{\sqrt{\sum_{l=1}^{\Ncl} u_l}}\right)
-
\exp\left(-\frac{1}{\sqrt{\sum_{l=1}^{\Ncl} u_l}}\right)
\right].
\end{equation}
Here, $u_l$ is the inverse-variance weight for the $l$th cluster, including the contribution from intrinsic scatter,
\begin{equation}
u_l^{-1}= S_\mathrm{BI}^2(\ln M_{\mathrm{WL},l}) + \sigma^2_\mathrm{int}(\ln \Mwl),
\end{equation}
where $S_\mathrm{BI}(\ln M_{\mathrm{WL},l})$ denotes the biweight scale (Section~\ref{subsec:modelling}) of the marginalised posterior distribution of $\ln\Mwl$ for the $l$th cluster, and $\sigma_\mathrm{int}(\ln \Mwl)$ is the lognormal intrinsic scatter in \WL mass at fixed true mass, $\Mtrue$. We adopted a fixed value of $\sigma_\mathrm{int}(\ln\Mwl)=0.2$, corresponding to $\approx 0.09$~dex, motivated by synthetic \WL data from cosmological simulations (Appendix~\ref{appendix:test}). An analogous estimator was used for the geometric mean concentration, $\langle c_{200,\mathrm{WL}}\rangle_\mathrm{g}$, for which we adopted a fixed intrinsic scatter of $\sigma_\mathrm{int}(\ln c_{200,\mathrm{WL}})=0.16\ln(10)$, corresponding to 0.16~dex, motivated by cosmological numerical simulations \citep{Diemer2015}. This assumption affects only the geometric-mean comparison quantities quoted in Table~\ref{tab:stack}, and not the stacked-profile modelling itself.

An additional advantage of the geometric mean is its symmetry under inversion, $\langle X/Y\rangle_\mathrm{g}=\langle Y/X\rangle_\mathrm{g}^{-1}$. This makes the estimator well suited for quantifying mean mass ratios between two cluster samples \citep{Donahue2014clash,Umetsu2014clash,Umetsu2016clash,Umetsu2020xxl}.

To assess the internal consistency of the two background-selection methods, we compared the \WL masses obtained with the conservative CC selection and with the \smem-threshold selection for the subset of 30 clusters for which both estimates are available. For each overdensity, we modelled the relation between the two mass estimates with a log-linear regression of the form
\begin{equation}
\log\left(\frac{M_{\Delta,\mathrm{smem}}}{M_{\Delta,\mathrm{piv}}}\right) = A + B \log\left(\frac{M_{\Delta,\mathrm{CC}}}{M_{\Delta,\mathrm{piv}}}\right),
\end{equation}
where $A$ and $B$ denote the intercept and slope of this mass-comparison relation, respectively. Exact agreement with no mass-dependent trend corresponds to $A=0$ and $B=1$. We included an intrinsic scatter term and fitted the relation using the LInear Regression in Astronomy (\lira) package \citep{Sereno2016lira,Sereno2016lirapackage}. We further accounted for the covariance between the two mass estimates arising from the partial overlap of the background source samples. For each cluster, we approximated the cross-correlation coefficient by the overlap fraction between the source samples selected with the \smem-threshold selection and with the CC selection, with a mean overlap fraction of $\sim 60\percent$ over the 30-cluster subsample (Section~\ref{subsec:back}).

The fitted relations are shown in Figure~\ref{fig:msys}. For $\Delta=500$ and $\Delta=200$, the slope is consistent with unity and the intercept is consistent with zero within the uncertainties, indicating no significant deviation from a one-to-one relation between the two mass estimators. Throughout this work, we adopted fixed reference pivot masses of $M_{500,\mathrm{piv}}=7\times 10^{14}\Msun$ and $M_{200,\mathrm{piv}}=10^{15}\Msun$, chosen to represent the characteristic high-mass scale of the CHEX-MATE sample. Specifically, $M_{500,\mathrm{piv}}$ lies close to the CHEX-MATE Tier-2 selection threshold, $\Msz>7.25\times10^{14}\Msun$, and nearly matches the value adopted in the dynamical mass calibration of \citet{Sereno2025chex}. The corresponding choice of $M_{200,\mathrm{piv}}$ is consistent with this mass scale under the typical conversion between $M_{500}$ and $M_{200}$, and matches the $M_{200}$ pivot used by \citet{Sereno2025chex}. At these pivot masses, the inferred mass ratios are $M_{500,\mathrm{smem}}/M_{500,\mathrm{CC}}=0.96\pm0.09$ and $M_{200,\mathrm{smem}}/M_{200,\mathrm{CC}}=0.96\pm0.14$. These results indicate an average offset of about $4\percent$, with the masses obtained from the \smem-threshold selection being slightly lower than those obtained from the CC selection.

As an external consistency check, we compared our \WL $M_{500}$ estimates with published \WL masses compiled in version 3.9 of the Literature Catalogs of weak Lensing Clusters (LC2) meta-catalogues\footnote{\url{http://pico.oabo.inaf.it/~sereno/CoMaLit/LC2/}} \citep{CoMaLit3}. Such a cluster-by-cluster comparison places our measurements in the context of previous \WL studies and provides a direct test of consistency across different datasets and analysis pipelines.

\begin{figure*}[tbp]
 \begin{center}
  \includegraphics[width=0.9\textwidth,angle=0,clip]{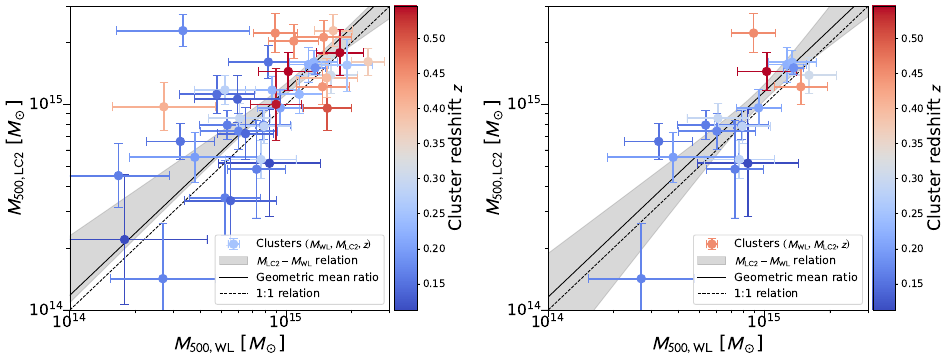}
 \end{center}
 \caption{Comparison of $M_{500}$ estimates from this study with literature \WL masses from version 3.9 of the LC2 meta-catalogues \citep{CoMaLit3} for the CHEX-MATE--\amalgam sample. The left and right panels show comparisons with and without including LC2 mass estimates from the WtG program \citep{WtG3,Herbonnet2020}, respectively. The circles with error bars represent the measured masses $(M_{500,\mathrm{WL}}, M_{500,\mathrm{LC2}})$ along with their $1\sigma$ uncertainties for individual clusters. The cluster redshifts are colour-coded according to the accompanying colour bar. In each panel, the solid black line represents the weighted geometric mean ratio, $\langle M_{\mathrm{LC2}}/\Mwl\rangle_\mathrm{g}$, while the shaded grey region denotes the marginalised $1\sigma$ credible interval for the mean $M_\mathrm{LC2}$--$\Mwl$ relation. The inferred mean mass ratios at $M_{500,\mathrm{WL}}=7\times 10^{14}\Msun$, with and without WtG mass estimates, are $M_\mathrm{LC2}/\Mwl=1.24\pm 0.11$ (41 clusters) and $1.12\pm 0.15$ (19 clusters), respectively.}
 \label{fig:lc2}
\end{figure*}

Figure~\ref{fig:lc2} shows the comparison between our $\Mwl$ estimates and the corresponding LC2 literature masses at $\Delta=500$. As in the internal comparison between the CC selection and the \smem-threshold selection, we used \lira to perform a direct regression between the two sets of measured masses, accounting for their statistical uncertainties. In this comparison, we assumed that covariances between our and LC2 mass estimates are zero. The left panel shows the full sample of 41 matched clusters. At the pivot mass $M_{500,\mathrm{WL}}=7\times10^{14}\Msun$, we find $M_{\mathrm{LC2}}/\Mwl = 1.24\pm0.11$. The fitted slope, $B=0.88\pm0.16$, is consistent with unity, indicating no clear evidence for a mass-dependent trend in the mass ratio over the range probed by the data. Thus, the main difference in the full-sample comparison is a modest normalisation offset, with the published LC2 masses tending to be higher than our \WL masses on average.

This full-sample comparison is, however, strongly influenced by measurements from the Weighing the Giants program \citep[WtG;][]{WtG3,Herbonnet2020}, which contribute 22 of the 41 LC2 mass estimates matched to our sample. In particular, the WtG masses were derived using NFW fits with the concentration fixed to $c_{200}=4$ and with a boost-factor correction for residual contamination by cluster members, whereas in our analysis $c_{200}$ was allowed to vary over a broad range, $c_{200}\in [1,20]$, with a log-uniform prior for each cluster, and the background sample was defined using the CC or \smem-threshold selection without applying a boost-factor correction \citep[see also][for a discussion of methodological differences among cluster \WL analyses]{Okabe+Smith2016}.

To examine their impact, the right panel of Figure~\ref{fig:lc2} shows the corresponding comparison with the WtG measurements excluded. In this case, the inferred mass ratio at the same pivot mass becomes $M_{\mathrm{LC2}}/\Mwl = 1.12\pm0.15$, indicating a substantially reduced offset. The fitted slope, $B=1.07\pm0.29$, is likewise fully consistent with unity. Given the heterogeneous nature of the LC2 compilation, which combines results from different datasets and analysis pipelines, we regard this comparison primarily as an external validation test rather than as a calibration dataset for our main analysis.

\subsection{Stacked cluster-lensing measurements}
\label{subsec:stack}

Stacking an ensemble of clusters suppresses statistical fluctuations in the \WL measurements of individual systems and thereby improves the precision of the mean lensing signal. The resulting stacked analysis is complementary to our primary approach based on individual-cluster \WL mass measurements, and a comparison of the two provides a useful consistency check based on different averaging and inference procedures. Interpreting the effective mass associated with a stacked lensing signal, however, requires some caution, because the signal amplitude is weighted by the lensing sensitivity and does not scale linearly with halo mass \citep[e.g.][]{Melchior2017,Sereno2017psz2lens}.

For a given subsample of clusters, we constructed the stacked $\Delta\Sigma_+$ profile by inverse-covariance weighting of the individual-cluster measurements. Denoting by $\Delta\Sigma_{+,l}$ the binned profile vector of cluster $l$ and by $C_l$ its associated covariance matrix (Section~\ref{subsec:cmat}), the stacked profile is given by \citep{Umetsu2020rev}
\begin{equation}
\label{eq:gt_stack}
\Delta\Sigma^{\mathrm{stack}}_+ =
\left(\sum_{l=1}^{\Ncl} C_l^{-1}\right)^{-1}
\left(\sum_{l=1}^{\Ncl} C_l^{-1}\Delta\Sigma_{+,l}\right),
\end{equation}
with the covariance matrix
\begin{equation}
\label{eq:cov_stack}
C^{\mathrm{stack}} = \left(\sum_{l=1}^{\Ncl} C_l^{-1}\right)^{-1}.
\end{equation}
Thus, the stacked analysis was carried out at the level of the individual-cluster lensing profiles, with each cluster weighted by its full covariance matrix. Likewise, using the same inverse-covariance weighting, we constructed the stacked cross-component profile $\Delta\Sigma_\times^\mathrm{stack}$, with covariance matrix $C^\mathrm{stack}$. The subsequent model fits to the stacked tangential-component profile used the full stacked covariance matrix, $C^\mathrm{stack}$, defined in Equation~\eqref{eq:cov_stack}.

For the tangential- and cross-component stacked profiles, we estimated the linear S/N using the full stacked covariance matrix. For a stacked profile $d_i$ with covariance matrix $C^{\mathrm{stack}}_{ij}$, we define
\begin{equation}
\label{eq:snr_stack}
\mathrm{S/N} =
\frac{\sum_{i,j=1}^{N} (C^{\mathrm{stack}})^{-1}_{ij} d_j}
     {\left[\sum_{i,j=1}^{N} (C^{\mathrm{stack}})^{-1}_{ij}\right]^{1/2}}.
\end{equation}
Here, $d_i=\Delta\Sigma_{+,i}^{\mathrm{stack}}$ for the tangential component and $d_i=\Delta\Sigma_{\times,i}^{\mathrm{stack}}$ for the cross component. This expression is the full-covariance analogue of the linear estimator in Equation~\eqref{eq:snr} and reduces to the diagonal inverse-variance form when $C^{\mathrm{stack}}$ is diagonal.\footnote{In matrix notation, Equation~\eqref{eq:snr_stack} can be written as
$\mathrm{S/N}=
\mathbf{1}^{T}(C^\mathrm{stack})^{-1}\mathbf{d}/
\left[\mathbf{1}^{T}(C^\mathrm{stack})^{-1}\mathbf{1}\right]^{1/2}$,
where $\mathbf{d}$ is the stacked profile vector and $\mathbf{1}$ denotes an $N$-dimensional column vector with all entries equal to unity.}

\begin{figure}[tbp]
 \begin{center}
  \includegraphics[width=0.9\columnwidth,angle=0,clip]{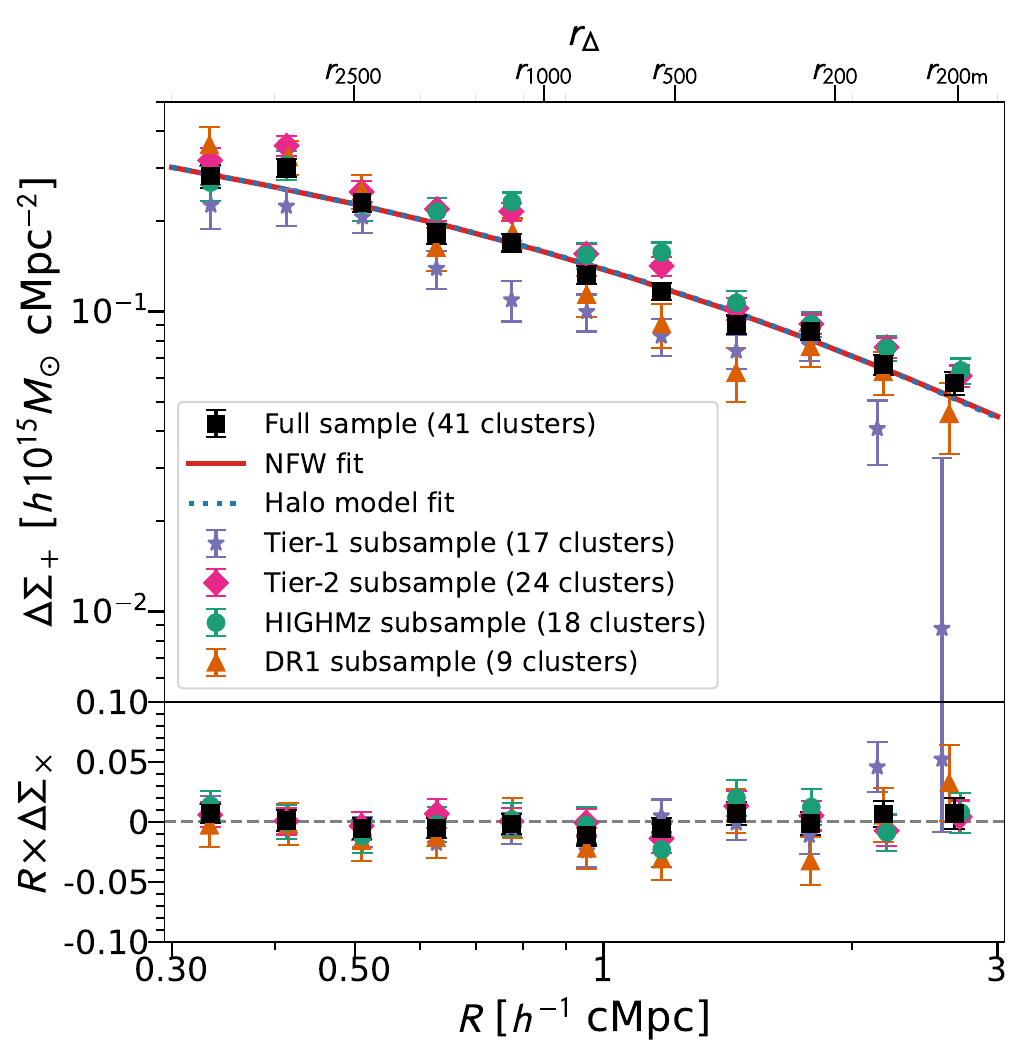}
 \end{center}
 \caption{Stacked excess surface mass density $\Delta\Sigma_+^\mathrm{stack}$, shown as a function of cluster-centric comoving radius, $R$ (upper panel; see also Figure~\ref{fig:kmap_stack}). The black squares show the stacked profile for the full sample of 41 clusters. The solid and dotted lines show the best-fit NFW and halo-model profiles, respectively. The circles, triangles, stars, and diamonds denote the HIGHMz, DR1, Tier-1, and Tier-2 subsamples, respectively; the tier-based stacks are defined to be mutually exclusive as described in the text. The lower panel shows the corresponding $45^\circ$-rotated cross component, $R \Delta\Sigma_\times^\mathrm{stack}$, which provides a null test.}
 \label{fig:stackgt}
\end{figure}

Figure~\ref{fig:stackgt} shows the stacked $\Delta\Sigma_+^\mathrm{stack}$ profile for the full sample of 41 clusters. In addition to the full sample, we show stacked measurements for several subsamples: HIGHMz, DR1, Tier-1, and Tier-2. Here HIGHMz denotes the high-mass ($\Msz>7.75\times10^{14}\Msun$), high-redshift ($z>0.2$) subsample \citep{Riva2024}; DR1 denotes the ``Data Release 1'' subsample of CHEX-MATE clusters, designed to be representative of the full CHEX-MATE sample \citep{Rossetti2024chex}; Tier-1 denotes the low-redshift northern-hemisphere selection; and Tier-2 denotes the high-mass CHEX-MATE selection (Section~\ref{sec:intro}).

For this stacked comparison, the tier-based subsamples are taken to be mutually exclusive: Tier-1 includes Tier-1-only and Tier-1+2 clusters, while Tier-2 includes only Tier-2-only clusters. The HIGHMz and DR1 subsamples are defined independently and therefore overlap with the tier-based subsamples. We note that the DR1 subsample used here contains only 9 clusters with available \amalgam \WL measurements, and therefore does not necessarily preserve the representativeness of the full CHEX-MATE DR1 sample. The slight inward shift of the outer effective bin centres for the Tier-1 and DR1 subsamples reflects the harmonic-bin-centre definition adopted for the lensing profiles (Equation~\ref{eq:binradius}) and the finite imaging field of view for the lower-redshift clusters, which limits their radial coverage at large projected radii.

\begin{table*}
\caption{Summary of the stacked-subsample lensing analysis.}
\label{tab:stack}
\centering
\normalsize
\setlength{\tabcolsep}{3.0pt}
\begin{tabular}{lccccccccccc}
\hline\hline
Sample & $N_{\mathrm{cl}}$ & $z_{l,\mathrm{med}}$ & $\langle z_l\rangle_{\mathrm{WL}}$ & $\mathrm{S/N}_+$ & $\mathrm{S/N}_\times$ & $c_{200,\mathrm{WL}}^{\mathrm{stack}}$ & $M_{200,\mathrm{WL}}^{\mathrm{stack}}$ & $M_{500,\mathrm{WL}}^{\mathrm{stack}}$ & $\langle c_{200,\mathrm{WL}}\rangle_{\mathrm{g}}$ & $\langle M_{200,\mathrm{WL}}\rangle_{\mathrm{g}}$ & $\langle M_{500,\mathrm{WL}}\rangle_{\mathrm{g}}$ \\
 &  &  &  &  &  &  & ($10^{14}\Msun$) & ($10^{14}\Msun$) & & ($10^{14}\Msun$) & ($10^{14}\Msun$)\\
\hline
Full & 41 & 0.23 & 0.28 & 38.1 & 0.1 & $3.11 \pm 0.27$ & $14.6 \pm 1.1$ & $9.6 \pm 0.6$ & $3.30 \pm 0.33$ & $14.6 \pm 1.0$ & $9.7 \pm 0.6$ \\
Tier-1 & 17 & 0.16 & 0.16 & 18.2 & -0.1 & $3.24 \pm 0.67$ & $8.4 \pm 1.1$ & $5.5 \pm 0.6$ & $3.10 \pm 0.57$ & $7.6 \pm 1.0$ & $5.3 \pm 0.6$ \\
Tier-2 & 24 & 0.34 & 0.33 & 33.6 & 0.2 & $3.35 \pm 0.31$ & $18.0 \pm 1.4$ & $12.0 \pm 0.8$ & $3.39 \pm 0.40$ & $18.2 \pm 1.4$ & $12.1 \pm 0.8$ \\
HIGHMz & 18 & 0.39 & 0.36 & 28.1 & 0.3 & $2.77 \pm 0.30$ & $20.2 \pm 1.9$ & $12.8 \pm 1.0$ & $2.89 \pm 0.40$ & $19.0 \pm 1.8$ & $12.2 \pm 1.0$ \\
DR1 & 9 & 0.18 & 0.25 & 16.9 & -1.0 & $5.54 \pm 1.27$ & $9.6 \pm 1.3$ & $7.0 \pm 0.8$ & $5.04 \pm 1.14$ & $11.2 \pm 1.6$ & $8.1 \pm 1.0$ \\
\hline
\end{tabular}
\tablefoot{Columns 1--4 list the stacked sample, number of clusters, median cluster redshift, and lensing-sensitivity-weighted mean cluster redshift, $\langle\zl\rangle_\mathrm{WL}$, used in the stacked NFW modelling. For the tier-based stacks, Tier-1 includes the Tier-1+2 clusters, whereas Tier-2 is restricted to Tier-2-only clusters. Columns 5 and 6 give the linear S/N values of the stacked tangential- and cross-component profiles, respectively, computed using Equation~\eqref{eq:snr_stack} with the full stacked covariance matrix. Columns 7--9 list the NFW quantities inferred from the stacked-profile analysis, and Columns 10--12 give the weighted geometric means and symmetrised uncertainties computed from the corresponding individual-cluster logarithmic posterior summaries for the same subsample. No correction for the \WL modelling bias was applied to the reported masses and concentrations.}
\end{table*}

The solid line in the upper panel shows the best-fit projected NFW model for the full sample, while the dotted line shows the corresponding halo-model prediction, constructed following the standard prescription of \citet{Oguri+Hamana2011} with a truncated NFW halo profile \citep{BMO} plus a 2-halo term describing the contribution from correlated large-scale structure. Over the radial range considered here, the two predictions are nearly indistinguishable (see Appendix~\ref{appendix:halomodel}). The lower panel displays the stacked cross-shear component, which serves as a null test and remains much smaller than the tangential signal. 

The best-fit NFW parameters, together with the linear S/N values $\mathrm{S/N}_+$ and $\mathrm{S/N}_\times$, for each subsample are summarised in Table~\ref{tab:stack}. We also list the effective cluster redshift, $\langle\zl\rangle_\mathrm{WL}$, adopted in the stacked NFW modelling, defined as
\begin{equation} 
\langle\zl\rangle_\mathrm{WL} = \frac{\sum_{l=1}^{\Ncl} \mathrm{Tr}(C_l^{-1})\, \zl} {\sum_{l=1}^{\Ncl}\mathrm{Tr}(C^{-1}_l)},
\end{equation}
with weights proportional to the trace of the inverse covariance matrix of the individual-cluster $\Delta\Sigma_+$ profiles. 

Overall, the stacked-subsample analysis yields NFW mass and concentration estimates that are broadly consistent with the corresponding geometric means of the individual-cluster constraints within the quoted uncertainties. In all cases, the tangential stacked signal is significantly detected, while the cross-shear signal is consistent with zero, with $|\mathrm{S/N}_\times|\simlt 1$.

\subsection{Systematic uncertainties in ensemble modelling}
\label{subsec:syst}

We accounted for the statistical uncertainties of the cluster \WL measurements using the total covariance matrix, $C=C^\mathrm{shape}+C^\mathrm{lss}$ (Equation~\ref{eq:cmat}), of the binned excess surface mass density profile, $\{\Delta\Sigma_+(R_i)\}_{i=1}^{\Nbin}$ (Section~\ref{subsec:DSigma}). These terms describe the measurement uncertainty of the observed \WL profiles for individual clusters, including shape noise and cosmic noise from uncorrelated large-scale structure.

In addition to these statistical uncertainties, the ensemble \WL modelling is subject to systematic effects that can bias the mapping between the observed \WL signal and the underlying halo properties. In our analysis, it is useful to distinguish between two classes of such effects. The first comprises simulation-calibrated NFW modelling biases in the recovered mass and concentration. For the mass calibration, we explicitly modelled the $\Mwl$--$\Mtrue$ relation described in Section~\ref{subsubsec:massbias} and Appendix~\ref{appendix:test}; for the concentration calibration, we applied the corresponding correction described below. The second is a set of residual observational calibration uncertainties, discussed in Section~\ref{subsubsec:residual}, that are not captured by this simulation-based treatment. A summary of our systematic treatment is given in Section~\ref{subsubsec:systsummary}.

\subsubsection{Simulation-calibrated mass and concentration biases}
\label{subsubsec:massbias}

We characterised simulation-calibrated biases in the \WL-inferred mass and concentration for the CHEX-MATE--\amalgam analysis using synthetic \WL data generated from a dark-matter-only realisation of the BAryons and HAloes of MAssive Systems (BAHAMAS) simulations \citep{McCarthy2017}, following \citet{Umetsu2020xxl}.
Appendix~\ref{appendix:test} describes these tests of our shear-to-mass modelling pipeline.

We used a subsample of 116 simulated high-mass haloes with $M_{500}>2\times10^{14}\Msun$ at $z=0.25$, selected from the BAHAMAS simulation. This snapshot redshift is close to the median redshift of the current sample, $\zlmed=0.23$. The BAHAMAS volume provides a useful calibration over the cluster-mass interval $2\times10^{14}\Msun \simlt M_{500}\simlt 10^{15}\Msun$, but does not fully sample the extreme high-mass tail, $M_{500}\gtrsim10^{15}\Msun$, reached by the most massive Tier-2 clusters. We therefore regard these tests as a calibration of the NFW modelling bias over the mass range directly sampled by this simulated subsample, with the uncertainty in the fitted $\Mwl$--$\Mtrue$ relation propagated in the population-level analysis (Section~\ref{sec:scaling}).

Over this mass range, the direct recovery tests show no statistically significant mean offset in the \WL mass estimates. The binned trends and weighted geometric mean ratios of $M_{\mathrm{WL}}/M_{\mathrm{true}}$ are consistent with unbiased mass recovery within the current statistical uncertainties (Appendix~\ref{appendix:test}). We then quantified the mass calibration by modelling the relation between the \WL-inferred mass, $\Mwl$, and the latent true halo mass, $\Mtrue$, as a power law with lognormal intrinsic scatter. After transforming the simulation-calibrated relation to the pivot masses adopted in our analysis (Section~\ref{subsec:modelling}), we find $\Mwl/\Mtrue = 0.95 \pm 0.07$ at $M_{500,\mathrm{true}}=7\times10^{14}\Msun$, corresponding to a $5\percent$ underestimate with a $7\percent$ uncertainty in the multiplicative normalisation. For the corresponding $\Delta=200$ calibration, we find $\Mwl/\Mtrue=1.04\pm 0.07$ at $M_{200,\mathrm{true}}=10^{15}\Msun$, corresponding to a $4\percent$ overestimate with a $7\percent$ uncertainty in the multiplicative normalisation. 

The fitted mass slope of the $\Mwl$--$\Mtrue$ relation is also consistent with unity, indicating no evidence for a statistically significant mass dependence of the \WL modelling bias over the BAHAMAS-calibrated mass range. The near-unity mass normalisation found here is consistent with the BAHAMAS-based tests of \citet{Umetsu2020xxl}, where the strongest mass-dependent \WL mass bias appeared only in the low-mass group regime, well below the cluster masses considered here. In the hierarchical analysis (Section~\ref{sec:scaling}), we explicitly forward-modelled this $\Mwl$--$\Mtrue$ relation to connect the latent true halo mass to the observed \WL mass.

In addition to the mass calibration, the same simulation tests allow us to assess possible systematic offsets in the recovered halo concentration. Over the same BAHAMAS-calibrated mass range, we find that the \WL-inferred concentrations, $\cwl$, exhibit a significant but approximately mass-independent offset relative to the true values, with a weighted geometric mean ratio of $\langle \cwl/c_{200,\mathrm{true}}\rangle_\mathrm{g} = 0.889\pm0.047$, corresponding to an underestimate of $\approx 11\percent$. A similar underestimation of \WL-inferred concentration was found in the BAHAMAS-based tests of \citet{Umetsu2020xxl} (see also Appendix~\ref{appendix:NFW_BAHAMAS}). In our analysis, we applied a constant multiplicative correction factor of $1/0.89$ to the individual \WL-inferred concentrations before fitting the $c$--$M$--$z$ relation. The uncertainty in this correction, $\approx 0.02$~dex in $\log c_{200}$, was not propagated separately in the regression because it is subdominant to the current statistical uncertainty.

\subsubsection{Residual calibration uncertainties}
\label{subsubsec:residual}

We next consider residual observational calibration terms that are not included in the simulation-based $\Mwl$--$\Mtrue$ calibration. The dominant contribution is the uncertainty associated with background-source selection and possible residual dilution of the \WL signal. Motivated by the CC--\smem comparison presented in Section~\ref{subsec:single}, we assigned a $4\percent$ fractional mass uncertainty to the background-source-selection systematic.
 
Additional residual calibration uncertainties arise from the shear calibration and from the source-redshift calibration. The \amalgam shape catalogues provide calibrated galaxy shape measurements for the \WL analysis. The validation tests described in Section~\ref{subsec:amalgam} demonstrated sub-percent-level shear-calibration performance for high-S/N sources with $\texttt{snr\_win}>20$. In our cluster-\WL analysis, we adopted a slightly lower threshold of $\texttt{snr\_win}>15$ to increase the usable background-source density. To allow for possible residual multiplicative shear-calibration uncertainty associated with this source selection, we adopted a conservative residual uncertainty of order $1\percent$ in the lensing signal amplitude, that is, in $\Delta\Sigma_+$.

As a robustness check, we repeated the individual-cluster NFW analysis for the full CHEX-MATE--\amalgam sample using a more stringent source selection, $\texttt{snr\_win}>20$, while keeping the remaining analysis choices fixed. This removed the subset of fiducial sources with $15<\texttt{snr\_win}<20$, reducing the mean selected-source density from $\langle\ngal\rangle=7.7$ to $6.3$~arcmin$^{-2}$. The resulting geometric mean mass over the full sample remains consistent with the fiducial value (Table~\ref{tab:stack}); for example, $\langle M_{500,\mathrm{WL}}\rangle_\mathrm{g}$ changes by only $\approx 3\percent$\footnote{Using the response approximation introduced below, $\delta\ln\Delta\Sigma_+ \simeq \Gamma_{500}\,\delta\ln M_{500}$ with $\Gamma_{500}=0.67$, a $3\percent$ shift in $M_{500}$ corresponds to a signal-level shift of approximately $2\percent$.} 
from $\langle M_{500,\mathrm{WL}}\rangle_\mathrm{g}=9.7\times10^{14}\Msun$ to $10.0\times10^{14}\Msun$. Since the two analyses differ only by the inclusion or exclusion of this lower-\texttt{snr\_win} subset, which represents about $20\percent$ of the selected source number density, the observed offset reflects the combined effect of its shape measurements, lensing-efficiency estimates, statistical weights, and random shape-noise contribution. The fiducial geometric mean mass uncertainty is $6\percent$, and under a simple shape-noise-dominated estimate for the two nested source samples, the expected statistical shift between the fiducial and $\texttt{snr\_win}>20$ estimates is of order $\sqrt{1.4/6.3}\times 6\percent\approx 3\percent$, comparable to the measured shift. We therefore find no evidence for a significant coherent mass-scale sensitivity to the adopted $\texttt{snr\_win}>15$ threshold.

For the source-redshift calibration, the HSC \WL analysis of the XMM-XXL cluster sample by \citet{Umetsu2020xxl} found a residual photo-$z$ calibration bias of $\approx 0.68\percent$ in $\Delta\Sigma_+$ for the full XXL sample at $\zlmed=0.30$. Here we adopted a more conservative value motivated by recent HSC Year~3 \WL cosmology analyses. In particular, \citet{Miyatake2023} introduced a residual source-redshift shift parameter $\Delta \zph$, defined by $p_\mathrm{true}(z)=p(z+\Delta \zph)$, and found a baseline posterior centred at $\Delta \zph \approx -0.05$. We adopted $|\Delta\zph|=0.05$ as a representative scale for residual photo-$z$ bias. For the median cluster redshift of our sample, $\zlmed=0.23$, and the mean photo-$z$ of the selected source galaxies, $\langle\zs\rangle\approx0.99$, a first-order propagation of this shift in our fiducial cosmology yields a fractional calibration uncertainty of approximately $1.5\percent$ in $\langle\beta\rangle$, and hence in $\Delta\Sigma_+$.

To first order, signal-level calibration biases propagate to the inferred \WL mass as $\delta\ln M_\Delta \simeq \delta\ln\Delta\Sigma_+/\Gamma_\Delta$, where $\Gamma_\Delta$ is the local logarithmic mass response of the reduced tangential shear profile, defined by $\Gamma_\Delta \equiv \partial \ln{\Delta\Sigma_+}/\partial\ln{M_\Delta}$. For a representative NFW model, this quantity was evaluated at the reference radius $r_\Delta$ while holding $c_\Delta$ fixed. For the best-fit NFW model of the full sample (Section~\ref{subsec:stack}), including the reduced-shear correction, we find $\Gamma_{200}=0.71$ and $\Gamma_{500}=0.67$. The adopted $1\percent$ shear-calibration and $1.5\percent$ source-redshift calibration uncertainties in the lensing signal therefore correspond to mass-level uncertainties of approximately $1.4$--$1.5\percent$ and $2.1$--$2.2\percent$, respectively. Combined in quadrature with the $4\percent$ source-selection term, these residual observational calibration uncertainties correspond to approximately $4.8\percent$ at the mass level.

\subsubsection{Summary of the systematic treatment}
\label{subsubsec:systsummary}

For the ensemble \WL modelling, we distinguish between simulation-calibrated NFW modelling biases and residual observational calibration uncertainties. For the mass calibration, we forward-modelled the BAHAMAS-calibrated $\Mwl$--$\Mtrue$ relation in the Bayesian population analysis (Section~\ref{sec:scaling}), adopting a $7\percent$ uncertainty in its normalisation (Section~\ref{subsubsec:massbias}; Appendix~\ref{appendix:test}).
The same simulation tests also indicate a systematic underestimate of the NFW concentration, which we accounted for by applying a constant multiplicative correction factor of $1/0.89$ to the \WL-inferred concentrations before the population-level concentration analysis.

The residual observational calibration uncertainties are not captured by the simulation-based $\Mwl$--$\Mtrue$ mapping. These include the source-selection uncertainty inferred from the comparison between the CC selection and the \smem-threshold selection (Section~\ref{subsec:single}), together with residual shear-calibration and source-redshift calibration uncertainties. These terms correspond to an observational calibration uncertainty of approximately $4.8\percent$ at the mass level (Section~\ref{subsubsec:residual}).
In applying this residual observational calibration budget, we treated these terms as sample-wide multiplicative uncertainties in the mass scale; possible mass- or redshift-dependent residuals are not constrained by our data and were therefore not modelled separately.

Combining the simulation-based mass-normalisation uncertainty of $7\percent$ with the residual observational mass-calibration budget yields a total systematic uncertainty of order $8\percent$ in the multiplicative normalisation of the \WL mass calibration. This defines the mass-calibration uncertainty model adopted in the population-level scaling analysis of Section~\ref{sec:scaling}.

\section{Scaling relations}
\label{sec:scaling}

In this section, we examine the concentration--mass--redshift ($c_{200}$--$M_{200}$--$z$) relation and the \Planck MMF3 SZ mass-proxy--mass--redshift ($\Msz$--$M_{500}$--$z$) relation for the CHEX-MATE--\amalgam sample using the \WL and MMF3 mass estimates described above. In the following analysis, the hierarchical regression was performed on the cluster-level \WL constraints derived above, rather than directly on the $\Delta\Sigma_+(R)$ profiles. The fitted \WL masses and concentrations, together with their measurement uncertainties and relevant covariances, were used as input observables and combined with the \Planck MMF3 mass proxy $\Msz$ to model the population-level scaling relations.

\subsection{Bayesian regression framework}
\label{subsec:bayes}

To model the population-level scaling relations of our cluster sample, we used the Bayesian hierarchical regression package \lira \citep{Sereno2016lira,Sereno2016lirapackage}. This formalism accounts simultaneously for measurement uncertainties, intrinsic scatter, and latent variables, and can incorporate sample-selection effects. We applied it to the $c_{200}$--$M_{200}$--$z$ relation and to the $\Msz$--$M_{500}$--$z$ relation. We modelled the latent true-mass distribution of the selected sample in both analyses and, for the $\Msz$--$M_{500}$--$z$ relation, included an effective treatment of the CHEX-MATE tier selection, as detailed in Section~\ref{subsec:selection}.

\begin{table*}
\caption{Mapping between the generic \lira variables and the physical quantities used in the two hierarchical regressions.}
\label{tab:lira_notation}
\centering
\small
\begin{tabular}{lll}
\hline\hline
Quantity & $\Msz$--$M_{500}$--$z$ relation & $c_{200}$--$M_{200}$--$z$ relation \\
\hline
Latent true halo mass &
$Z=\log\left(M_{500,\mathrm{true}}/M_{500,\mathrm{piv}}\right)$ &
$Z=\log\left(M_{200,\mathrm{true}}/M_{200,\mathrm{piv}}\right)$ \\
\WL-inferred halo mass &
$X=\log\left(M_{500,\mathrm{WL}}/M_{500,\mathrm{piv}}\right)$ &
$X=\log\left(M_{200,\mathrm{WL}}/M_{200,\mathrm{piv}}\right)$ \\
Response variable &
$Y=\log\left(\Msz/M_{500,\mathrm{piv}}\right)$ &
$Y=\log c_{200,\mathrm{WL}}$ \\
Pivot mass &
$M_{500,\mathrm{piv}}=7\times10^{14}\Msun$ &
$M_{200,\mathrm{piv}}=10^{15}\Msun$ \\
Population-level relation &
$\Msz$ as a function of $M_{500,\mathrm{true}}$ and $z$ &
$c_{200}$ as a function of $M_{200,\mathrm{true}}$ and $z$ \\
\hline
\end{tabular}
\tablefoot{The redshift dependence is parametrised in both regressions using $F_z(z)=(1+z)/(1+\zref)$ with $\zref=0.25$. The variables $X$ and $Y$ denote observationally inferred quantities entering the regression, while $Z$ denotes the corresponding latent true-mass variable. The simulation-calibrated $\Mwl$--$\Mtrue$ relation connects $X$ to $Z$ in the hierarchical model; the separate correction applied to the \WL-inferred concentration is described in Section~\ref{subsec:cMR}.}
\end{table*}

We describe a generic scaling relation of a cluster observable, $O$, with halo mass and redshift as $O \propto 10^\alpha M_\Delta^\beta F_z(z)^\gamma$, where $\alpha$, $\beta$, and $\gamma$ are the intercept, mass-trend, and redshift-trend parameters, respectively. We adopted $F_z(z)=(1+z)/(1+\zref)$ with $\zref = 0.25$, which is close to the median redshift of our sample, $\zlmed=0.23$, and coincides with the snapshot redshift at which the BAHAMAS-based \WL modelling-bias relations were calibrated (Appendix~\ref{appendix:test}).

Following the \lira formalism, we express these power-law relations as linear relations in base-10 logarithmic variables. We define the latent true-mass variable and the \WL-inferred mass variable, respectively, as
\begin{equation}
Z = \log\left(\frac{M_{\Delta,\mathrm{true}}}{M_{\Delta,\mathrm{piv}}}\right), \qquad
X = \log\left(\frac{M_{\Delta,\mathrm{WL}}}{M_{\Delta,\mathrm{piv}}}\right).
\end{equation}
The relation-specific choices of $M_{\Delta,\mathrm{piv}}$ and of the response variable $Y$ are summarised in Table~\ref{tab:lira_notation}. In brief, we use $Y=\log c_{200,\mathrm{WL}}$ for the $c_{200}$--$M_{200}$--$z$ relation and $Y=\log(\Msz/M_{500,\mathrm{piv}})$ for the $\Msz$--$M_{500}$--$z$ relation, with the simulation-calibrated \WL modelling-bias relations connecting the observed \WL quantities to the latent halo variables.

For $X$ and $Y$, we distinguish between the population mean relation and the intrinsically scattered realisation about that relation. We denote the mean relations by $X_Z(Z)$ and $Y_Z(Z)$, while $X$ and $Y$ denote the corresponding intrinsically scattered quantities. We modelled the intrinsic scatter of the \WL-inferred mass and of the secondary observable as Gaussian in log space,
\begin{equation}
\label{eq:lira_scatter}
X \sim \mathcal{N}(X_Z,\sigma_{X|Z}^2), \qquad
Y \sim \mathcal{N}(Y_Z,\sigma_{Y|Z}^2),
\end{equation}
where $\sigma_{X|Z}$ and $\sigma_{Y|Z}$ are the intrinsic scatters at fixed latent mass. The corresponding mean relations are written as
\begin{equation}
\label{eq:lira_mean}
\begin{aligned}
X_Z(Z) &= \alpha_{X|Z} + \beta_{X|Z} Z,\\
Y_Z(Z) &= \alpha_{Y|Z} + \beta_{Y|Z} Z + \gamma_{Y|Z}\log F_z(z).
\end{aligned}
\end{equation}
Here, $X_Z(Z)$ specifies the mean $\Mwl$--$\Mtrue$ calibration relation, while $Y_Z(Z)$ specifies the mean scaling relation of the secondary observable at fixed $\Mtrue$ and redshift. For the $c_{200}$--$M_{200}$--$z$ relation, $Y_Z$ describes the mean logarithmic concentration at fixed latent $M_{200}$ and redshift. For the $\Msz$--$M_{500}$--$z$ relation, $Y_Z$ describes the mean logarithmic SZ mass proxy at fixed latent $M_{500}$ and redshift. 

As detailed in Appendix~\ref{appendix:test}, we assigned informative Gaussian priors to the \WL calibration parameters $\alpha_{X|Z}$ and $\beta_{X|Z}$ based on the simulation-based calibration of the $\Mwl$--$\Mtrue$ relation established at the snapshot redshift $z=0.25$, close to the median redshift of our sample, $\zlmed=0.23$; the explicit prior distributions are given in Section~\ref{subsec:priors}. We fixed the intrinsic scatter of the \WL-inferred mass at fixed latent mass to $\sigma_{X|Z}=0.2/\ln(10)$, motivated by the same simulation-based calibration.

\subsection{Observational uncertainties}
\label{subsec:obs_uncert}

For each cluster, the \lira regression includes the measurement uncertainties associated with the input observables in log space. For the $\Msz$--$M_{500}$--$z$ regression, the uncertainty on $\Msz$ was taken from the quoted \Planck MMF3 mass-proxy uncertainty, while the uncertainty on $M_{500,\mathrm{WL}}$ was derived from the NFW posterior samples. We assumed zero measurement-error covariance between $\Msz$ and $M_{500,\mathrm{WL}}$, since the \Planck SZ mass proxy and the \WL-inferred mass are obtained from independent data and measurement procedures. We note that $\Msz$ is the MMF3 mass proxy as reported by \Planck and is not evaluated within the \WL-inferred $r_{500}$. We therefore did not apply an aperture correction; the regression calibrates the reported MMF3 mass proxy against the latent true $M_{500}$. 
For the $c_{200}$--$M_{200}$--$z$ regression, the measurement-error covariance between $M_{200,\mathrm{WL}}$ and $c_{200,\mathrm{WL}}$ was computed for each cluster from the joint MCMC posterior samples of the NFW parameters and supplied to the \lira regression.

\subsection{Selection effects}
\label{subsec:selection}

Proper modelling of the latent true-mass distribution $P(Z)$ is crucial in Bayesian regression. For an observable-selected cluster sample, $P(Z)$ is shaped by the interplay between the declining halo mass function and the sample selection function, and is therefore expected to be approximately unimodal and redshift dependent. In this context, the adopted $P(Z)$ is an empirical model for the latent mass distribution of the analysed sample \citep{Sereno2016lira}; it is not a cosmological model of the parent halo mass function or an explicit model of the \amalgam data-availability selection. In this analysis, all masses entering $Z$ and the effective truncation thresholds were evaluated in the fixed fiducial cosmology adopted throughout this work; cosmological parameters were not varied in the regression.

For the $\Msz$--$M_{500}$--$z$ regression, we modelled the sample selection through the truncation formalism implemented in \lira, following the same strategy adopted by \citet{Sereno2025chex}. In CHEX-MATE, the Tier-1 and Tier-2 clusters are drawn from the \Planck MMF3 sample subject to a detection threshold in $(S/N)_\mathrm{MMF3}$, while Tier-2 clusters satisfy the additional requirement $\Msz>7.25\times10^{14}\Msun$ (Section~\ref{sec:intro}). We approximated this selection in terms of an effective truncation in the observed $\Msz$ proxy, which captures the dominant Malmquist-type bias in the $\Msz$--$M_{500}$--$z$ relation. Accordingly, for each cluster $i$, we define the truncation threshold in the same base-10 logarithmic space used by \lira as
\begin{equation}
Y_{\mathrm{th},i} = \log\left(\frac{M_{\mathrm{SZ,th},i}}{M_{500,\mathrm{piv}}}\right),
\end{equation}
where $M_{\mathrm{SZ,th},i}=2\times10^{14}\Msun$ for Tier-1-only and Tier-1+2 clusters, and $M_{\mathrm{SZ,th},i}=7.25\times10^{14}\Msun$ for Tier-2-only clusters. For Tier-1, this lower threshold should be regarded as an effective approximation to the underlying $(S/N)_\mathrm{MMF3}$-limited selection, rather than as a formal mass cut. We assigned a cluster-dependent log-space uncertainty to the effective truncation threshold, $\delta Y_{\mathrm{th},i} = \sigma(M_{\mathrm{SZ},i}) / (M_{\mathrm{SZ},i}\ln 10)$, where $\sigma(M_{\mathrm{SZ},i})$ is the quoted uncertainty on the \Planck MMF3 mass proxy.

We modelled the latent true-mass distribution $P(Z)$ of the selected sample as a redshift-dependent single Gaussian characterised by the mean $\mu_Z(z)$ and the dispersion $\sigma_Z(z)$. This provides a good approximation for a regular unimodal distribution \citep{Kelly2007,Andreon+Berge2012,CoMaLit4,Sereno2016lira}. We parametrise the mean and dispersion of $P(Z)$ as
\begin{equation}
\mu_Z(z) = \mu_{Z,0} + \gamma_{\mu_Z,D} \log \mathcal{D}(z),
\qquad
\sigma_Z(z) = \sigma_{Z,0},
\end{equation}
where $\mathcal{D}(z) = D_L(z)/D_L(\zref)$, with $D_L(z)$ the luminosity distance at redshift $z$ evaluated in our fiducial cosmology. Here $\mu_{Z,0}$ is the mean of the selected-sample mass distribution at the reference redshift $\zref$, $\gamma_{\mu_Z,D}$ describes its redshift trend, and $\sigma_{Z,0}$ is the dispersion, assumed to be constant with redshift.

\subsection{Priors}
\label{subsec:priors}

In our regression analysis, we considered seven population-level parameters,
\begin{equation}
(\alpha_{Y|Z}, \beta_{Y|Z}, \gamma_{Y|Z}, \sigma_{Y|Z}, \mu_{Z,0}, \gamma_{\mu_Z,D}, \sigma_{Z,0}),
\end{equation}
which describe the secondary-observable scaling relation, $Y_Z(Z)$, and the latent true-mass distribution of the selected sample, $P(Z)$. In addition, the hierarchical model includes the two \WL calibration parameters $(\alpha_{X|Z}, \beta_{X|Z})$, which specify the $X_Z(Z)$ relation. 

For the seven population-level parameters, we adopted the default weakly informative priors implemented in \lira,
\begin{equation}
 \begin{aligned}
  &\alpha_{Y|Z},\, \mu_{Z,0} \sim \mathcal{U}(-1/\epsilon,+1/\epsilon),\\
  &\beta_{Y|Z},\, \gamma_{Y|Z},\, \gamma_{\mu_Z,D} \sim t_1,\\
  &\sigma^{-2}_{Y|Z}, \, \sigma^{-2}_{Z,0} \sim \Gamma(\epsilon,\epsilon),
 \end{aligned}
\end{equation}
where $\mathcal{U}(a,b)$ denotes a uniform prior between $a$ and $b$, $t_1$ is a Student's $t$ distribution with one degree of freedom, $\Gamma$ denotes a gamma distribution used as a prior on the inverse variance, and $\epsilon$ is a small number set to $\epsilon=10^{-4}$. These priors are sufficiently weak to allow the data to constrain the scaling relation and the selected-sample mass distribution with minimal prior regularisation.

The \WL mass calibration parameters were assigned informative Gaussian priors. Specifically, the intercept and slope of $X_Z(Z)$, which represents the log-space $\Mwl$--$\Mtrue$ calibration relation defined in Equation~\eqref{eq:lira_mean}, were constrained using the simulation-based calibration described in Appendix~\ref{appendix:test}. For the $\Delta=200$ analysis, we adopted
\begin{equation}
 \alpha_{X|Z}^{(200)} \sim \mathcal{N}(0.019, 0.035^2),\qquad
 \beta_{X|Z}^{(200)}  \sim \mathcal{N}(1.024, 0.081^2),
\end{equation}
and for the $\Delta=500$ analysis, we adopted
\begin{equation}
\alpha_{X|Z}^{(500)} \sim \mathcal{N}(-0.021, 0.035^2),\qquad
\beta_{X|Z}^{(500)}  \sim \mathcal{N}(0.969, 0.084^2).
\end{equation}
Here, $\mathcal{N}(\mu,\sigma^2)$ denotes a Gaussian distribution with mean $\mu$ and variance $\sigma^2$. The prior means were set by the simulation-based \WL mass calibration at the pivot masses adopted in the main analysis. The prior widths of the intercept parameters were set to $0.035$ in $\log_{10}$ space, corresponding to a total uncertainty of $\approx 8\percent$ in the multiplicative normalisation of the \WL mass calibration (Section~\ref{subsec:syst}). This uncertainty combines in quadrature the simulation-based intercept uncertainty and the additional residual observational calibration uncertainty. By contrast, the priors on the slope parameters were taken directly from the simulation-based calibration without further broadening.

\subsection{\Planck mass-proxy scaling relation and posterior mass estimates}
\label{subsec:mm}

We constrained the $\Msz$--$M_{500}$--$z$ relation using the hierarchical Bayesian framework developed in the preceding subsections. The analysis jointly modelled the latent $\Msz$--$\Mtrue$--$z$ scaling relation, the $\Mwl$--$\Mtrue$ calibration relation, the latent true-mass distribution of the selected sample, and the effective truncation in the observed SZ mass proxy used to represent the CHEX-MATE tier selection.

We considered two configurations for the $\Msz$--$\Mtrue$--$z$ relation: a restricted constant-bias model with $\beta_{Y|Z}=1$ and $\gamma_{Y|Z}=0$, corresponding to a mass- and redshift-independent multiplicative mass bias, and a flexible model in which the intercept, mass trend, and redshift trend of the $\Msz$ scaling relation, $(\alpha_{Y|Z}, \beta_{Y|Z}, \gamma_{Y|Z})$, are all allowed to vary. We adopted the flexible model as our baseline analysis, while using the restricted model for comparison with the conventional constant-bias parametrisation. The corresponding posterior summary statistics are listed in Table~\ref{tab:MM}.

\begin{table*}
\centering
\caption{Posterior summary statistics of the population-level parameters for the $M_\mathrm{SZ}$--$M_{500}$--$z$ relation.\label{tab:MM}}
\small
\begin{tabular}{cccccccc}
\hline\hline
Model & $\alpha_{Y|Z}$ & $\beta_{Y|Z}$ & $\gamma_{Y|Z}$ & $\sigma_{Y|Z}$ & $\mu_{Z,0}$ & $\gamma_{\mu_{Z},D}$ & $\sigma_{Z,0}$ \\
\hline
Free slopes & $-0.08 \pm 0.04$ & $0.50 \pm 0.29$ & $0.30 \pm 0.86$ & $0.10 \pm 0.02$ & $0.13 \pm 0.05$ & $0.60 \pm 0.15$ & $0.11 \pm 0.05$ \\
Fixed slopes & $-0.14 \pm 0.06$ & $1$ & $0$ & $0.08 \pm 0.03$ & $0.14 \pm 0.05$ & $0.46 \pm 0.10$ & $0.07 \pm 0.03$ \\
\hline
\end{tabular}
\tablefoot{The free-slopes case is the baseline regression model, whereas the fixed-slopes case characterises the mass-bias parameter $1-b_\mathrm{SZ}$.}
\end{table*}

\begin{figure}[tbp]
 \begin{center}
  \includegraphics[width=0.9\columnwidth,angle=0,clip]{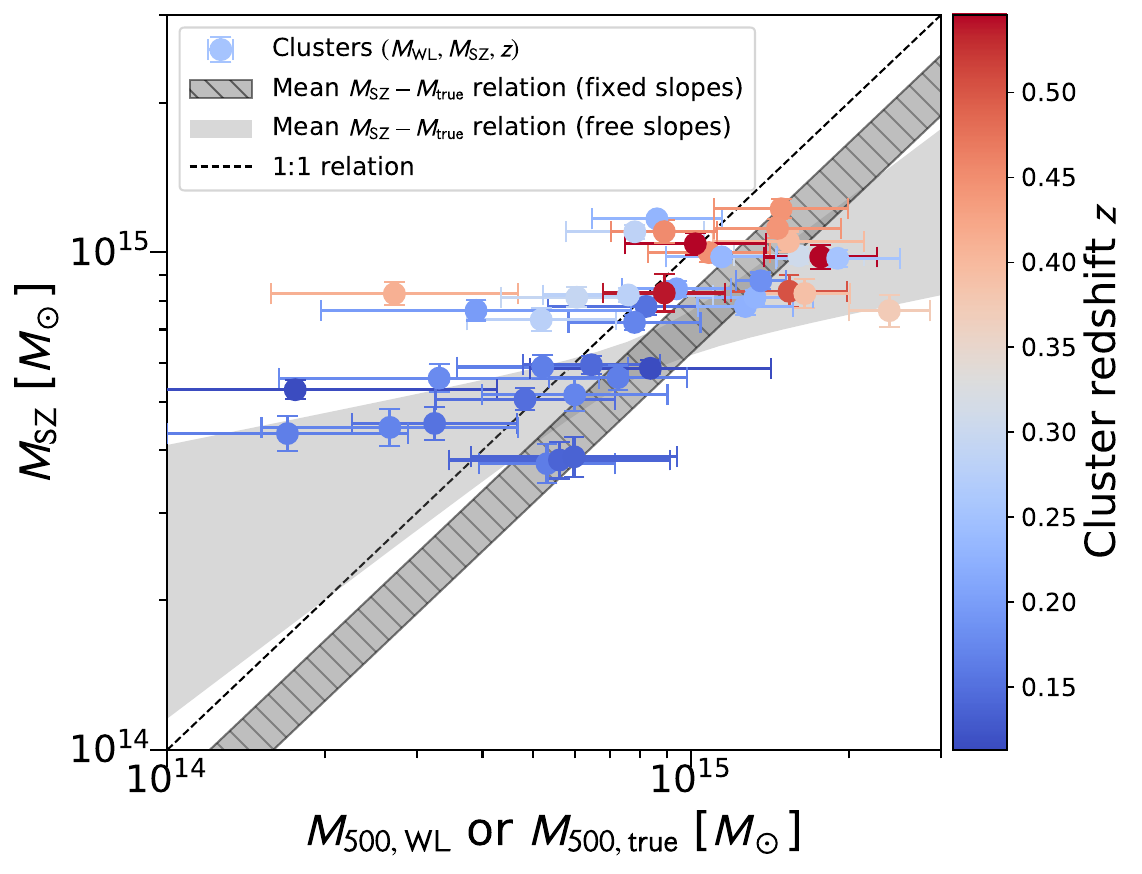}
 \end{center}
 \caption{Joint regression of the $\Msz$--$\Mtrue$--$z$ and $\Mwl$--$\Mtrue$ relations for the full sample. The circles with error bars represent \WL and \Planck MMF3 mass estimates $(M_{500,\mathrm{WL}}, \Msz)$ along with their $1\sigma$ uncertainties for individual clusters. The cluster redshifts are colour-coded according to the colour bar. The shaded grey region shows the marginalised $1\sigma$ credible interval for the population mean $\Msz$--$\Mtrue$ relation at $\zref=0.25$, inferred after accounting for the simulation-calibrated $\Mwl$--$\Mtrue$ relation, the latent true-mass distribution of the selected sample, and the effective truncation in the observed SZ mass proxy used to represent the CHEX-MATE tier selection. The hatched grey region shows the regression result for the restricted constant-bias model with $\beta_{Y|Z}=1$ and $\gamma_{Y|Z}=0$, which yields $1-\bsz=\Msz/\Mtrue=0.72 \pm 0.11$. The dashed black line represents the one-to-one relation.}
 \label{fig:MM}
\end{figure}

Figure~\ref{fig:MM} shows the resulting joint regression for the CHEX-MATE--\amalgam sample. The larger fractional uncertainties seen for some lower-mass clusters mainly reflect their lower lensing signal amplitudes. Since the lensing signal scales approximately, and sub-linearly, with halo mass (Section~\ref{subsec:syst}), a given profile-level uncertainty translates into a larger fractional uncertainty in $M_\Delta$ for lower-mass systems. 
We find no evidence for redshift evolution, with $\gamma_{Y|Z} = 0.30 \pm 0.86$, and only a marginal indication of mass dependence, with $\beta_{Y|Z} = 0.50\pm0.29$. At the pivot latent mass $M_{500,\mathrm{true}}=7\times10^{14}\Msun$ and $\zref = 0.25$, the inferred mass ratio is $\Msz/\Mtrue = 0.83\pm0.09$. The inferred intrinsic scatter is $\sigma_{Y|Z}=0.10\pm 0.02$~dex.

For direct comparison with the conventional \Planck mass-bias parametrisation, we also considered a restricted model with $\beta_{Y|Z}=1$ and $\gamma_{Y|Z}=0$. In this case, $\Msz$ is related to $\Mtrue$ by a constant multiplicative factor, so the intercept can be expressed as
\begin{equation}
1-\bsz = \Msz/\Mtrue.
\end{equation}
The corresponding fit is shown by the grey hatched region in Figure~\ref{fig:MM}. Under this restricted parametrisation, we obtain $1-\bsz = 10^{\alpha_{Y|Z}} = 0.72 \pm 0.11$, with an intrinsic scatter of $\sigma_{Y|Z}=0.08\pm 0.03$~dex.

Hence, for this sample, the \Planck mass proxy underestimates the \WL-calibrated halo mass by $(28\pm11)\percent$. We use this restricted result as a compressed summary for comparison with the \Planck cluster-cosmology literature, while the baseline fit with free $(\beta_{Y|Z}, \gamma_{Y|Z})$ provides the more general description adopted in our hierarchical analysis.

\begin{figure}[tbp]
 \begin{center}
  \includegraphics[width=0.9\columnwidth,angle=0,clip]{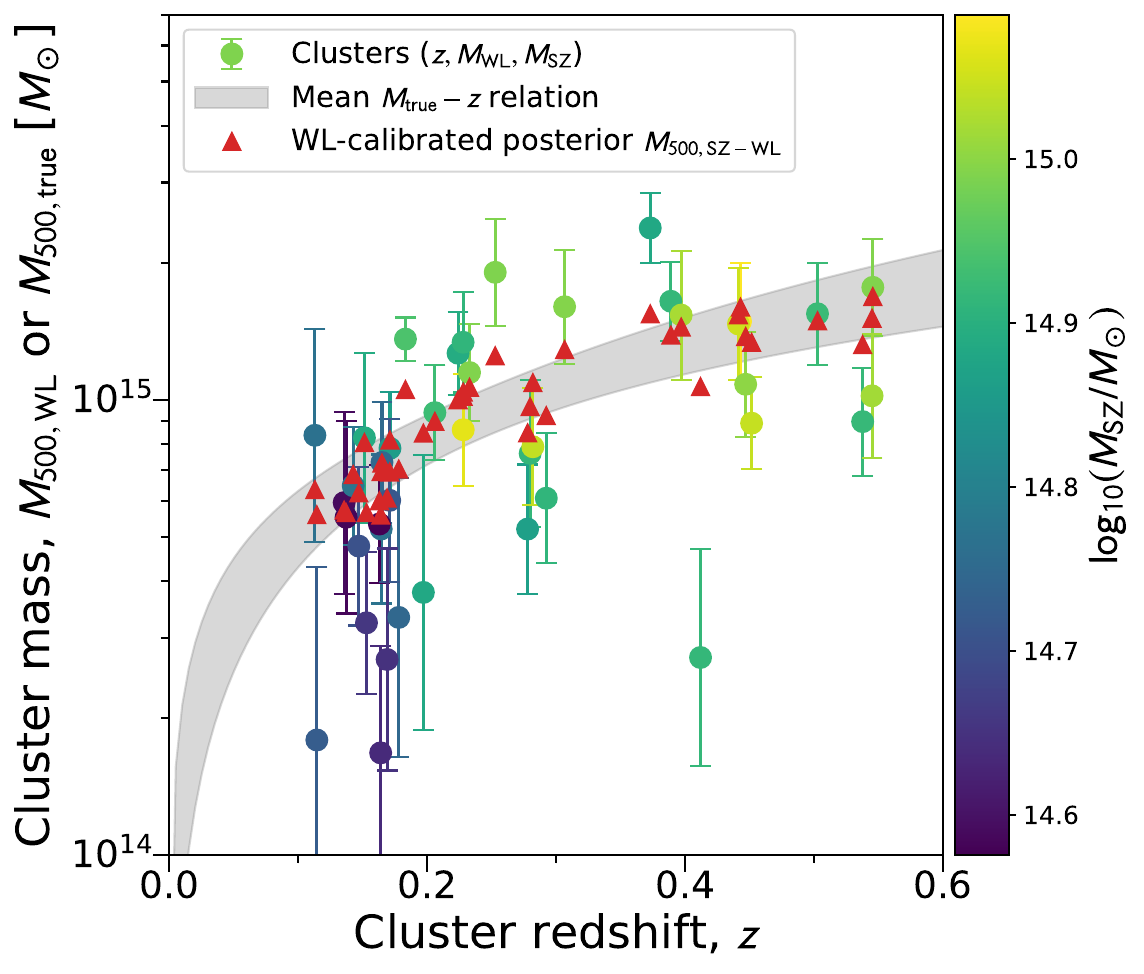}
 \end{center}
 \caption{Cluster mass vs. redshift diagram for the full sample of 41 clusters. The circles with error bars represent \WL mass estimates $M_{500,\mathrm{WL}}$ with their $1\sigma$ uncertainties as a function of redshift for individual clusters. The \Planck SZ mass proxy $\Msz$ of each cluster is colour-coded according to the colour bar. The shaded grey region shows the marginalised $1\sigma$ credible interval for the redshift-dependent mean of the latent true-mass distribution, $M_{500,\mathrm{true}}(z)$ (Equation~\ref{eq:Mz}), inferred from the baseline joint regression of the $\Msz$--$\Mtrue$--$z$ and $\Mwl$--$\Mtrue$ relations shown in Figure~\ref{fig:MM}. The red triangles represent the corresponding \WL-calibrated posterior estimates, $M_{500,\mathrm{SZ\text{--}WL}}$, for individual clusters.}
 \label{fig:Mz}
\end{figure}

Figure~\ref{fig:Mz} shows the cluster mass versus redshift diagram for the CHEX-MATE--\amalgam sample. 
The grey shaded region shows the marginalised $1\sigma$ credible interval for the population mean halo mass as a function of redshift, $M_{500,\mathrm{true}}(z)$, defined as
\begin{equation}
\label{eq:Mz}
\frac{M_{500,\mathrm{true}}(z)}{M_{500,\mathrm{piv}}}
\equiv 10^{\mu_Z(z)}
=10^{\mu_{Z,0}}\,\left[\frac{D_L(z)}{D_L(\zref)}\right]^{\gamma_{\mu_Z,D}},
\end{equation}
which corresponds to the redshift-dependent mean of the latent true-mass distribution $P(Z)$ inferred from the baseline joint regression analysis. 

From the baseline regression, we infer a population mean mass of $M_{500,\mathrm{true}}(\zref)=(9.5\pm1.1)\times10^{14}\Msun$ at $\zref=0.25$, a redshift-trend parameter of $\gamma_{\mu_Z,D}=0.60\pm0.15$, and a scale parameter of $\sigma_{Z,0}=0.11\pm0.05$ for the latent true-mass distribution $P(Z)$, corresponding to a characteristic scatter of about 30\% in linear mass. The positive value of $\gamma_{\mu_Z,D}$ indicates that the redshift dependence of the population mean halo mass for the selected sample is well captured by our adopted power-law parametrisation in $D_L(z)$.

The joint regression of the $\Msz$--$\Mtrue$--$z$ and $\Mwl$--$\Mtrue$ relations provides a posterior predictive framework for inferring the latent halo mass from an observed \Planck SZ mass proxy. In this context, the posterior mass estimate is obtained by inferring $\Mtrue$ from $\Msz$, using the calibrated forward scaling relation together with the inferred population model for $\Mtrue$ at a given redshift \citep{Sereno2016lira,Umetsu2020xxl,Tam2026}. This estimate is obtained by marginalising over the measurement uncertainty in $\Msz$, the intrinsic scatter of the scaling relation, and the posterior uncertainty in the regression and population-level parameters. It therefore represents a \WL-calibrated probabilistic estimate of the true halo mass, conditioned on the observed $\Msz$ and on the hierarchical model inferred from our cluster sample. We denote this posterior estimate of $M_{500}$ by $M_{500,\mathrm{SZ\text{--}WL}}$.

The red triangles in Figure~\ref{fig:Mz} represent the posterior $M_{500,\mathrm{SZ\text{--}WL}}$ estimates for individual clusters based on the baseline \WL-calibrated $\Msz$--$M_{500}$--$z$ relation. The corresponding values for the full sample are listed in Table~\ref{tab:mass}.

\subsection{Concentration--mass--redshift scaling relation}
\label{subsec:cMR}

We next constrain the $c_{200}$--$M_{200}$--$z$ relation using the \lira hierarchical Bayesian framework, by jointly modelling the latent $c_{200}$--$\Mtrue$--$z$ relation and the $\Mwl$--$\Mtrue$ calibration relation at overdensity $\Delta=200$ within the selected-sample population model described in Section~\ref{subsec:selection}.
Before performing the \lira regression, we corrected the individual \WL-inferred concentration constraints for the approximately mass-independent offset identified in our simulation tests by applying a constant multiplicative factor of $1/0.89$ to $c_{200}$ (see Section~\ref{subsubsec:massbias} and Appendix~\ref{appendix:test}). This correction is approximate and valid under the assumption that the modelling bias in halo concentration is independent of the latent halo mass, $M_{200,\mathrm{true}}$. More generally, a fully self-consistent treatment of \WL-inferred masses and concentrations would require forward modelling in the joint $c$--$M$ parameter space \citep[see][]{Okabe2025}. Since the concentrations and \WL masses of clusters are inferred from the same $\Delta\Sigma_+(R)$ profile, we accounted for the covariance between the two NFW parameters using the covariance matrices derived from the MCMC posterior samples (Section~\ref{subsec:modelling}).

\begin{figure*}[t]
 \begin{center}
  \includegraphics[width=0.9\textwidth,angle=0,clip]{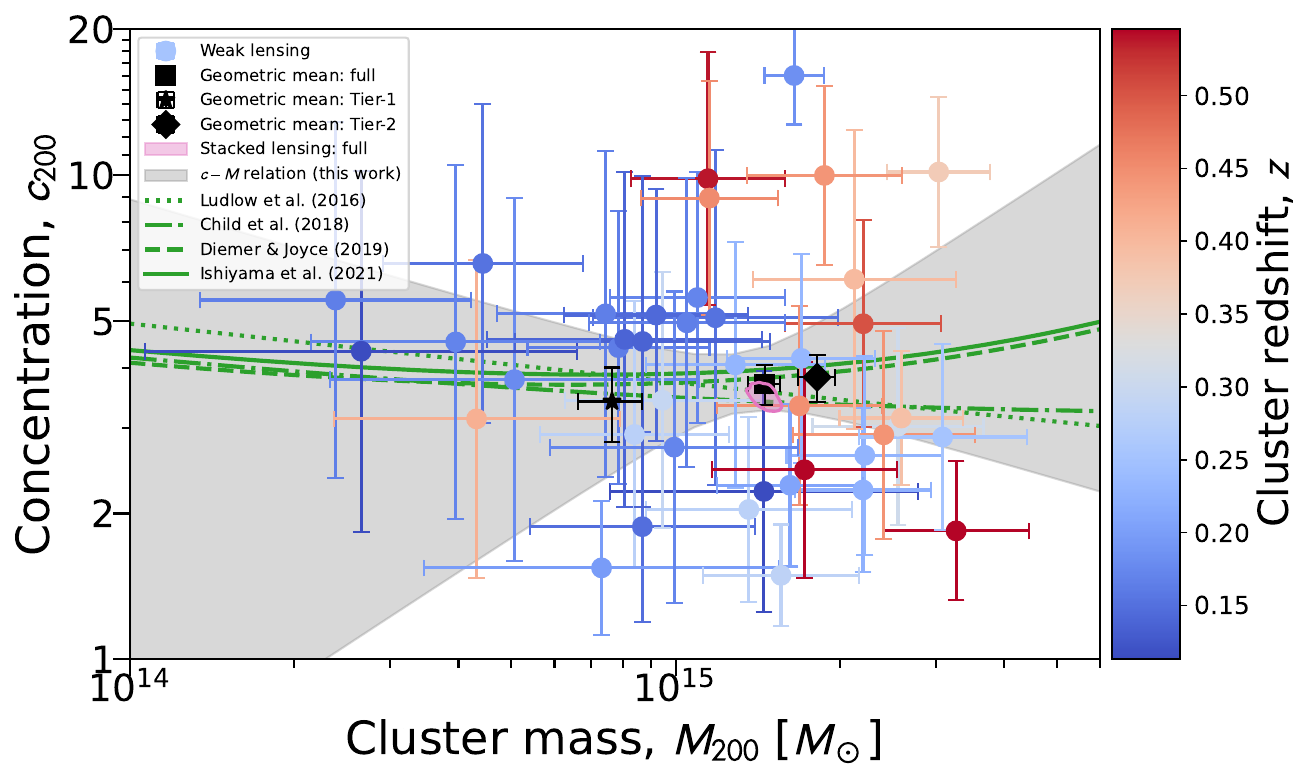}
 \end{center}
 \caption{Concentration--mass relation for the CHEX-MATE--\amalgam sample of 41 galaxy clusters from our \WL analysis. The circles with error bars represent the \WL-inferred parameters and their $1\sigma$ uncertainties for individual clusters. The covariance between $M_{200}$ and $c_{200}$ is not shown but was accounted for in the Bayesian inference. The cluster redshifts are colour-coded according to the colour bar. The shaded grey region denotes the marginalised $1\sigma$ credible interval for the population mean relation at $\zref=0.25$. The black symbols with error bars indicate the weighted geometric-mean values of $M_{200}$ and $c_{200}$ computed from the individual-cluster \WL constraints for the full sample, the combined Tier-1 and Tier-1+2 subsample, and the Tier-2-only subsample (Table~\ref{tab:stack}). The shaded magenta contour shows the joint $1\sigma$ credible region in the $c$--$M$ plane derived from the NFW fit to the stacked $\Delta\Sigma_+$ profile for the full sample (Figure~\ref{fig:stackgt}). All \WL-inferred concentrations shown have been corrected for the simulation-calibrated \WL bias by applying a constant multiplicative factor of $1/0.89$. The results are compared with theoretical $c$--$M$ relations evaluated at $\zref=0.25$ (green lines) from numerical simulations of \LCDM cosmologies \citep{Ludlow2016,Child2018cm,Diemer2019,Ishiyama2021}.}
 \label{fig:cM}
\end{figure*}

Posterior summaries for the population-level parameters are listed in Table~\ref{tab:cM}. The corresponding regression results for the CHEX-MATE--\amalgam sample are summarised in Figure~\ref{fig:cM}. The grey shaded region shows the marginalised $1\sigma$ credible interval for the population mean $c_{200}$--$M_{200}$ relation at $\zref=0.25$, inferred within the selected-sample population model after accounting for \WL modelling bias. We find no evidence for a dependence on either mass, with $\beta_{Y|Z}=0.20 \pm 0.54$, or redshift, with $\gamma_{Y|Z}=-0.13 \pm 1.18$. At the pivot latent mass $M_{200}=10^{15}\Msun$ and $\zref=0.25$, the inferred population mean concentration is $c_{200}=3.53 \pm 0.71$. The resulting population mean relation can thus be summarised as
\begin{equation}
  c_{200} = (3.53\pm0.71)
  \left(\frac{M_{200}}{10^{15}\Msun}\right)^{0.20\pm0.54}
  \left(\frac{1+z}{1+\zref}\right)^{-0.13\pm1.18}.
\end{equation}
The inferred intrinsic scatter is $0.22\pm0.04$~dex, corresponding to $(51\pm10)\percent$. Overall, the data are consistent with a mean concentration that is approximately constant over the mass and redshift range probed by our sample.

\begin{table*}
\centering
\caption{Posterior summary statistics of the population-level parameters for the $c_{200}$--$M_{200}$--$z$ relation.\label{tab:cM}}
\small
\begin{tabular}{ccccccc}
\hline\hline
$\alpha_{Y|Z}$ & $\beta_{Y|Z}$ & $\gamma_{Y|Z}$ & $\sigma_{Y|Z}$ & $\mu_{Z,0}$ & $\gamma_{\mu_{Z},D}$ & $\sigma_{Z,0}$ \\
\hline
$0.55 \pm 0.09$ & $0.20 \pm 0.54$ & $-0.13 \pm 1.18$ & $0.22 \pm 0.04$ & $0.12 \pm 0.05$ & $0.54 \pm 0.15$ & $0.07 \pm 0.04$ \\
\hline
\end{tabular}
\end{table*}

From the regression, we infer a population mean mass of $M_{200,\mathrm{true}}(\zref)=(13.3\pm1.5)\times10^{14}\Msun$ at $\zref=0.25$, a redshift-trend parameter of $\gamma_{\mu_Z,D}=0.54\pm0.15$, and a scale parameter of $\sigma_{Z,0}=0.07\pm0.04$ for the latent true-mass distribution $P(Z)$. These constraints are consistent with those obtained from the $\Msz$--$M_{500}$--$z$ regression of the same sample (Table~\ref{tab:MM}), as expected given the typical ratio $M_{200}/M_{500}\sim1.4$ for massive clusters with concentrations characteristic of our sample.

Figure~\ref{fig:cM} also shows, as black symbols with error bars, the weighted geometric-mean values of $(M_{200,\mathrm{WL}},c_{200,\mathrm{WL}})$ computed from the individual-cluster \WL constraints for the full sample, the combined Tier-1 and Tier-1+2 subsample, and the Tier-2-only subsample. For consistency with the other \WL-inferred concentrations shown in the figure, the plotted concentrations of these geometric-mean points were multiplied by the simulation-calibrated correction factor $1/0.89$; the corresponding values listed in Table~\ref{tab:stack} are the uncorrected NFW-fit values. These points lie close to the inferred mean $c_{200}$--$M_{200}$--$z$ relation and are broadly consistent with the theoretical \LCDM predictions evaluated at $\zref=0.25$.

The magenta shaded contour shows the joint $1\sigma$ credible region in the $c$--$M$ plane derived from the stacked $\Delta\Sigma_+$ profile for the full sample (Figure~\ref{fig:stackgt}), likewise corrected by the same constant multiplicative factor. The stacked-lensing constraint for the full sample is also consistent with the corresponding weighted geometric mean of the individual-cluster constraints. This agreement is non-trivial, because the two procedures combine the lensing information in different ways: one first stacks the lensing signal and then infers the NFW parameters, whereas the other first infers the NFW parameters for individual clusters and then averages the results. 
Taken together, these comparisons support the internal consistency of the inferred $c_{200}$--$M_{200}$--$z$ relation and its broad agreement with expectations from numerical simulations of \LCDM cosmologies \citep{Ludlow2016,Child2018cm,Diemer2019,Ishiyama2021}.

\section{Discussion}
\label{sec:discussion}

\subsection{Mass calibration of the \Planck SZ mass proxy}

For comparison with the standard \Planck mass-bias parametrisation, we consider the restricted model with $\beta_{Y|Z}=1$ and $\gamma_{Y|Z}=0$, in which the \Planck mass proxy is related to the latent halo mass by a constant multiplicative factor. In this case, we find $1-\bsz=0.72\pm 0.11$. This result indicates that for our cluster sample, $\Msz$ underestimates the \WL-calibrated halo mass at $\Delta=500$ by $(28\pm11)\percent$ on average. A deviation of $1-\bsz$ from unity is expected on physical grounds, as the \Planck SZ mass proxy is tied to X-ray-calibrated scaling relations based on hydrostatic mass estimates; additional observational and calibration systematics may also contribute to the net bias \citep{Donahue2014clash,CoMaLit5}.

Comparison with previous \WL recalibrations of \Planck cluster masses requires some care, because the published estimates of the conventional mass-bias parameter $1-b$ are not all based on the same statistical treatment. As summarised by \citet{Miyatake2026}, the available \WL-based mass calibrations of SZ-selected cluster samples are broadly consistent with $1-b \sim 0.7$--$0.8$, without significant trends with mass or redshift. At the same time, \WL studies based on subsamples of \Planck clusters have reported values of $1-b$ ranging from $\sim0.6$ to nearly unity, depending on the sample selection, redshift range, and statistical treatment \citep[e.g.][]{vonderLinden2014calib,Hoekstra2015,Smith2016locuss,Sereno2017psz2lens,PennaLima2017,Medezinski2018planck,Zubeldia2019}. 

Analyses that do not account for statistical effects such as Eddington bias can infer higher values of $1-b$ than analyses that explicitly model the latent mass distribution and selection effects \citep[see][]{Battaglia2016,Miyatake2019}. Our estimate, $1-\bsz = 0.72\pm0.11$, lies within the broad range of previous \WL-based calibrations. The analysis modelled the latent true-mass distribution of the selected sample, included an effective treatment of the CHEX-MATE tier selection through truncation in the observed SZ mass proxy, and propagated uncertainty in the $\Mwl$--$\Mtrue$ calibration.

A recent \WL mass calibration of 19 \Planck SZ-selected clusters based on the HSC-SSP Year~3 shape catalogue found a closely consistent conventional mass-bias constraint, $1-b=0.73^{+0.10}_{-0.11}$ \citep{Malagon2026}. The overlap with our CHEX-MATE--\amalgam sample is limited to one cluster, PSZ2~G087.03$-$57.37, indicating that the agreement is not driven by a common cluster subset.

This interpretation is consistent with the CHEX-MATE analysis of \citet{Sereno2025chex}, who found that the \Planck SZ masses are biased low with respect to LC2 \WL-calibrated mass estimates. The same study also provided a dynamical calibration of the CHEX-MATE sample and, under the same restricted setup, found $1-\bsz = 0.62\pm0.04$ in our notation. The corresponding bias inferred from galaxy dynamics is somewhat larger than our \WL-based estimate, but the two results remain statistically compatible given the current uncertainty in our lensing calibration.

It is important, however, to distinguish this restricted-fit result from our baseline hierarchical regression. In Section~\ref{subsec:mm}, we adopted the model with free $(\beta_{Y|Z},\gamma_{Y|Z})$ as the more general description of the $\Msz$--$M_{500}$--$z$ relation, while the restricted case is used here as a compressed summary for comparison with the literature. Our data therefore indicate that $\Msz$ is biased low relative to \WL-calibrated mass, while leaving the detailed form of the $\Msz$--$M_{500}$--$z$ relation only weakly constrained.

\subsection{The concentration--mass--redshift relation}
\label{subsec:disc_cM}

The internal structure of dark matter haloes, characterised by the concentration parameter $c_{200}$, provides a useful probe of halo assembly in \LCDM cosmology \citep{Bullock2001,Wechsler2002}. 
For the CHEX-MATE--\amalgam sample, the fitted population mean relation has a normalisation of $c_{200}=3.53\pm0.71$ at the pivot mass $M_{200}=10^{15}\Msun$ and $\zref=0.25$. As shown in Figure~\ref{fig:cM}, the inferred concentration is consistent with recent simulation-based predictions for the $c_{200}$--$M_{200}$--$z$ relation in standard \LCDM cosmology \citep{Ludlow2016,Child2018cm,Diemer2019,Ishiyama2021}. These predictions are calibrated directly in the massive cluster regime relevant to our sample, rather than relying on an extrapolation from lower halo masses. For massive clusters, comparisons of dark-matter-only and hydrodynamical runs in the combined BAHAMAS and MAssive ClusterS and Intercluster Structures (MACSIS) simulations indicate that baryonic effects on the $c$--$M$ relation are modest compared with our observational uncertainties, although the detailed impact remains model-dependent \citep[e.g.][]{Henson2017}. Our finding of no significant mass dependence is likewise consistent with these theoretical expectations, since the predicted slope of the $c$--$M$ relation is weak over the mass range probed here.

This normalisation is also consistent with lensing-inferred concentrations for X-ray-selected massive clusters \citep[e.g.][]{Okabe+Smith2016,Umetsu2016clash}. In particular, \citet{Umetsu+Diemer2017} obtained $c_{200}=3.66\pm0.11$ at $M_{200}\approx 1.4\times 10^{15}\Msun$ from a joint \WL and strong-lensing analysis of 16 X-ray-selected clusters from the Cluster Lensing And Supernova survey with Hubble (CLASH).

Our lensing-based measurement of the scatter at fixed halo mass, $0.22\pm0.04$~dex, is broadly consistent with theoretical expectations for the full halo population, including relaxed and unrelaxed systems. In particular, \citet{Diemer2015} found a $68\percent$ rms scatter of $\approx0.16$~dex in $\log c_{200}$, consistent with earlier numerical studies \citep{Bullock2001,Wechsler2002,Duffy2008,Bhattacharya2013}. This comparison should, however, be interpreted with caution, because the simulated relations quoted here are defined for three-dimensional halo concentrations, whereas lensing-inferred concentrations can be affected by projection effects from halo triaxiality, orientation, and correlated structure along the line of sight. In our analysis, these projection effects were not modelled as a separate intrinsic covariance term for $c_{200}$; the inferred scatter should therefore be regarded as an effective scatter in the lensing-derived concentrations, including physical halo-to-halo variation and residual projection-induced contributions.

For comparison, earlier lensing studies of X-ray-selected clusters found smaller concentration scatters of $\lesssim20\percent$, corresponding to $\simlt 0.09$~dex \citep{Okabe+Smith2016,Umetsu2016clash,Umetsu2020xxl}. A plausible explanation is that X-ray selection preferentially favours cool-core or dynamically more relaxed systems, which are expected to exhibit a narrower concentration distribution \citep{Buote2007,Ettori2010,Eckert2011cc,Meneghetti2014clash}.

Taken together, the agreement of the inferred normalisation and scatter with theoretical expectations suggests that our cluster sample is broadly consistent with the halo structure expected for massive clusters in \LCDM cosmology. However, given the limited sample size and the complex selection function, this result should not be interpreted as demonstrating the absence of structural selection effects. 

Some simulation-based models predict a flattening or upturn of the $c$--$M$ relation for high-peak-height systems, associated with the high-mass and/or high-redshift regime and possibly linked to the non-equilibrium structure of rapidly accreting massive haloes \citep{Diemer2019,Ishiyama2021,Scofield2026}. The high-mass measurements in Figure~\ref{fig:cM}, including the Tier-2 geometric mean, lie in the regime where such behaviour may become relevant, but their current uncertainties do not allow us to test this feature robustly. A robust assessment of such a feature will require a larger statistical sample and more precise validation of the \WL mass and concentration modelling biases.

\subsection{Systematics and future prospects}

An important limitation of our analysis is that the CHEX-MATE--\amalgam sample is effectively defined by the availability of suitable ground-based lensing data from the \amalgam project \citep{Gavazzi2026} within the full CHEX-MATE sample. This necessarily limits the sample size and, potentially, its representativeness. In particular, the current subsample shows substantial overlap with earlier well-studied cluster samples targeted for lensing analyses \citep[e.g.][]{WtG3,Umetsu2014clash,Hoekstra2015,Herbonnet2020}, which may introduce implicit selection effects beyond the nominal parent-sample selection of CHEX-MATE. Our work should therefore be regarded as an initial \WL mass calibration of the CHEX-MATE programme, rather than as a definitive calibration of the full CHEX-MATE sample.

Looking ahead, the collaboration is constructing new shape and multiband photometric catalogues based on the modern \textsc{SourceXtractor++} \citep{Bertin2020,Kummel2025} pipeline \citep[see also][]{Schrabback2026}, incorporating the COSMOS2020 photometry \citep{COSMOS2020}, for a substantially larger sample of $\sim 90$ CHEX-MATE clusters. This will provide a more homogeneous basis for future \WL analyses and will improve the statistical precision and representativeness of the CHEX-MATE lensing sample.

For the \WL mass calibration, the current simulation-based assessment of the $\Mwl$--$\Mtrue$ relation is based on synthetic observations constructed from a BAHAMAS-based simulated cluster sample at $z=0.25$ (Appendix~\ref{appendix:test}). These tests provide a useful calibration over the mass range directly sampled by this simulated subsample, $2\times10^{14}\Msun \simlt M_{500,\mathrm{true}}\simlt 10^{15}\Msun$, and show no evidence for a statistically significant mass dependence of the \WL modelling bias over this interval. However, the finite BAHAMAS simulation volume limits the statistics in the extreme high-mass tail, $M_{500,\mathrm{true}}\simgt 10^{15}\Msun$, reached by the most massive Tier-2 clusters. In our analysis, we accounted for this limitation at first order by propagating the fitted normalisation and slope uncertainties of the $\Mwl$--$\Mtrue$ relation in the population-level modelling (Section~\ref{subsec:priors}). A more direct calibration of this high-mass regime will require larger simulated cluster samples. In addition, the \Planck SZ selection can introduce orientation-dependent effects in the \WL calibration \citep{Saxena2025}.

To address these issues, an ongoing effort within the CHEX-MATE collaboration is developing multi-probe synthetic observations based on The Three Hundred Project \citep{Cui2018}. This should enable a more realistic validation of the modelling of cluster observables, including \WL mass calibration, selection-induced biases, and the dependence of the \WL modelling biases on the adopted baryonic-physics treatment, using a larger and more representative simulated cluster sample. Such validation will be particularly important for future attempts to test subtle features of the concentration--mass relation, such as the possible upturn at high peak height.

\section{Summary and conclusions}
\label{sec:summary}

We have presented a \WL analysis of 41 \Planck SZ-selected galaxy clusters at $0.11 \le z \le 0.55$ (Table~\ref{tab:sample}), drawn from the CHEX-MATE sample and covered by the \amalgam project \citep{Gavazzi2026}. Using wide-field imaging from Subaru/Suprime-Cam and CFHT/MegaPrime, we measured azimuthally averaged reduced tangential shear profiles centred on the X-ray peak of each cluster over the comoving radial range $R\in[0.3,3]\Mpch$ (Section~\ref{sec:data}). The \WL signal has a median S/N of $6.5$ per cluster, while the $45^\circ$-rotated component is statistically consistent with zero, with a median S/N of $-0.1$ (Figure~\ref{fig:WLSN}). These results support the overall quality of the \WL measurements for the CHEX-MATE--\amalgam sample (see Figures~\ref{fig:gt_all} and \ref{fig:kmap_all}).

We modelled the azimuthally averaged excess surface mass density profile $\Delta\Sigma_+(R)$ of each cluster with a spherical NFW profile, using the full covariance matrix, including statistical shape noise and projections of uncorrelated large-scale structure (Section~\ref{sec:mass}). This yields homogeneous constraints on $(M_{200,\mathrm{WL}}, c_{200,\mathrm{WL}})$ and the related overdensity masses for the full sample (Table~\ref{tab:mass}). A complementary stacked-lensing analysis provides an ensemble-level consistency check (Figure~\ref{fig:stackgt}). The resulting NFW parameters and derived masses agree well with the weighted geometric means of those inferred from the individual-cluster analyses (Table~\ref{tab:stack}).

We then used a hierarchical Bayesian framework to constrain the population-level $c_{200}$--$M_{200}$--$z$ and $\Msz$--$M_{500}$--$z$ relations for the sample, accounting for measurement uncertainty, intrinsic scatter, and simulation-calibrated modelling biases in the \WL mass and concentration (Section~\ref{sec:scaling}). We modelled the latent true-mass distribution of the selected sample in both analyses, and for the $\Msz$--$M_{500}$--$z$ relation, we additionally included an effective treatment of the CHEX-MATE tier selection through truncation in the observed SZ mass proxy.

For the concentration--mass--redshift relation, the fitted population mean relation has a normalisation of $c_{200}=3.53\pm0.71$ at the pivot mass $M_{200}=10^{15}\Msun$ and a reference redshift of $\zref=0.25$, with an intrinsic scatter of $0.22\pm0.04$~dex in $\log c_{200}$ (Table~\ref{tab:cM}). We find no significant mass or redshift dependence over the range probed by our sample. The normalisation and the scatter are consistent with recent simulation-based expectations for massive haloes in \LCDM (Figure~\ref{fig:cM}), suggesting that our sample is broadly consistent with the halo structure expected for massive clusters. The current sample, however, is not yet large enough to test more subtle features, such as a possible upturn in concentration at the highest peak heights.

For the \Planck SZ mass proxy $\Msz$, the hierarchical regression with free slopes $(\beta_{Y|Z},\gamma_{Y|Z})$ yields a \WL-calibrated $\Msz$--$M_{500}$--$z$ relation without evidence of redshift evolution and only a marginal indication of a mass dependence (Table~\ref{tab:MM} and Figure~\ref{fig:MM}). At the pivot mass $M_{500}=7\times10^{14}\Msun$ and $\zref=0.25$, the inferred ratio is $\Msz/M_{500}=0.83\pm0.09$, with an intrinsic scatter of $0.10\pm0.02$~dex. For a direct comparison with the conventional \Planck mass-bias parametrisation, we also considered a restricted model with $\beta_{Y|Z}=1$ and $\gamma_{Y|Z}=0$, for which we find $1-\bsz=0.72\pm0.11$. Under this restricted parametrisation, the \Planck mass proxy underestimates the \WL-calibrated halo mass by $(28\pm11)\percent$. This restricted result serves mainly as a compact summary for comparison with the literature, whereas the free-slope model provides the more general description we adopted in our main analysis. From the posterior predictive distribution implied by the baseline $\Msz$--$M_{500}$--$z$ relation, we further derived \WL-calibrated posterior estimates of $M_{500}$ for all clusters in our sample (Table~\ref{tab:mass} and Figure~\ref{fig:Mz}).

Overall, this work provides an initial \WL mass calibration for multi-probe studies within the CHEX-MATE programme. Several limitations should nevertheless be kept in mind (Section~\ref{sec:discussion}). The current \amalgam subsample is defined by the availability of suitable ground-based lensing data and therefore does not yet constitute a fully homogeneous \WL follow-up of the parent CHEX-MATE sample. In addition, our calibration of the \WL mass scale and residual observational systematics is adequate for the current statistical precision, but will need to be revisited for future higher-precision applications.

An extension of the analysis to a substantially larger and more homogeneous CHEX-MATE lensing sample, together with improved simulation-based validation of the \WL mass calibration, should enable tighter constraints on the \Planck mass scale, a more robust characterisation of the concentration distribution of massive clusters, and a stronger \WL anchor for future CHEX-MATE cosmological and astrophysical analyses. In this broader context, joint analyses combining \WL measurements with galaxy kinematics from complementary spectroscopic observations \citep[e.g.][]{Umetsu2025,Pizzuti2026} will provide an additional route to constrain the cluster mass distributions, identify dynamically complex systems, and test gravity on cluster scales in the CHEX-MATE sample.

\begin{acknowledgements}
We dedicate this paper to the memory of our friend and colleague Mario Nonino. 

We are grateful to members of the CHEX-MATE Lensing Working Group for useful discussions during regular teleconferences. We thank Stefano Andreon for providing helpful comments on the manuscript. K.U. acknowledges support from the National Science and Technology Council, Taiwan (grants NSTC 112-2112-M-001-027-MY3 and NSTC 115-2112-M-001-027-) and from the Academia Sinica Investigator Award (grant AS-IA-112-M04).
R.G. acknowledges support from the ANR grant ``AMALGAM'' (PI: R. Gavazzi). Part of this work was conducted on the morpho and infinity computing facilities at IAP. R.G. thanks S. Rouberol and V. de Lapparent for their help in ensuring the smooth operation of these computing facilities.
M.S. acknowledges financial support from the INAF mainstream project 1.05.01.86.10, INAF Theory Grant 2023 ``Gravitational lensing detection of matter distribution at galaxy cluster boundaries and beyond'' (1.05.23.06.17), and the INAF Guest Observer Grant 2024 ``Towards anchoring the mass scale of galaxy clusters with galaxy kinematics'' (1.05.24.02.15).
S.E., F.G., C.G., L.L., M.R., and M.S. acknowledge PRIN-MUR 2022 supported by Next Generation EU (n. 20227RNLY3, ``The concordance cosmological model: stress-tests with galaxy clusters'').
J.K. acknowledges support from the Basic Science Research Program through the National Research Foundation of Korea (NRF), funded by the Ministry of Education (2019R1A6A1A10073887), and from the National Research Foundation of Korea (NRF) grant funded by the Korean government (MSIT; RS-2025-16302968).
B.J.M. acknowledges support from the Science and Technology Facilities Council grant ST/Y002008/1.
L.P. acknowledges support from the Italian Ministry of University and Research (MUR) under Grant ``Progetto Dipartimenti di Eccellenza 2023--2027'' (BiCoQ).
E.P. acknowledges support from CNRS/INSU and CNES, the French space agency.
G.W.P. acknowledges long-term support from CNES, the French space agency.
E.R. and J.S. acknowledge support from NASA grants 80NSSC25K0006 and 80NSSC25K8009.
This research was supported by the International Space Science Institute (ISSI) in Bern through ISSI International Team project \#565 ({\it Multi-Wavelength Studies of the Culmination of Structure Formation in the Universe}). This work is based in part on data collected at Subaru Telescope, which is operated by the National Astronomical Observatory of Japan. 
\end{acknowledgements}

\bibliographystyle{aa}

\begin{appendix}

\section{Individual cluster properties and weak-lensing measurements}
\label{appendix:individual}

This appendix summarises the basic properties and individual \WL measurements of the 41 galaxy clusters in the CHEX-MATE--\amalgam sample. Table~\ref{tab:sample} lists the cluster properties, imaging information,
and background-source selections used in the analysis. Table~\ref{tab:mass} summarises the individual NFW-based \WL mass and concentration measurements.

\begin{figure*}[tbp]
 \begin{center}
  \includegraphics[width=\textwidth,angle=0,clip]{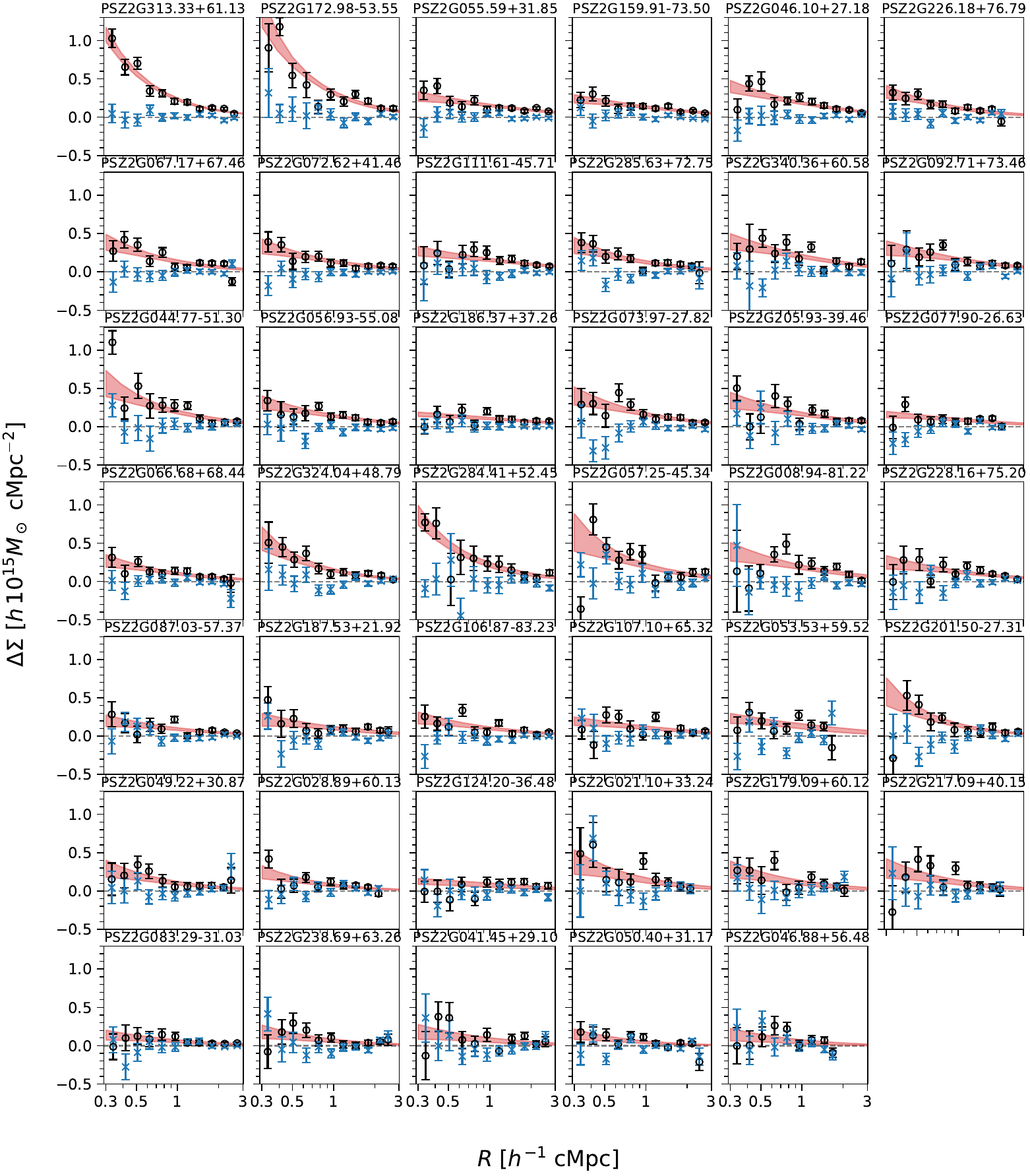}
 \end{center}
 \caption{Excess surface mass density profiles measured over the comoving radial range $R\in[0.3,3]\Mpch$, centred on the X-ray peak position for the full sample of 41 clusters. In each panel, the black circles with error bars represent the $\Delta\Sigma_+$ signal derived from reduced tangential shear measurements, while the blue crosses with error bars denote the $45^\circ$-rotated cross component, $\Delta\Sigma_\times$, which is expected to be statistically consistent with zero. The shaded red region indicates the $1\sigma$ credible interval of the NFW fit to the $\Delta\Sigma_+$ profile (Table~\ref{tab:mass}). The clusters are displayed in descending order of \WL S/N ($\mathrm{S/N}_+$; see Table~\ref{tab:sample}) in the $\Delta\Sigma_+(R)$ profile measurements.}
 \label{fig:gt_all}
\end{figure*}

\begin{figure*}[tbp]
 \begin{center}
  \includegraphics[width=\textwidth,angle=0,clip]{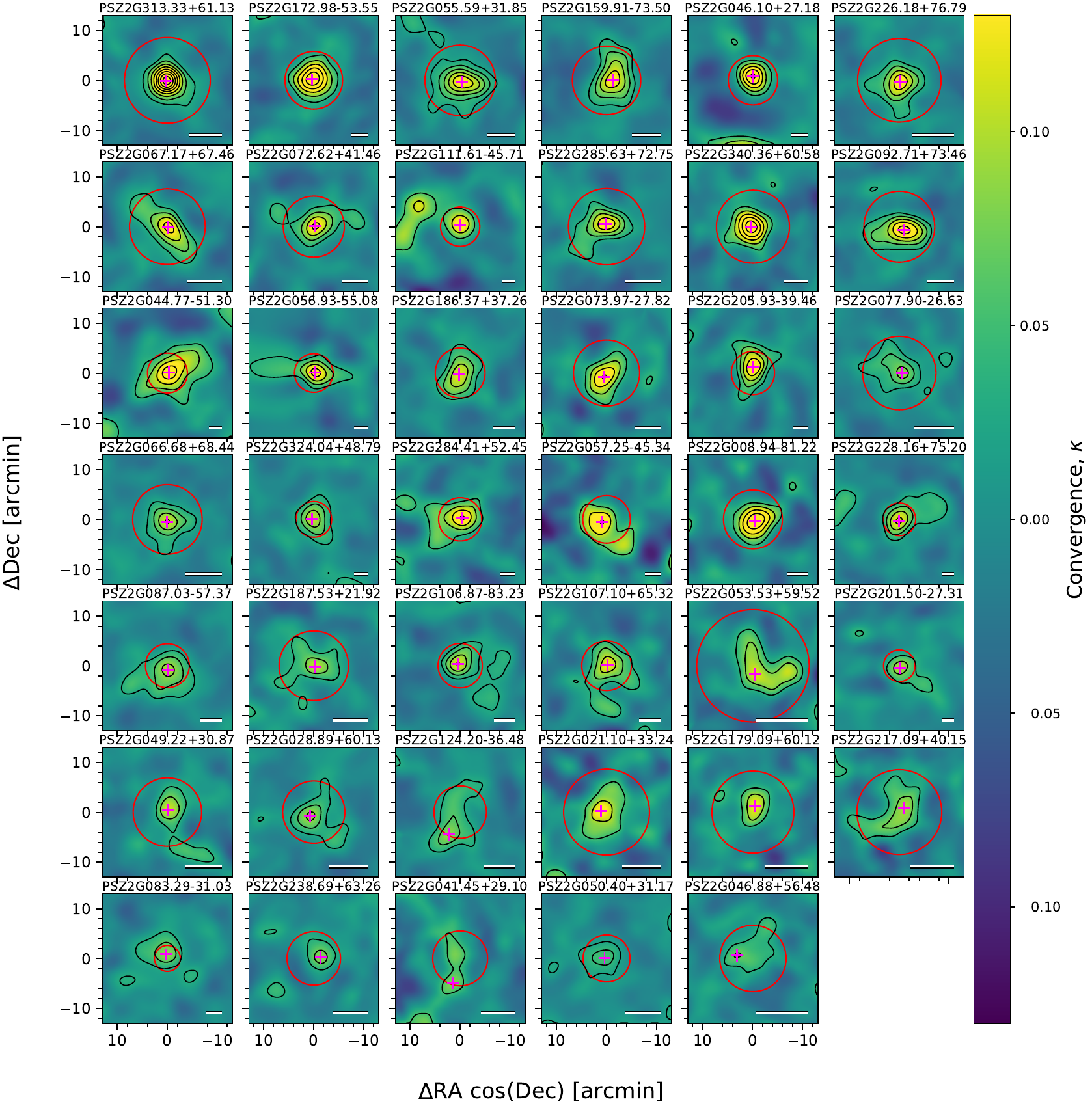}
 \end{center}
 \caption{Two-dimensional \WL mass maps for the full sample of 41 clusters, reconstructed using wide-field multiband imaging data from the \amalgam project. The maps were reconstructed from reduced-shear fields smoothed with a circular Gaussian kernel of FWHM $4.0\arcmin$. Each panel covers a fixed $26\arcmin\times 26\arcmin$ field centred on the respective X-ray peak position. The colour bar indicates the reconstructed $E$-mode convergence field, $\kappa(\btheta)$. The overlaid contours show the $E$-mode convergence, with the lowest contour level and contour interval both set to $2\sigma_B$, where $\sigma_B$ is the standard deviation of the corresponding $B$-mode map. The magenta plus sign marks the maximum of the reconstructed $E$-mode convergence field in each panel. A horizontal white bar in each panel indicates a scale of $1\Mpch$ at the cluster redshift. The red circle centred on each image indicates the cluster radius $r_{500,\mathrm{WL}}$ estimated from NFW modelling of the $\Delta\Sigma_+$ profile (Table~\ref{tab:mass}). The clusters are displayed in descending order of \WL S/N, $\mathrm{S/N}_+$ (Table~\ref{tab:sample}), measured from the $\Delta\Sigma_+$ profiles shown in Figure~\ref{fig:gt_all}. North is up and east is to the left.}
 \label{fig:kmap_all}
\end{figure*}

The azimuthally averaged excess surface mass density profiles, $\Delta\Sigma_+(R)$ and $\Delta\Sigma_\times(R)$, are shown in Figure~\ref{fig:gt_all}. The corresponding two-dimensional projected mass maps, reconstructed from Gaussian-smoothed reduced-shear fields, are shown in Figure~\ref{fig:kmap_all}.

\begin{table*}
\caption{Basic properties of the CHEX-MATE--\amalgam cluster sample.}
\label{tab:sample}
\centering
\normalsize
\setlength{\tabcolsep}{3.0pt}
\begin{tabular}{lccccccccccc}
\hline\hline
Name & RA & Dec & $\zl$ & Tier & Band & Filters & BG sel. & $\ngal$ & $\langle\beta\rangle$ & $\mathrm{S/N}_+$ \\
\hline
PSZ2~G008.94$-$81.22 & 3.5796 & -30.3912 & 0.307 & 2 & $R_\mathrm{C}$ & \underline{$B_\mathrm{J}$} \underline{$R_\mathrm{C}$} $i^\star$ \underline{$z^+$} & CC & 4.1 & 0.62 & 6.2 \\
PSZ2~G021.10$+$33.24 & 248.1958 & 5.5751 & 0.151 & 1+2 & $i^+$ & $g^\star$ \underline{$V_\mathrm{J}$} \underline{$r^\star$} \underline{$i^+$} & CC & 2.3 & 0.73 & 4.2 \\
PSZ2~G028.89$+$60.13 & 225.0817 & 21.3692 & 0.153 & 1 & $i^+$ & $V_\mathrm{J}$ $i^+$ & \smem & 9.4 & 0.76 & 4.4 \\
PSZ2~G041.45$+$29.10 & 259.4366 & 19.6766 & 0.178 & 1 & $i^+$ & $V_\mathrm{J}$ $i^+$ & \smem & 5.4 & 0.71 & 3.1 \\
PSZ2~G044.77$-$51.30 & 333.7390 & -14.0032 & 0.503 & 2 & $R_\mathrm{C}$ & \underline{$B_\mathrm{J}$} $V_\mathrm{J}$ $r^\star$ \underline{$R_\mathrm{C}$} $I_\mathrm{C}$ \underline{$z^+$} & CC & 9.5 & 0.49 & 7.6 \\
PSZ2~G046.10$+$27.18 & 262.9124 & 22.8633 & 0.389 & 2 & $R_\mathrm{C}$ & $B_\mathrm{J}$ $r^\star$ $R_\mathrm{C}$ & \smem & 7.8 & 0.57 & 10.5 \\
PSZ2~G046.88$+$56.48 & 231.0341 & 29.8851 & 0.115 & 1 & $r^\star$ & $g^\star$ $r^\star$ & \smem & 4.2 & 0.77 & 2.3 \\
PSZ2~G049.22$+$30.87 & 260.0411 & 26.6253 & 0.164 & 1 & $i^+$ & \underline{$B_\mathrm{J}$} \underline{$V_\mathrm{J}$} \underline{$R_\mathrm{C}$} \underline{$i^+$} & CC & 5.1 & 0.73 & 4.6 \\
PSZ2~G050.40$+$31.17 & 260.0358 & 27.6705 & 0.164 & 1 & $i^+$ & $V_\mathrm{J}$ $i^+$ & \smem & 9.0 & 0.74 & 3.0 \\
PSZ2~G053.53$+$59.52 & 227.5525 & 33.5083 & 0.113 & 1 & $z^+$ & \underline{$B_\mathrm{J}$} $g^+$ \underline{$R_\mathrm{C}$} \underline{$z^+$} & CC & 3.8 & 0.83 & 5.5 \\
PSZ2~G055.59$+$31.85 & 260.6134 & 32.1324 & 0.224 & 2 & $i^+$ & $u^\star$ \underline{$B_\mathrm{J}$} $g^\star$ \underline{$V_\mathrm{J}$} $r^\star$ \underline{$R_\mathrm{C}$} \underline{$i^+$} & CC & 12.9 & 0.68 & 11.9 \\
PSZ2~G056.93$-$55.08 & 340.8415 & -9.5954 & 0.447 & 2 & $V_\mathrm{J}$ & \underline{$g^\star$} \underline{$V_\mathrm{J}$} $r^\star$ \underline{$i^\star$} $z^\star$ \underline{$z^+$} & CC & 11.9 & 0.51 & 7.6 \\
PSZ2~G057.25$-$45.34 & 332.9406 & -3.8294 & 0.397 & 2 & $z^+$ & \underline{$u^\star$} \underline{$B_\mathrm{J}$} $V_\mathrm{J}$ $r^\star$ \underline{$i^+$} \underline{$z^+$} & CC & 2.3 & 0.52 & 6.2 \\
PSZ2~G066.68$+$68.44 & 215.4183 & 37.2917 & 0.163 & 1 & $i^+$ & $V_\mathrm{J}$ $i^+$ & \smem & 9.7 & 0.74 & 6.6 \\
PSZ2~G067.17$+$67.46 & 216.5085 & 37.8260 & 0.171 & 1 & $R_\mathrm{C}$ & $g^\star$ \underline{$g^+$} $r^\star$ \underline{$R_\mathrm{C}$} \underline{$i^+$} & CC & 7.4 & 0.78 & 8.4 \\
PSZ2~G072.62$+$41.46 & 250.0840 & 46.7087 & 0.228 & 2 & $i^+$ & \underline{$B_\mathrm{J}$} $V_\mathrm{J}$ \underline{$R_\mathrm{C}$} \underline{$i^+$} & CC & 7.8 & 0.72 & 8.2 \\
PSZ2~G073.97$-$27.82 & 328.4032 & 17.6950 & 0.233 & 2 & $R_\mathrm{C}$ & $u^\star$ \underline{$B_\mathrm{J}$} $V_\mathrm{J}$ \underline{$R_\mathrm{C}$} $I_\mathrm{C}$ \underline{$z^+$} & CC & 6.1 & 0.68 & 7.4 \\
PSZ2~G077.90$-$26.63 & 330.2217 & 20.9720 & 0.147 & 1 & $i^+$ & \underline{$g^\star$} \underline{$r^\star$} \underline{$i^+$} & CC & 9.2 & 0.79 & 6.6 \\
PSZ2~G083.29$-$31.03 & 337.1387 & 20.6201 & 0.412 & 2 & $R_\mathrm{C}$ & \underline{$B_\mathrm{J}$} $g^\star$ $V_\mathrm{J}$ $r^\star$ \underline{$R_\mathrm{C}$} $I_\mathrm{C}$ \underline{$z^+$} & CC & 11.1 & 0.57 & 3.7 \\
PSZ2~G087.03$-$57.37 & 354.4071 & 0.2675 & 0.278 & 2 & $i^+$ & \underline{$B_\mathrm{J}$} \underline{$V_\mathrm{J}$} \underline{$R_\mathrm{C}$} \underline{$i^+$} $z^\star$ & CC & 10.0 & 0.61 & 5.9 \\
PSZ2~G092.71$+$73.46 & 203.8248 & 41.0003 & 0.228 & 2 & $i^+$ & $V_\mathrm{J}$ $i^+$ & \smem & 6.1 & 0.64 & 7.8 \\
PSZ2~G106.87$-$83.23 & 10.8534 & -20.6235 & 0.292 & 2 & $i^+$ & $V_\mathrm{J}$ $i^+$ & \smem & 7.8 & 0.59 & 5.6 \\
PSZ2~G107.10$+$65.32 & 203.1623 & 50.5584 & 0.280 & 2 & $i^+$ & $u^\star$ \underline{$B_\mathrm{J}$} $g^\star$ $g^+$ $r^\star$ \underline{$R_\mathrm{C}$} $i^\star$ $i^+$ \underline{$z^+$} & CC & 6.5 & 0.64 & 5.6 \\
PSZ2~G111.61$-$45.71 & 4.6397 & 16.4365 & 0.546 & 2 & $i^+$ & \underline{$B_\mathrm{J}$} $g^\star$ $V_\mathrm{J}$ \underline{$R_\mathrm{C}$} $i^+$ $I_\mathrm{C}$ \underline{$z^+$} & CC & 7.0 & 0.44 & 8.2 \\
PSZ2~G124.20$-$36.48 & 13.9598 & 26.4098 & 0.197 & 1+2 & $i^+$ & \underline{$V_\mathrm{J}$} \underline{$r^\star$} \underline{$i^+$} & CC & 5.4 & 0.63 & 4.3 \\
PSZ2~G159.91$-$73.50 & 22.9723 & -13.6118 & 0.206 & 2 & $R_\mathrm{C}$ & \underline{$B_\mathrm{J}$} \underline{$R_\mathrm{C}$} \underline{$z^+$} & CC & 18.1 & 0.75 & 11.0 \\
PSZ2~G172.98$-$53.55 & 39.9725 & -1.5789 & 0.373 & 2 & $I_\mathrm{C}$ & \underline{$B_\mathrm{J}$} $g^\star$ $g^+$ $V_\mathrm{J}$ $r^\star$ \underline{$R_\mathrm{C}$} $I_\mathrm{C}$ $z^\star$ \underline{$z^+$} & CC & 7.3 & 0.53 & 12.2 \\
PSZ2~G179.09$+$60.12 & 160.1861 & 39.9529 & 0.137 & 1 & $r^\star$ & $g^\star$ $r^\star$ & \smem & 3.1 & 0.73 & 3.9 \\
PSZ2~G186.37$+$37.26 & 130.7381 & 36.3658 & 0.282 & 2 & $i^+$ & $V_\mathrm{J}$ $i^+$ & \smem & 7.8 & 0.60 & 7.4 \\
PSZ2~G187.53$+$21.92 & 113.0846 & 31.6328 & 0.171 & 1 & $i^+$ & \underline{$g^\star$} \underline{$V_\mathrm{J}$} \underline{$r^\star$} \underline{$i^+$} & CC & 6.0 & 0.73 & 5.9 \\
PSZ2~G201.50$-$27.31 & 73.5456 & -3.0149 & 0.538 & 2 & $R_\mathrm{C}$ & $u^\star$ \underline{$B_\mathrm{J}$} $g^\star$ $V_\mathrm{J}$ $r^\star$ \underline{$R_\mathrm{C}$} $I_\mathrm{C}$ \underline{$z^+$} & CC & 6.1 & 0.44 & 5.1 \\
PSZ2~G205.93$-$39.46 & 64.3948 & -11.9093 & 0.443 & 2 & $R_\mathrm{C}$ & \underline{$V_\mathrm{J}$} $r^\star$ \underline{$R_\mathrm{C}$} \underline{$I_\mathrm{C}$} & CC & 8.5 & 0.47 & 6.8 \\
PSZ2~G217.09$+$40.15 & 141.0240 & 14.1736 & 0.136 & 1 & $r^\star$ & \underline{$u^\star$} \underline{$g^\star$} \underline{$r^\star$} \underline{$z^\star$} & CC & 3.0 & 0.76 & 3.8 \\
PSZ2~G226.18$+$76.79 & 178.8243 & 23.4047 & 0.143 & 1 & $R_\mathrm{C}$ & \underline{$B_\mathrm{J}$} $g^\star$ \underline{$V_\mathrm{J}$} $r^\star$ \underline{$R_\mathrm{C}$} \underline{$i^+$} & CC & 16.0 & 0.80 & 9.2 \\
PSZ2~G228.16$+$75.20 & 177.3973 & 22.4027 & 0.545 & 2 & $R_\mathrm{C}$ & $u^\star$ \underline{$B_\mathrm{J}$} $r^\star$ \underline{$R_\mathrm{C}$} \underline{$z^+$} & CC & 9.8 & 0.48 & 6.2 \\
PSZ2~G238.69$+$63.26 & 168.2267 & 13.4346 & 0.169 & 1 & $i^+$ & $V_\mathrm{J}$ $i^+$ & \smem & 5.9 & 0.72 & 3.2 \\
PSZ2~G284.41$+$52.45 & 181.5507 & -8.8011 & 0.441 & 2 & $z^+$ & \underline{$B_\mathrm{J}$} $g^\star$ $V_\mathrm{J}$ $r^\star$ \underline{$R_\mathrm{C}$} $i^\star$ $I_\mathrm{C}$ \underline{$z^+$} & CC & 3.3 & 0.46 & 6.4 \\
PSZ2~G285.63$+$72.75 & 187.6972 & 10.5521 & 0.165 & 1 & $i^+$ & \underline{$u^\star$} $g^\star$ \underline{$V_\mathrm{J}$} $r^\star$ $i^\star$ \underline{$i^+$} \underline{$z^\star$} & CC & 10.9 & 0.76 & 8.1 \\
PSZ2~G313.33$+$61.13 & 197.8725 & -1.3414 & 0.183 & 2 & $R_\mathrm{C}$ & \underline{$B_\mathrm{J}$} $V_\mathrm{J}$ \underline{$R_\mathrm{C}$} $i^+$ \underline{$z^+$} & CC & 13.8 & 0.77 & 15.4 \\
PSZ2~G324.04$+$48.79 & 206.8778 & -11.7520 & 0.452 & 2 & $R_\mathrm{C}$ & \underline{$g^\star$} $V_\mathrm{J}$ $r^\star$ \underline{$R_\mathrm{C}$} \underline{$I_\mathrm{C}$} \underline{$z^+$} & CC & 10.9 & 0.53 & 6.6 \\
PSZ2~G340.36$+$60.58 & 210.2585 & 2.8783 & 0.253 & 2 & $i^+$ & $g^\star$ \underline{$V_\mathrm{J}$} \underline{$r^\star$} \underline{$i^+$} & CC & 4.3 & 0.60 & 8.0 \\
\hline
\end{tabular}
\tablefoot{Columns 1--5 list the PSZ2 identifier, the J2000 coordinates of the adopted X-ray peak, the cluster redshift, and the CHEX-MATE tier assignment; "1+2" denotes clusters common to both tiers. Columns 6 and 7 give the shape-measurement band used for lensing and the broad-band filters entering the \amalgam source-redshift products we used, respectively (see Table~\ref{tab:filters}). These filters are used to derive photo-$z$-based quantities, including the lensing efficiency, $\beta=D_{ls}/D_s$, and the cluster membership score, \smem. Column 8 gives the background-source selection method, with entries CC and \smem for the CC and \smem-threshold selections, respectively. For CC-selected clusters, the underlined filters in Column 7 identify the subset used to define the CC selection; all filters listed in Column 7 enter the \amalgam source-redshift products. Columns 9--11 list the mean background-source number density used in the \WL analysis, the weighted mean lensing efficiency, and the linear S/N of the $\Delta\Sigma_+$ profile estimated with Equation~\eqref{eq:snr}.}
\end{table*}

\begin{table*}
\caption{Summary of the mass and concentration measurements for the CHEX-MATE--\amalgam cluster sample.}
\label{tab:mass}
\centering
\normalsize
\setlength{\tabcolsep}{3.0pt}
\begin{tabular}{lccccccc}
\hline\hline
Name & $\log{c_{200,\mathrm{WL}}}$ & $\log{M_{200,\mathrm{WL}}}$ & $\log{M_{500,\mathrm{WL}}}$ & $c_{200,\mathrm{WL}}$ & $M_{200,\mathrm{WL}}$ & $M_{500,\mathrm{WL}}$ & $M_{500,\mathrm{SZ\text{--}WL}}$ \\
 & & & & & ($10^{14}\Msun$) & ($10^{14}\Msun$) & ($10^{14}\Msun$) \\
\hline
PSZ2~G008.94$-$81.22 & $0.43 \pm 0.20$ & $15.41 \pm 0.16$ & $15.20 \pm 0.13$ & $2.75 \pm 1.27$ & $25.9 \pm 9.3$ & $16.1 \pm 4.6$ & $12.9 \pm 2.9$ \\
PSZ2~G021.10$+$33.24 & $0.65 \pm 0.35$ & $15.07 \pm 0.22$ & $14.92 \pm 0.19$ & $5.00 \pm 4.20$ & $12.0 \pm 6.2$ & $8.3 \pm 3.5$ & $8.0 \pm 1.9$ \\
PSZ2~G028.89$+$60.13 & $0.77 \pm 0.33$ & $14.65 \pm 0.18$ & $14.51 \pm 0.16$ & $6.78 \pm 4.91$ & $4.5 \pm 1.9$ & $3.3 \pm 1.2$ & $5.6 \pm 1.4$ \\
PSZ2~G041.45$+$29.10 & $0.53 \pm 0.37$ & $14.70 \pm 0.34$ & $14.52 \pm 0.31$ & $3.33 \pm 3.32$ & $5.2 \pm 3.8$ & $3.4 \pm 2.2$ & $7.0 \pm 1.7$ \\
PSZ2~G044.77$-$51.30 & $0.64 \pm 0.21$ & $15.34 \pm 0.14$ & $15.19 \pm 0.11$ & $4.44 \pm 2.15$ & $22.3 \pm 7.4$ & $15.6 \pm 4.0$ & $14.9 \pm 3.5$ \\
PSZ2~G046.10$+$27.18 & $0.45 \pm 0.14$ & $15.41 \pm 0.11$ & $15.22 \pm 0.09$ & $2.83 \pm 0.89$ & $26.1 \pm 6.8$ & $16.5 \pm 3.3$ & $13.9 \pm 3.0$ \\
PSZ2~G046.88$+$56.48 & $0.59 \pm 0.37$ & $14.42 \pm 0.40$ & $14.25 \pm 0.38$ & $4.13 \pm 4.05$ & $2.8 \pm 2.3$ & $1.9 \pm 1.4$ & $5.6 \pm 1.5$ \\
PSZ2~G049.22$+$30.87 & $0.66 \pm 0.34$ & $14.87 \pm 0.20$ & $14.72 \pm 0.16$ & $5.05 \pm 4.11$ & $7.5 \pm 3.5$ & $5.3 \pm 2.0$ & $6.9 \pm 1.5$ \\
PSZ2~G050.40$+$31.17 & $0.69 \pm 0.37$ & $14.38 \pm 0.25$ & $14.22 \pm 0.23$ & $5.94 \pm 4.96$ & $2.4 \pm 1.3$ & $1.7 \pm 0.9$ & $5.6 \pm 1.6$ \\
PSZ2~G053.53$+$59.52 & $0.30 \pm 0.25$ & $15.16 \pm 0.28$ & $14.92 \pm 0.23$ & $1.93 \pm 1.07$ & $15.4 \pm 9.7$ & $8.7 \pm 4.6$ & $6.3 \pm 1.6$ \\
PSZ2~G055.59$+$31.85 & $0.30 \pm 0.17$ & $15.34 \pm 0.13$ & $15.10 \pm 0.09$ & $2.03 \pm 0.80$ & $22.3 \pm 6.4$ & $12.7 \pm 2.7$ & $10.0 \pm 2.1$ \\
PSZ2~G056.93$-$55.08 & $0.47 \pm 0.21$ & $15.23 \pm 0.15$ & $15.03 \pm 0.12$ & $3.03 \pm 1.41$ & $17.1 \pm 5.9$ & $10.9 \pm 2.9$ & $13.8 \pm 3.1$ \\
PSZ2~G057.25$-$45.34 & $0.73 \pm 0.31$ & $15.33 \pm 0.19$ & $15.19 \pm 0.14$ & $5.87 \pm 4.27$ & $21.4 \pm 9.4$ & $15.5 \pm 5.1$ & $14.4 \pm 3.3$ \\
PSZ2~G066.68$+$68.44 & $0.59 \pm 0.28$ & $14.90 \pm 0.17$ & $14.73 \pm 0.13$ & $3.96 \pm 2.59$ & $8.0 \pm 3.1$ & $5.4 \pm 1.6$ & $6.0 \pm 1.5$ \\
PSZ2~G067.17$+$67.46 & $0.70 \pm 0.26$ & $15.04 \pm 0.16$ & $14.89 \pm 0.12$ & $5.07 \pm 3.06$ & $11.1 \pm 4.1$ & $7.9 \pm 2.3$ & $8.1 \pm 1.7$ \\
PSZ2~G072.62$+$41.46 & $0.56 \pm 0.25$ & $15.11 \pm 0.16$ & $14.93 \pm 0.12$ & $3.66 \pm 2.11$ & $13.1 \pm 4.9$ & $8.7 \pm 2.4$ & $10.4 \pm 2.6$ \\
PSZ2~G073.97$-$27.82 & $0.57 \pm 0.22$ & $15.23 \pm 0.14$ & $15.06 \pm 0.11$ & $3.78 \pm 1.84$ & $17.2 \pm 5.3$ & $11.5 \pm 2.8$ & $10.6 \pm 2.4$ \\
PSZ2~G077.90$-$26.63 & $0.22 \pm 0.20$ & $14.94 \pm 0.21$ & $14.68 \pm 0.17$ & $1.64 \pm 0.72$ & $8.9 \pm 4.1$ & $4.9 \pm 1.9$ & $6.2 \pm 1.5$ \\
PSZ2~G083.29$-$31.03 & $0.45 \pm 0.33$ & $14.64 \pm 0.26$ & $14.43 \pm 0.24$ & $2.69 \pm 2.09$ & $4.4 \pm 2.5$ & $2.8 \pm 1.4$ & $10.7 \pm 3.1$ \\
PSZ2~G087.03$-$57.37 & $0.41 \pm 0.27$ & $14.92 \pm 0.17$ & $14.72 \pm 0.14$ & $2.58 \pm 1.62$ & $8.6 \pm 3.3$ & $5.3 \pm 1.7$ & $8.4 \pm 2.1$ \\
PSZ2~G092.71$+$73.46 & $0.37 \pm 0.21$ & $15.35 \pm 0.14$ & $15.13 \pm 0.11$ & $2.40 \pm 1.15$ & $22.5 \pm 7.4$ & $13.5 \pm 3.4$ & $10.2 \pm 2.2$ \\
PSZ2~G106.87$-$83.23 & $0.48 \pm 0.26$ & $14.98 \pm 0.18$ & $14.78 \pm 0.14$ & $3.09 \pm 1.85$ & $9.7 \pm 3.9$ & $6.2 \pm 2.0$ & $9.2 \pm 2.2$ \\
PSZ2~G107.10$+$65.32 & $0.26 \pm 0.19$ & $15.13 \pm 0.19$ & $14.88 \pm 0.16$ & $1.83 \pm 0.82$ & $13.9 \pm 5.9$ & $7.7 \pm 2.8$ & $9.6 \pm 2.2$ \\
PSZ2~G111.61$-$45.71 & $0.21 \pm 0.14$ & $15.51 \pm 0.13$ & $15.25 \pm 0.11$ & $1.66 \pm 0.57$ & $33.1 \pm 10.0$ & $17.8 \pm 4.3$ & $16.8 \pm 3.9$ \\
PSZ2~G124.20$-$36.48 & $0.14 \pm 0.14$ & $14.86 \pm 0.33$ & $14.58 \pm 0.30$ & $1.35 \pm 0.41$ & $7.7 \pm 5.1$ & $3.9 \pm 2.5$ & $8.4 \pm 2.0$ \\
PSZ2~G159.91$-$73.50 & $0.31 \pm 0.17$ & $15.21 \pm 0.13$ & $14.97 \pm 0.10$ & $2.07 \pm 0.81$ & $16.4 \pm 5.0$ & $9.4 \pm 2.2$ & $9.0 \pm 1.9$ \\
PSZ2~G172.98$-$53.55 & $0.96 \pm 0.16$ & $15.48 \pm 0.09$ & $15.38 \pm 0.08$ & $9.20 \pm 3.37$ & $30.6 \pm 6.7$ & $24.0 \pm 4.3$ & $15.4 \pm 3.8$ \\
PSZ2~G179.09$+$60.12 & $0.61 \pm 0.35$ & $14.91 \pm 0.25$ & $14.74 \pm 0.21$ & $4.31 \pm 3.73$ & $8.3 \pm 4.8$ & $5.6 \pm 2.7$ & $5.6 \pm 1.5$ \\
PSZ2~G186.37$+$37.26 & $0.12 \pm 0.10$ & $15.19 \pm 0.14$ & $14.90 \pm 0.13$ & $1.32 \pm 0.33$ & $15.8 \pm 5.1$ & $8.0 \pm 2.3$ & $10.9 \pm 2.6$ \\
PSZ2~G187.53$+$21.92 & $0.39 \pm 0.32$ & $15.00 \pm 0.23$ & $14.78 \pm 0.18$ & $2.24 \pm 1.64$ & $10.4 \pm 5.4$ & $6.1 \pm 2.5$ & $6.9 \pm 1.6$ \\
PSZ2~G201.50$-$27.31 & $0.94 \pm 0.26$ & $15.06 \pm 0.14$ & $14.95 \pm 0.12$ & $9.55 \pm 5.10$ & $11.5 \pm 3.7$ & $9.0 \pm 2.4$ & $13.2 \pm 3.4$ \\
PSZ2~G205.93$-$39.46 & $0.41 \pm 0.22$ & $15.38 \pm 0.17$ & $15.17 \pm 0.13$ & $2.65 \pm 1.31$ & $24.5 \pm 9.4$ & $15.0 \pm 4.4$ & $16.0 \pm 3.8$ \\
PSZ2~G217.09$+$40.15 & $0.61 \pm 0.34$ & $14.94 \pm 0.23$ & $14.77 \pm 0.20$ & $4.29 \pm 3.63$ & $8.8 \pm 4.7$ & $6.0 \pm 2.7$ & $5.7 \pm 1.5$ \\
PSZ2~G226.18$+$76.79 & $0.66 \pm 0.26$ & $14.96 \pm 0.17$ & $14.81 \pm 0.13$ & $4.63 \pm 2.81$ & $9.4 \pm 3.6$ & $6.5 \pm 2.0$ & $6.8 \pm 1.5$ \\
PSZ2~G228.16$+$75.20 & $0.34 \pm 0.22$ & $15.24 \pm 0.17$ & $15.01 \pm 0.14$ & $2.21 \pm 1.14$ & $17.6 \pm 6.7$ & $10.3 \pm 3.2$ & $15.1 \pm 3.7$ \\
PSZ2~G238.69$+$63.26 & $0.60 \pm 0.37$ & $14.60 \pm 0.26$ & $14.43 \pm 0.24$ & $4.42 \pm 4.12$ & $4.0 \pm 2.3$ & $2.7 \pm 1.4$ & $6.1 \pm 1.6$ \\
PSZ2~G284.41$+$52.45 & $0.95 \pm 0.18$ & $15.27 \pm 0.14$ & $15.17 \pm 0.12$ & $9.23 \pm 3.98$ & $19.0 \pm 6.4$ & $14.8 \pm 4.2$ & $15.4 \pm 3.5$ \\
PSZ2~G285.63$+$72.75 & $0.64 \pm 0.30$ & $15.02 \pm 0.18$ & $14.86 \pm 0.13$ & $4.54 \pm 3.22$ & $10.6 \pm 4.5$ & $7.4 \pm 2.3$ & $7.3 \pm 1.5$ \\
PSZ2~G313.33$+$61.13 & $1.15 \pm 0.10$ & $15.22 \pm 0.05$ & $15.13 \pm 0.05$ & $14.46 \pm 3.23$ & $16.5 \pm 2.1$ & $13.6 \pm 1.5$ & $10.5 \pm 2.3$ \\
PSZ2~G324.04$+$48.79 & $0.90 \pm 0.24$ & $15.06 \pm 0.13$ & $14.95 \pm 0.10$ & $8.60 \pm 4.63$ & $11.6 \pm 3.4$ & $8.9 \pm 2.1$ & $13.4 \pm 3.1$ \\
PSZ2~G340.36$+$60.58 & $0.41 \pm 0.19$ & $15.49 \pm 0.15$ & $15.28 \pm 0.12$ & $2.62 \pm 1.13$ & $31.3 \pm 11.1$ & $19.2 \pm 5.2$ & $12.5 \pm 3.1$ \\
\hline
\end{tabular}
\tablefoot{Column 1 lists the PSZ2 name. Columns 2--4 give symmetrised posterior summaries in logarithmic variables for the NFW-based \WL measurements of $c_{200,\mathrm{WL}}$, $M_{200,\mathrm{WL}}$, and $M_{500,\mathrm{WL}}$, respectively. Here, $\log$ denotes the base-10 logarithm, and the logarithmic masses are expressed as $\log(M_{\Delta,\mathrm{WL}}/M_\odot)$. Columns 5--7 give symmetrised posterior summaries evaluated directly for the corresponding linear variables. The quantities in Columns 2--7 are quoted without correction for the \WL modelling bias. Column 8 lists the posterior estimate $M_{500,\mathrm{SZ\text{--}WL}}$ inferred from the \Planck MMF3 SZ mass proxy using the \WL-calibrated $\Msz$--$M_{500}$--$z$ relation derived in Section~\ref{subsec:mm}.}
\end{table*}

\section{Simulation tests of weak-lensing halo modelling}
\label{appendix:test}

In this appendix, we use the synthetic \WL data presented in \citet{Umetsu2020xxl} to test the \WL halo modelling adopted for the CHEX-MATE--\amalgam analysis. While \citet{Umetsu2020xxl} performed a comprehensive analysis of 639 \LCDM haloes spanning $\log{(M_{500}/\Msun)} \in [13,15]$, here we focus on the high-mass regime relevant to the CHEX-MATE sample. We used these tests to calibrate the \WL mass recovery through the $\Mwl$--$\Mtrue$ relation and to assess possible systematic offsets in the recovered halo concentration.

\subsection{Simulated haloes and synthetic weak-lensing data} 
\label{appendix:bahamas}

We assessed the accuracy of our \WL mass estimates using synthetic observations of \LCDM haloes selected from a dark-matter-only run of the BAHAMAS simulations \citep{McCarthy2017,McCarthy2018}. The simulation snapshot is at $z=0.25$, which closely matches the median redshift of the CHEX-MATE--\amalgam sample ($\zlmed=0.23$). The simulation adopts a flat \LCDM cosmology with \WMAP 9-year parameters in a $(400\Mpch)^3$ box.

A full description of the synthetic catalogue construction is provided in Appendix~A of \citet{Umetsu2020xxl}; here we provide a summary. To efficiently sample the mass-dependent bias, haloes were selected to achieve a uniform distribution in logarithmic mass, with up to 100 haloes randomly drawn per bin of width $\Delta\log{M_{500}}=0.25$ over the range $\log{(M_{500}/\Msun)}\in[13,15]$. This selection yielded a total sample of 639 haloes.

Synthetic lensing maps were generated by projecting the particle distribution within a cube of side length $30\,\mathrm{pMpc}$ centred on each halo. For each halo, a single projection was constructed along the simulation $z$-axis. Convergence and reduced shear maps were computed following \citet{McCarthy2018}, assuming a single source plane at $\zs=0.829$. The maps were sampled to a mean source density of $\ngal=20$~galaxies~arcmin$^{-2}$, and Gaussian shape noise was added to each source galaxy with a dispersion of $\sigma_g \approx 0.20$ per component.

For our analysis, we utilised a high-mass subsample of 116 haloes with $M_{500}>2\times 10^{14}\Msun$. This threshold corresponds approximately to the minimum $\Msz$ mass of the CHEX-MATE Tier-1 subsample limited by $(\mathrm{S/N})_\mathrm{MMF3}>6.5$ \citep{Chexmate2021}.

\subsection{NFW mass modelling}
\label{appendix:NFW_BAHAMAS}

We analysed the synthetic \WL data using the same NFW modelling pipeline employed for the observational analysis of the CHEX-MATE--\amalgam sample. For each simulated halo, we computed the excess surface mass density profile, $\{\Delta\Sigma_+(R_i)\}_{i=1}^{\Nbin}$, in $\Nbin=11$ logarithmic bins spanning the comoving radial range $R\in[0.3,3]\Mpch$. We then fitted the \WL signal with a spherical NFW profile to infer the halo mass, $M_{200,\mathrm{WL}}$, and concentration, $\cwl$, treating both as free parameters with the same log-uniform priors as in the observational analysis (Section~\ref{subsec:modelling}).

To assess biases in the \WL-inferred mass and concentration in a manner directly applicable to the observational analysis, we reproduced, as closely as possible, the radial weighting applied to the binned profile data vector $\{\Delta\Sigma_+(R_i)\}_{i=1}^{\Nbin}$ in the real-data fits. This matching is essential because the projected lensing profiles of simulated haloes are not expected to follow an exact NFW lensing profile. When the NFW model is only an approximation to the true projected signal, the inferred parameters depend on how deviations from the model are weighted across radius. The effective radial weighting therefore controls how non-NFW features, including halo asphericity and correlated structure, are projected onto the fitted values of $M_{200,\mathrm{WL}}$ and $\cwl$.

In our observational analysis, the weighting is governed by the inverse covariance matrix, $C^{-1}$, with $C=C^\mathrm{shape}+C^\mathrm{lss}$ (Section~\ref{subsec:cmat}). Because the mean source density in the synthetic data is roughly twice that of the CHEX-MATE--\amalgam observations, the relative importance of shape noise and cosmic noise would otherwise differ, leading to a different effective radial weighting.

Although the synthetic observations do not contain an explicit realisation of uncorrelated large-scale-structure noise, we therefore included a representative $C^\mathrm{lss}$ contribution in the fitting likelihood, together with the shape-noise term realised in the synthetic data, in order to approximate the covariance weighting used for the observations. Specifically, we adopted the $C^\mathrm{lss}$ matrix evaluated for the mean source depth of the CHEX-MATE--\amalgam sample, $\zs\approx0.99$, and scaled its normalisation by the ratio of the observed to simulated mean source densities, $7.7/20$. This effective scaling was applied only to reproduce approximately the same radial transition between the shape-noise-dominated and cosmic-noise-dominated regimes as in the observational fits.

We applied this fitting procedure to the high-mass subsample of 116 haloes with $M_{500}>2\times 10^{14}\Msun$ defined in Appendix~\ref{appendix:bahamas}. Figure~\ref{fig:bahamas_comp} compares the recovered parameters, $M_{500,\mathrm{WL}}$, $M_{200,\mathrm{WL}}$, and $\cwl$, with their true input values. The weighted geometric mean ratios for this subsample are $\langle M_{200,\mathrm{WL}}/M_{200,\mathrm{true}}\rangle_\mathrm{g}=1.027\pm 0.032$, $\langle M_{500,\mathrm{WL}}/M_{500,\mathrm{true}}\rangle_\mathrm{g}=0.986\pm 0.028$, and $\langle \cwl/c_{200,\mathrm{true}}\rangle_\mathrm{g}=0.889\pm 0.047$. Thus, the recovered \WL masses are nearly unbiased on average, whereas the \WL-inferred concentrations exhibit a significant underestimate.

The near-unity mass recovery found here for the high-mass subsample is consistent with the BAHAMAS-based tests of \citet{Umetsu2020xxl}, where the strongest mass-dependent \WL mass bias appeared only in the low-mass group regime \citep[see also][]{Akino2022}, well below the cluster masses considered here. Their control tests using synthetic NFW lenses showed no corresponding mass bias from low lensing S/N alone; the low-mass negative bias was instead interpreted as likely arising from departures of the projected $\Delta\Sigma_+(R)$ profiles of \LCDM haloes from the NFW form, plausibly caused by correlated surrounding structures projected around the haloes.

We quantified the fidelity of our mass recovery by modelling the $\Mwl$--$\Mtrue$ relation with a log-linear regression of the form
\begin{equation}
\label{eq:bias_model}
 \log\left(\frac{M_{\mathrm{WL}}}{M_{\mathrm{piv}}}\right) =
\alpha_\mathrm{WL} + \beta_\mathrm{WL} \log\left(\frac{M_{\mathrm{true}}}{M_{\mathrm{piv}}}\right),
\end{equation}
with intrinsic scatter $\sigma_\mathrm{WL}$ in $\log\Mwl$ at fixed $\Mtrue$. Here $\alpha_\mathrm{WL}$ and $\beta_\mathrm{WL}$ are the intercept and slope parameters, respectively, and $M_{\mathrm{piv}}$ is the reference pivot mass adopted for the high-mass BAHAMAS regression.

From the \lira regression of the synthetic measurements for the high-mass subsample, we constrain the bias parameters at $\Delta=500$ to be:
\begin{equation}
\begin{aligned}
  \alpha_{\mathrm{WL}} &= -0.016 \pm 0.022,\\
  \beta_{\mathrm{WL}}  &=  0.969 \pm 0.084,\\
  \sigma_{\mathrm{WL}} &= (9.9\pm 1.2)\times 10^{-2}\,\text{dex},
\end{aligned}
\end{equation}
with a pivot mass of $M_{500,\mathrm{piv}} = 5\times 10^{14}\Msun$. The measured scatter, $\sigma_{\mathrm{WL}}$, corresponds to a fractional scatter of $(23 \pm 3) \percent$. These results indicate that the \WL masses at $\Delta=500$ are recovered with high accuracy, showing a mean bias factor of $10^{\alpha_{\mathrm{WL}}} \approx 0.96$ at the pivot mass.

Similarly, for the overdensity $\Delta=200$, we find the following constraints:
\begin{equation}
\begin{aligned}
  \alpha_{\mathrm{WL}} &= 0.015 \pm 0.020,\\
  \beta_{\mathrm{WL}}  &= 1.024 \pm 0.081,\\
  \sigma_{\mathrm{WL}} &= (7.7\pm 1.4)\times 10^{-2}\,\text{dex},
\end{aligned}
\end{equation}
with a pivot mass of $M_{200,\mathrm{piv}} = 7\times 10^{14}\Msun$. The measured scatter, $\sigma_{\mathrm{WL}}$, corresponds to a fractional scatter of $(18 \pm 3) \percent$. Again, the \WL masses at $\Delta=200$ are recovered with high accuracy, showing a mean bias factor of $10^{\alpha_{\mathrm{WL}}} \approx 1.03$ at the pivot mass.

For the CHEX-MATE--\amalgam analysis, we express the simulation-based \WL mass calibration at the pivot masses adopted in the main population modelling, namely $M^\prime_{500,\mathrm{piv}}=7\times10^{14}\Msun$ for $\Delta=500$ and $M^\prime_{200,\mathrm{piv}}=10^{15}\Msun$ for $\Delta=200$ (Section~\ref{subsec:single}). Under this pivot transformation, the slope and intrinsic scatter are unchanged, and the corresponding intercepts inferred from the posterior samples are
\begin{equation}
\label{eq:mbias}
\begin{aligned}
 \alpha_{\mathrm{WL}}^{\prime(500)} &= -0.021 \pm 0.032,\\
 \alpha_{\mathrm{WL}}^{\prime(200)} &= +0.019 \pm 0.030,
\end{aligned}
\end{equation}
for the $\Delta=500$ and $\Delta=200$ overdensities, respectively. These correspond to mean \WL mass bias factors of $0.95\pm0.07$ at $M_{500,\mathrm{true}}=7\times10^{14}\Msun$ and $1.04\pm0.07$ at $M_{200,\mathrm{true}}=10^{15}\Msun$.

For each overdensity, we adopted for our main analysis the $\Mwl$--$\Mtrue$ relation specified by $(\alpha^\prime_\mathrm{WL}, \beta_\mathrm{WL})$ at the pivot mass $M^\prime_{\mathrm{piv}}$. Within our Bayesian population modelling framework, we marginalised over the uncertainties in the intercept $\alpha^\prime_{\mathrm{WL}}$ and slope $\beta_{\mathrm{WL}}$ (see Section~\ref{subsec:priors}). For the intrinsic scatter, we adopted a fixed value of $\sigma_\mathrm{WL}=0.2/\ln(10)$ for both overdensities, corresponding to a fractional scatter of $20\percent$. This choice is broadly consistent with the results of our synthetic analysis ($\approx 18$--$23\percent$) as well as with independent simulation-based studies in the literature \citep[e.g.][]{Becker+Kravtsov2011,Gruen2015}.

The concentration offset was calibrated separately from the $\Mwl$--$\Mtrue$ relation, using the $c_{200,\mathrm{WL}}$--$c_{200,\mathrm{true}}$ comparison above. Although mass and concentration are inferred from the same NFW fit, the high-mass BAHAMAS tests show that nearly unbiased \WL masses can coexist with an $\approx 11\percent$ underestimate of the recovered concentration. Given the nearly unbiased recovery of $M_{200}$, the measured underestimate of $c_{200,\mathrm{WL}}=r_{200,\mathrm{WL}}/r_{\mathrm{s,WL}}$ instead points to a systematic overestimate of the fitted NFW scale radius $r_{\mathrm{s,WL}}$, driven by the radial shape of the projected halo lensing signal over the adopted fitting range. A similar underestimation was found in the BAHAMAS-based tests of \citet{Umetsu2020xxl}, which extended to substantially lower masses than the high-mass calibration used here; in that work, the effect was treated as a $16\percent$ systematic uncertainty on the inferred $c$--$M$ normalisation rather than corrected explicitly. 

For our high-mass sample, we applied a constant multiplicative correction factor of $1/0.89$ to the \WL-inferred concentrations before fitting the $c$--$M$--$z$ relation. The uncertainty in this correction corresponds to a multiplicative uncertainty of $\approx 5\percent$ in the concentration normalisation, or $\approx 0.02$~dex in $\log c_{200}$, and is subdominant to the current statistical uncertainty on the $c$--$M$ normalisation. We therefore did not propagate this uncertainty separately in the regression.

\subsection{Halo-model mass modelling}
\label{appendix:halomodel}
 
We also tested our shear-to-mass modelling pipeline by refitting the same individual synthetic $\Delta\Sigma_+(R)$ profiles with the halo model, using the same shape-noise realisations and fitting setup as for the baseline NFW modelling. In this model, the large-scale clustering contribution is included as a 2-halo term \citep{Oguri+Takada2011}, and we write the three-dimensional density profile as
\begin{equation}
\rho(r)=\rho_\mathrm{NFW}(r)f_\mathrm{trunc}(r)+\rho_\mathrm{2h}(r),
\end{equation}
where $f_\mathrm{trunc}(r)=[1+(r/r_\mathrm{t})^2]^{-2}$ denotes the Baltz--Marshall--Oguri (BMO) truncation function \citep{BMO}, $r_\mathrm{t}=3\,r_{200}$ is the truncation radius \citep{Oguri+Hamana2011}, and $\rho_\mathrm{2h}(r)$ is the 2-halo term. We modelled the projected 2-halo contribution to the lensing signal following \citet{Umetsu2016clash}, for a given halo mass and in our fiducial cosmology (see Section~\ref{subsec:cmat}). We described the projected halo model with $M_{200}$ and $c_{200}$ as fitting parameters and used the same priors as for the NFW model.

For the halo-model test, it is important to specify the mass convention, because the model profile is no longer a pure NFW density profile. For such a model, at least three different spherical-overdensity mass definitions can be considered: (i) the mass $M_\Delta^\mathrm{NFW}$ defined by the underlying NFW component $\rho_\mathrm{NFW}(r)$ \citep[e.g.][]{Umetsu2016clash}; (ii) the BMO-truncated NFW mass $M_\Delta^\mathrm{BMO}$ obtained from $\rho_\mathrm{NFW}(r)f_\mathrm{trunc}(r)$ \citep{BMO}; and (iii) the total-profile mass $M_\Delta^\mathrm{tot}$ obtained from the full density profile \citep[e.g.][]{colossus}, including the 2-halo term. These definitions are not generally identical. 

The quoted halo-model masses and concentrations are defined using the first convention, $M_\Delta \equiv M_\Delta^\mathrm{NFW}$ and $c_\Delta \equiv r_\Delta^\mathrm{NFW}/r_\mathrm{s}^\mathrm{NFW}$ (see Section~\ref{subsec:modelling}), so the halo-model results can be compared directly with the baseline NFW fits and with the input spherical-overdensity halo masses in the simulations. The truncation factor and the 2-halo term enter only the forward model for the projected lensing profile.

\begin{figure*}[tbp]
 \begin{center}
  \includegraphics[width=0.9\textwidth,angle=0,clip]{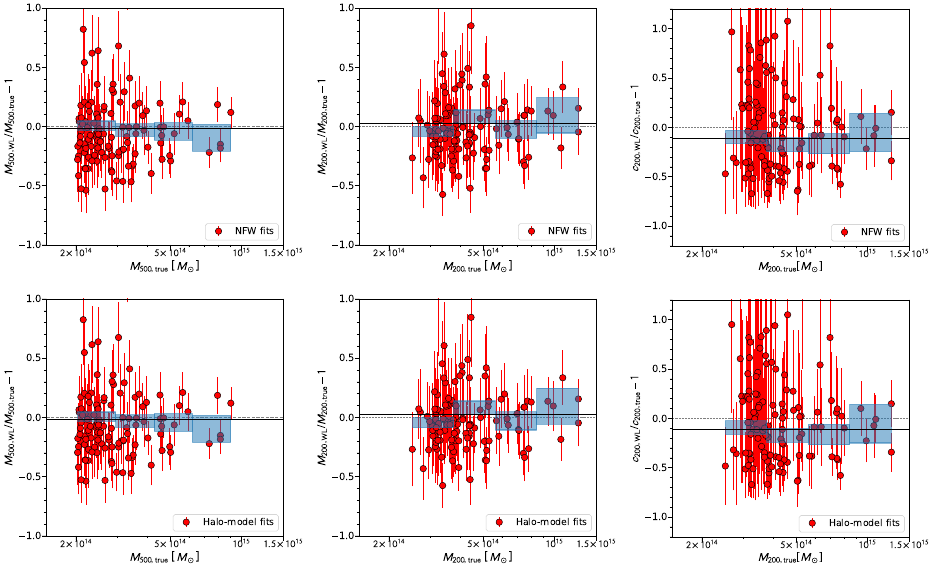} 
 \end{center}
 \caption{Comparison of true and \WL-inferred halo quantities from synthetic \WL observations of a high-mass subsample of 116 \LCDM haloes with $M_{500,\mathrm{true}} > 2\times 10^{14}\Msun$ at $z=0.25$, selected from a dark-matter-only BAHAMAS run in a \WMAP 9-year cosmology. The left, middle, and right panels show the fractional residuals in $M_{500}$, $M_{200}$, and $c_{200}$, respectively. The upper panels show the results obtained with the NFW model, while the lower panels show those obtained with the halo model. The red points with error bars show the individual halo fits. The horizontal dotted line indicates zero bias. In each panel, the horizontal solid line indicates the weighted geometric mean residual for the full high-mass subsample, and the shaded blue boxes show the weighted geometric mean residuals and associated $1\sigma$ uncertainties in four logarithmically spaced bins of the corresponding true halo mass.}
 \label{fig:bahamas_comp} 
\end{figure*}

In the lower panels of Figure~\ref{fig:bahamas_comp}, we compare the recovered parameters, $M_{500,\mathrm{WL}}$, $M_{200,\mathrm{WL}}$, and $\cwl$, with their true input values. The weighted geometric mean ratios for the high-mass subsample are $\langle M_{200,\mathrm{WL}}/M_{200,\mathrm{true}}\rangle_\mathrm{g}=1.025\pm 0.033$, $\langle M_{500,\mathrm{WL}}/M_{500,\mathrm{true}}\rangle_\mathrm{g}=0.985\pm 0.029$, and $\langle \cwl/c_{200,\mathrm{true}}\rangle_\mathrm{g}=0.891\pm 0.048$. The projected halo-model results are therefore consistent with the corresponding NFW results at the sub-percent level.

At the representative halo mass and redshift of the CHEX-MATE--\amalgam sample, the maximum fitting radius, $R_\mathrm{max}=3\Mpch$, corresponds to about $1.9\,r_{200}$ in comoving units and is therefore below the adopted truncation radius, $r_\mathrm{t}=3\,r_{200}$. Over this radial range, the projected NFW and BMO-truncated halo-model predictions for $\Delta\Sigma(R)$ are expected to be nearly indistinguishable, as illustrated in Figure~\ref{fig:stackgt} and consistent with \citet{Oguri+Hamana2011}. This accounts for the negligible differences between the NFW and halo-model fits shown in Figure~\ref{fig:bahamas_comp}, with sub-percent-level shifts in the inferred mass and concentration. For lower-mass group-scale haloes, such as those probed in the HSC-XXL analysis of \citet{Umetsu2020xxl}, the same fixed comoving fitting range extends to larger multiples of $r_{200}$, so the relative differences between the projected NFW and halo-model descriptions become more visible; even in that regime, however, the resulting differences in recovered mass and concentration were found to be only at the $1$--$3\percent$ level (see Table~6 of \citealt{Umetsu2020xxl}).

We therefore adopted the NFW model as the baseline description for the \WL mass and concentration measurements of our CHEX-MATE--\amalgam sample. The halo-model test shows that including the BMO truncation and 2-halo term does not materially change the inferred masses or concentrations over the radial range used here.

\begin{figure*}[tbp]
 \begin{center}
  \includegraphics[width=0.9\textwidth,angle=0,clip]{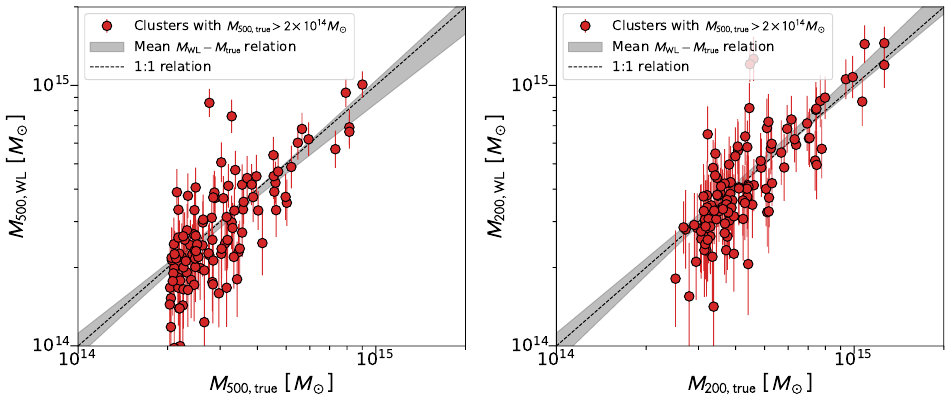} 
 \end{center}
 \caption{\WL-inferred mass vs. true mass relation at $z=0.25$, calibrated from NFW modelling of the synthetic \WL data shown in Figure~\ref{fig:bahamas_comp}. The left and right panels show the results for $M_{500}$ and $M_{200}$, respectively. In each panel, the red circles with error bars represent individual haloes, while the shaded grey region shows the marginalised $1\sigma$ credible interval for the mean relation inferred from the \lira regression.}
 \label{fig:bahamas_massbias}
\end{figure*}

\end{appendix} 
\end{document}